%% file: durbin_main.tex
\documentclass[12pt,a4paper,notitlepage,fleqn]{article}

\usepackage{setspace}
\usepackage[dvipdfmx,margin=1in]{geometry}
\usepackage[symbol]{footmisc}

\usepackage[T1]{fontenc}
\usepackage[utf8]{inputenc}
\usepackage{authblk}



\title{Finite-Sample Properties of Model Specification Tests for Multivariate Dynamic Regression Models}

\author{Koichiro Moriya$^{a}$\thanks{\scriptsize Corresponding Author. E-mail: moriya.koichiro@keio.jp, Tel/Fax: +81-466-49-3406.} \ and \ Akihiko Noda$^{b}$

{\scriptsize ${}^{a}$ \it Graduate School of Media and Governance, Keio University, 5322 Endo, Fujisawa, Kanagawa 252-0882, Japan} 

{\scriptsize ${}^{b}$ \it School of Commerce, Meiji University, 1-1 Kanda-Surugadai, Chiyoda-ku, Tokyo 101-8301, Japan}}

\date{The latest version of this paper is available at\\
\url{https://arxiv.org/pdf/2601.21272}}

\vspace{5mm}

\renewcommand\thefootnote{\arabic{footnote}}

\usepackage{graphicx,pdflscape}

\usepackage[]{natbib}
\usepackage{xcolor}
\usepackage{amsmath,amsthm,amssymb}
\usepackage{ascmac}
\usepackage[option]{colortbl}
\usepackage{multirow}
\usepackage{lscape}%
\usepackage{subfigmat}%
\usepackage{pifont}%
\usepackage{arydshln}
\usepackage{marvosym}
\usepackage{ulem}
\usepackage{url}

\PassOptionsToPackage{hyphens}{url}
\usepackage[
            bookmarkstype=toc,
	    colorlinks=true, 
	    pdfborder={0 0 1}, 
	    bookmarks=true, 
	    bookmarksnumbered=true, 
	    citecolor=blue]{hyperref}
\bibpunct{(}{)}{;}{a}{}{,}
\usepackage{here}
\usepackage{booktabs}
\usepackage{mathrsfs}
\usepackage{mathtools}
\usepackage{algorithm}
\usepackage{algpseudocode}
\usepackage{caption}

\newtheoremstyle{customstyle}
  {1em}
  {1em}
  {\itshape}
  {}
  {\bfseries}
  {:}
  {0.5em}
  {}
\theoremstyle{customstyle}

\newtheorem{theorem}{Theorem}
\newtheorem{lemma}{Lemma}
\newtheorem{proposition}{Proposition}

\newtheorem{corollary}{Corollary}

\newtheoremstyle{normalstyle}
  {1em}
  {1em}
  {\normalfont}
  {}
  {\bfseries}
  {:}
  {0.5em}
  {}
\theoremstyle{normalstyle}
\newtheorem{remark}{Remark}

\newtheorem{assumption}{Assumption}

\newcommand{\bm}[1]{\mbox{\boldmath{$#1$}}}

\newcommand{\citetapos}[1]{\citeauthor{#1}'s \citeyearpar{#1}}
\newcommand{\citeapos}[2]{\citeauthor{#1}'s (\citeyear{#2})}
\newcommand{\ex}{{\mathbb{E}}}
\newcommand{\cov}{{\mathbb{C}\rm{ov}}}
\newcommand{\var}{{\mathbb{V}\rm{ar}}}

\let\originalleft\left 
\let\originalright\right
\renewcommand{\left}{\mathopen{}\originalleft}
\renewcommand{\right}{\mathclose{}\originalright}

\def\ve #1{{\mbox{\boldmath $#1$}}}

\DeclareMathOperator*{\vecop}{vec}

\DeclareMathOperator{\rank}{rank}

\begin{document}

\begin{titlepage}

\renewcommand{\thepage}{}
\renewcommand{\thefootnote}{\fnsymbol{footnote}}

\maketitle

\vspace{-10mm}

\noindent
\hrulefill

\noindent
{\bfseries Abstract:} We propose a new model specification test for multiple-equation systems with cross-equation error and dynamic regressor--error dependences. Conventional tests often rely on exogeneity conditions strong enough to ensure consistency of the OLS estimator. These exogeneity conditions are violated when regressors and errors are dynamically dependent, rendering conventional model specification tests invalid. To address these limitations, we clarify the relationship among alternative exogeneity conditions, characterize the consistency of competing multiple-equation estimators, and propose a generalized Durbin estimator for multiple-equation systems with an intercept, cross-equation error and regressor--error dependences. We show that our estimator remains consistent under the weakest exogeneity condition. We then derive its asymptotic distribution and construct Wald tests. Our Monte Carlo experiments confirm that the bootstrap-based Wald test substantially improves finite-sample size control. An application of the bootstrap-based Wald test to the Fama--French multifactor models leaves the null hypothesis unrejected in cases where competing FGLS-based tests reject it.\\



\noindent
\hrulefill

\end{titlepage}

\bibliographystyle{asa}


\input{durbin_intro}

\input{durbin_model}

\input{durbin_sim}

\input{durbin_emp}

\input{durbin_con}

\input{durbin_ack}


\input{durbin_main.bbl}
\clearpage

\input{durbin_table}



\end{document}

%% file: durbin_intro.tex
\section{Introduction}\label{sec:durbin_intro}
Economic phenomena typically involve multiple variables that are simultaneously determined and interact with one another. To analyze these interdependencies, economists commonly employ multiple-equation regression frameworks, such as the seemingly unrelated regression (SUR) model introduced by \citet{zellner1962eme}. When error terms are correlated across equations, SUR and related multiple-equation frameworks provide a natural way to account for such dependence. For example, in systems of household demand equations, unobservable household-specific attributes may jointly influence the demand for multiple commodities, generating cross-equation dependence in the error terms (e.g., \citet{fiebig2001cte,henley1994tue}). Likewise, when a cost equation is estimated jointly with cost share equations, unobservable firm-specific conditions may simultaneously influence both input allocation and total costs, inducing cross-equation dependence in the error terms (e.g., \citet{kumbhakar1996tct,brown1995sse}). Ignoring such cross-equation dependence can lead to size distortions and misleading inference; see \citet{laitinen1978wdh,bewley1983trl,sarafidis2012csd}. Consequently, model specification tests that explicitly account for cross-equation dependence in the error terms are essential for rigorous evaluation.

This consideration is particularly relevant in the evaluation of asset pricing models. Asset returns are jointly affected by common economic and financial shocks, generating substantial co-movement and dependence across assets (e.g., \citet{cieslak2021css,asgharian2023eus,bali2017ieu}). Such cross-sectional dependence is often particularly pronounced during episodes of severe market stress, including financial crises, pandemics, and periods of heightened geopolitical risk (e.g., \citet{longin2001eci,bae2003nam,diebold2009mfa,zhang2020fmg,umar2022iru,jalloul2023emc}). When the common dependence across asset returns is not fully captured by the asset pricing model under evaluation, the unexplained common component remains in the residuals, causing the resulting error terms to exhibit substantial cross-equation dependence. Accordingly, model specification tests that explicitly account for the cross-equation error dependence are essential for evaluating asset pricing models.

To this end, \citet{gibbons1989teg} developed a multivariate test (the GRS test) to assess the validity of asset pricing models. The GRS test remains widely used in empirical studies; see, for example, \citet{fama1993crf,fama2015ffa,fama2016daf,fama2017itf,fama2018cf,fama2020ccs} and \citet{cakici2013svm}. However, it is derived under the assumption of independently and identically distributed multivariate normal errors. This assumption may be violated in practice, because asset returns often exhibit fat tails, skewness, and other departures from normality (e.g., \citet{cont2001epa,harvey2002csa,huang2012edr}). Accordingly, even after controlling for common risk factors, the error terms in asset pricing models may still exhibit departures from normality. Indeed, \citet{affleck1989nta} and \citet{zhou1993apt} show that violations of normality can lead to substantial size distortions in the GRS test by inflating its Type I error rate.

In response to the restrictive distributional assumptions underlying the GRS test, the literature has developed heteroskedasticity- and autocorrelation-robust (HAR) tests based on nonparametric estimators of the long-run covariance matrix, e.g., \citet{newey1987sps,newey1994als,andrews1991hac}. While attractive in empirical asset pricing because they do not require normality, conventional HAR tests can still exhibit substantial finite-sample size distortions, especially when serial dependence is strong; see \citet{haan1997pgr,muller2014hcs}. These distortions can become even more severe in multivariate asset-pricing applications when many intercept restrictions are tested jointly (\citet{ray2008pha,ray2009tcr}). This has led to a large literature on improved HAR inference, including fixed-$b$ methods of \citet{kiefer2000srt,kiefer2005nat}, exponentiated-kernel or fixed-$p$ approaches of \citet{phillips2006sde,phillips2007lrv}, loss-function-based bandwidth selection of \citet{sun2008obs}, and more recent practical and optimality-oriented refinements of \citet{lazarus2018hir,lazarus2021spt}.

Despite these improvements, HAR tests still suffer from a fundamental limitation. Because they are typically built on OLS estimators, their validity ultimately depends on exogeneity conditions strong enough to ensure consistency of the underlying OLS estimator. This issue is well known in the simultaneous-equations literature, where simultaneity induces correlation between regressors and error terms, rendering OLS estimator inconsistent (\citet{haavelmo1943sis,koopmans1945ses}). More generally, in time-series environments, error terms may exhibit both serial dependence and cross-equation dependence (\citet{wagner2020fmo}), while regressors may also be dynamically dependent on the error terms, thereby violating exogeneity conditions (\citet{engle1983exo,boswijk1997lamt}). When such dynamic regressor--error dependence is present, the consistency of the underlying OLS estimator is no longer guaranteed, and HAR tests may no longer deliver valid inference.

To address these limitations, several recent studies have developed alternative estimation frameworks for regressions with dynamic regressor--error dependence, including \citet{perron2026fgls,baillie2024rit}. However, the related literature relies on different notions of exogeneity, and the relationships among them have not been fully clarified. In particular, it remains unclear how these alternative conditions are related to one another, which of them are stronger than others, and how they determine the validity of competing estimators in multiple-equation settings. Moreover, these studies are confined to single-equation regressions and exclude an intercept term. Consequently, these frameworks do not directly apply to multivariate model specification tests that require joint restrictions across equations to be assessed while allowing for cross-equation error dependence. This issue arises in a wide range of settings, including, for example, asset pricing tests of whether pricing errors are jointly zero across assets.

This paper makes five contributions to the literature on multivariate model specification testing. First, we clarify the relationships among the alternative exogeneity conditions considered in the related literature. Second, within the unified two-step estimation and inference framework of \citet{murphy1985eit,pagan1986sre}, we clarify the asymptotic properties of competing estimators in multiple-equation settings under dynamic regressor--error dependence. Third, we generalize \citetapos{durbin1970tsc} regression to a multiple-equation system with an intercept, allowing for both cross-equation error dependence and dynamic regressor--error dependence. Within this generalized framework, we show that the generalized Durbin estimator remains consistent under the weakest exogeneity condition considered and derive its asymptotic distribution. Fourth, building on this asymptotic theory, we develop a Wald test for model specification testing. Last, to improve finite-sample performance, we propose a bootstrap Wald test based on the generalized Durbin estimator.

The remainder of the paper is organized as follows. Section \ref{sec:durbin_model} introduces a multiple-equation regression model with joint time-series dynamics for the regressors and error terms. It clarifies the relationships among the relevant exogeneity conditions, studies the inconsistency of OLS and alternative GLS-type estimators under dynamic regressor--error dependence, and establishes the asymptotic properties of the generalized Durbin estimator and develops a bootstrap-corrected Wald test for joint specification testing. Section \ref{sec:durbin_sim} presents Monte Carlo evidence on the finite-sample performance of the competing procedures. Section \ref{sec:durbin_emp} provides an empirical application. Section \ref{sec:durbin_con} concludes.

\subsection*{Related literature}
The problem considered in this paper, namely how to conduct model specification testing for a multiple-equation system under dynamic regressor--error dependence, does not fall within a single strand of literature but instead lies at the intersection of at least three closely related strands. First, in the presence of dynamic regressor--error dependence, the consistency of an estimator depends on the exogeneity condition imposed. We therefore begin by reviewing the literature on exogeneity conditions in time-series regressions. Second, differences in exogeneity conditions are closely linked to the validity regions of competing estimators. We then review the literature on estimation under dynamic regressor--error dependence. Third, these estimators often take the form of two-step procedures involving auxiliary parameter estimation. When estimation takes this form, valid statistical inference requires that first-step estimation error be properly taken into account. We therefore also review the literature on variance correction that properly accounts for first-step estimation error.

The consistency of an estimator depends fundamentally on the exogeneity condition assumed. In cross-sectional environments, it is often sufficient to focus on the contemporaneous relationship between regressors and error terms. In time-series environments, by contrast, one must also account for the dynamic dependence between regressors and error terms. Accordingly, the time-series regression literature has developed alternative exogeneity conditions based on orthogonality restrictions between regressors and error terms, such as strict exogeneity and present-and-past exogeneity, e.g. \citet{engle1983exo,stock2019ite,mayer2020cts,mikusheva2025lrw}. However, because the error term in a time-series regression can itself be represented as a linear combination of current and past innovations, imposing conditions on the error term alone does not necessarily make clear to which shocks the regressors are related.

In light of this observation, \citet{perron2026fgls} define exogeneity with respect to the underlying innovations rather than the error term itself. By contrast, \citet{baillie2024rit} analyze the joint process of regressors and error terms through a Wold/VAR representation and formulate exogeneity as restrictions on their joint dynamics, in particular on lagged feedback and the contemporaneous innovation covariance structure. Although a variety of exogeneity conditions have been proposed to determine estimator consistency under dynamic regressor--error dependence, their relative strength and hierarchical relationships remain unclear. Clarifying these relationships is therefore essential for characterizing the validity regions of existing estimators.

These differences in exogeneity conditions are directly relevant for the validity of competing estimators. In time-series regressions, whether the OLS estimator is consistent depends on the exogeneity condition imposed. Violations of the exogeneity condition have traditionally motivated the use of alternative estimators, most notably instrumental variables (IV) and generalized method of moments (GMM), to achieve consistent estimation under endogeneity; see, for example, \citet{hayashi1983nee,hansen1982lsp,hansen1982ivp,cumby1983tst}. More recently, the literature has also explored alternative approaches that seek to restore consistency by explicitly exploiting the dynamic dependence between regressors and error terms. \citet{perron2026fgls} reformulate exogeneity in terms of the innovations underlying the error process and show that their proposed FGLS estimator (hereafter, the FD estimator) remains consistent even when OLS is inconsistent. \citet{baillie2024rit} extend the analysis to the joint dynamics of regressors and error terms, and show that the \citetapos{durbin1970tsc} estimator remains consistent under a broader range of dependence structures than the FD estimator. Taken together, these studies suggest that, when endogeneity arises primarily from structured dynamic regressor--error dependence, consistent estimation may in some cases be recovered by exploiting that dependence directly, rather than by relying on IV- or GMM-based procedures. However, these studies are confined to single-equation regressions and do not accommodate an intercept term. Consequently, they do not directly address joint hypothesis testing in multiple-equation systems with cross-equation error dependence.

Estimators designed to handle dynamic regressor--error dependence often involve auxiliary parameter estimation. In such cases, the procedure can be viewed as a two-step procedure in which the first step estimates nuisance quantities and the second step uses them to construct the estimator of interest. In such two-step procedures, first-step estimation error can affect the asymptotic distribution of the second-step estimator at first order. Consequently, naive plug-in variance formulas that ignore first-step estimation error generally do not deliver valid inference (\citet{newey1984mmi,cattaneo2019tse}). For example, the FD estimator of \citet{perron2026fgls} relies on auxiliary parameter estimation, yet its asymptotic variance does not explicitly account for first-step estimation error. If that error contributes at first order, the resulting variance formula is misspecified, and inference based on it may therefore be invalid.\footnote{In the \href{https://at-noda.com/appendix/durbin_appendix.pdf}{Online Appendix} A.1, we investigate the finite-sample size consequences of ignoring first-step estimation error in inference under single-equation environments considered in the existing literature, and confirm that the resulting naive variance formulas can produce non-negligible over-rejection.} In this regard, \citet{murphy1985eit,pagan1986sre} provide a unified treatment of variance correction that properly accounts for first-step estimation error. Building on this literature, we develop estimators for regressions with dynamic regressor--error dependence in a unified two-step estimation and inference framework.

%% file: durbin_model.tex
\section{Model and Test}\label{sec:durbin_model}

\subsection{Setup}
We consider the following linear regressions:
\begin{equation}
 y_{i,t}=\alpha_i + \ve{x}_{i,t}^\prime\ve{\beta}_i + u_{i,t},\quad i=1,\ldots,N,\quad t=1,\ldots,T, \label{eq_2.1}
\end{equation}
where $y_{i,t}$ is the scalar dependent variable, $\ve{x}_{i,t}=(x_{i,t,1},x_{i,t,2},\ldots,x_{i,t,k_i})^\prime$ is a $k_i\times 1$ vector of regressors, and $u_{i,t}$ is a scalar disturbance. The number of regressors $k_i$ is allowed to vary across equations.

Let $\ve{y}_t=(y_{1,t},\ldots,y_{N,t})^\prime$, $\ve{u}_t=(u_{1,t},\ldots,u_{N,t})^\prime$ with $\ex[\ve{u}_t]=\ve{0}$, and define $\ve{\alpha}=(\alpha_{1},\ldots,\alpha_{N})^\prime$ and $\ve{\beta}=(\ve{\beta}_{1}^\prime,\ldots,\ve{\beta}_{N}^\prime)^\prime$, where $k:=\sum_{i=1}^N k_i$. Further, define the block-diagonal regressor matrix
\[
 \ve{X}_t:=\operatorname{blkdiag}(\ve{x}_{1,t},\ldots,\ve{x}_{N,t})
=
 \begin{bmatrix}
  \ve{x}_{1,t} & \ve{0} & \cdots & \ve{0}\\
  \ve{0} & \ve{x}_{2,t} & \cdots & \ve{0}\\
  \vdots & \vdots & \ddots & \vdots\\
  \ve{0} & \ve{0} & \cdots & \ve{x}_{N,t}
 \end{bmatrix}
 \in\mathbb{R}^{k\times N}.
\]
Stacking the $N$ equations, we can write the system compactly as
\[
\begin{bmatrix}
 y_{1,t}\\
 y_{2,t}\\
 \vdots\\
 y_{N,t}\\
\end{bmatrix}=
\begin{bmatrix}
 \alpha_{1}\\
 \alpha_{2}\\
 \vdots\\
 \alpha_{N}\\
\end{bmatrix}+
 \begin{bmatrix}
  \ve{x}_{1,t}^\prime & \ve{0} & \cdots & \ve{0}\\
  \ve{0} & \ve{x}_{2,t}^\prime & \cdots & \ve{0}\\
  \vdots & \vdots & \ddots & \vdots\\
  \ve{0} & \ve{0} & \cdots & \ve{x}_{N,t}^\prime
 \end{bmatrix}
\begin{bmatrix}
 \ve{\beta}_{1}\\
 \ve{\beta}_{2}\\
 \vdots\\
 \ve{\beta}_{N}\\
\end{bmatrix}+
\begin{bmatrix}
 u_{1,t}\\
 u_{2,t}\\
 \vdots\\
 u_{N,t}\\
\end{bmatrix},
\] 
equivalently
\begin{equation}
 \ve{y}_t=\ve{\alpha} + \ve{X}_t^\prime\ve{\beta} + \ve{u}_t ,\quad t=1,\ldots,T. \label{eq_2.2}
\end{equation}

To describe the joint second-order dynamics of regressors and disturbances, let $\ve{x}_t=(\ve{x}_{1,t}^\prime,\ldots,\ve{x}_{N,t}^\prime)^\prime$ with $\ex[\ve{x}_t]=\ve{\mu}_{x}$, and define
\[
 \ve{z}_t=
\begin{bmatrix}
 \ve{x}_t\\
 \ve{u}_t\\
\end{bmatrix} \in \mathbb{R}^{m}, \qquad m:= k + N,
\]
with mean $\ve{\mu}_z:=[\ve{\mu}_x^\prime,\ \ve{0}^\prime]^\prime = \ex[\ve{z}_t]$. Let $\bar{\ve{z}}_t:=\ve{z}_t-\ve{\mu}_z$.

We work in the Hilbert space $L^2(\Omega,\mathscr{F},\mathbb{P};\mathbb{R}^{m})$ equipped with inner product $\langle U,V\rangle=\ex[U^\prime V]$. To define the one-step-ahead linear innovation of $\{\bar{\ve z}_t\}$, define
\[
 \mathscr{H}_{t-1}
 :=
 \overline{
 \left\{
 \sum_{j=1}^{p} \ve{C}_j \bar{\ve z}_{t-j}
 \;:\;
 p<\infty,\;
 \ve{C}_j\in\mathbb{R}^{m\times m}
 \right\}
 },
\]
that is, the closed linear subspace generated by all finite matrix-coefficient linear combinations of past values of $\{\bar{\ve z}_t\}$. Let $\mathscr{P}_{t-1}:L^2(\Omega,\mathscr{F},\mathbb{P};\mathbb{R}^m)\to \mathscr{H}_{t-1}$ denote the $L^2$-orthogonal projection onto $\mathscr{H}_{t-1}$. The one-step-ahead linear innovation is then defined by
\[
 \ve{\varepsilon}_t
 :=
 \bar{\ve z}_t-\mathscr{P}_{t-1}[\bar{\ve z}_t]
 =
 \begin{bmatrix}
  \ve{\varepsilon}_{x,t}\\
  \ve{\varepsilon}_{u,t}
 \end{bmatrix}.
\]
By construction, $\ve{\varepsilon}_t$ is orthogonal to every element of $\mathscr{H}_{t-1}$ and therefore represents the component of $\bar{\ve z}_t$ that cannot be linearly predicted from the past of the system. Our definition formalizes the idea in \citet{sims1980mr} that an innovation is a new component of the system, namely, the part that is not predicted from its past.

In general, the closed linear subspace $\mathscr{H}_{t-1}$ does not exclude infinite-order linear combinations of past values of $\{\bar{\ve z}_t\}$. However, in a finite sample, an unrestricted infinite-order linear predictor is not directly operational. In our joint time-series representation, the innovation covariance matrix governs contemporaneous dependence between regressor-side and disturbance-side innovations, whereas lagged dependence in the system governs dynamic regressor--disturbance dependence. By treating $\ve{\varepsilon}_{x,t}$ and $\ve{\varepsilon}_{u,t}$ as distinct contemporaneous shock blocks, we separate instantaneous shock dependence from lagged transmission mechanisms. This interpretation is consistent with the literature in which distinct components of a joint system are modeled as separate shock blocks; see, for example, \citet{cushman1997imp,zha1999brs}. The following assumption therefore provides a transparent benchmark that makes the dependence structure of the joint system easier to interpret and analyze.

\begin{assumption}[Finite-predictor exactness at lag $p_0$] \label{assumption:1}\hfill
\begin{description}
\item[{[1]}] The centered joint process $\{\bar{\ve z}_t\}$ is covariance-stationary with $\ex[\|\bar{\ve{z}}_t\|^2]<\infty$ and is purely nondeterministic, and there exists a finite integer $p_0\ge 1$ such that
 \[
  \mathscr{P}_{t-1}[\bar{\ve z}_t]
  =
  \mathscr{P}_{t-1}^{(p_0)}[\bar{\ve z}_t],
 \]
where $\mathscr{P}_{t-1}^{(p_0)}$ denotes the $L^2$-orthogonal projection onto
 \[
  \mathscr{H}_{t-1}^{(p_0)}
  :=
  \left\{
   \sum_{j=1}^{p_0} \ve{C}_j \bar{\ve z}_{t-j}
   \;:\;
   \ve{C}_j\in\mathbb{R}^{m\times m}
  \right\}.
 \]
\item[{[2]}] The innovation sequence $\{\ve{\varepsilon}_t\}$ is i.i.d. with $\ve{\varepsilon}_t \sim i.i.d.\,(\ve{0},\ve{\Sigma})$ and $\ex[\|\ve{\varepsilon}_t\|^{4+2\delta}]<\infty$ for some $\delta>0$, where $\ve{\Sigma}$ is positive definite. In addition, for each $t$, $\ve{\varepsilon}_{x,t}$ and $\ve{\varepsilon}_{u,t}$ are independent, e.g. $\ve{\varepsilon}_{x,t} \perp \ve{\varepsilon}_{u,t}$.
\end{description}
\end{assumption}

Under Assumption~\ref{assumption:1}-[1], the centered joint process $\{\bar{\ve z}_t\}$ is covariance-stationary and purely nondeterministic. Hence, by the Wold decomposition, $\bar{\ve z}_t$ admits a unique orthogonal VMA($\infty$) representation with no deterministic component. Moreover, finite-predictor exactness at lag $p_0$ implies that $\bar{\ve z}_t$ admits a VAR($p_0$) innovations representation and can also be interpreted as practical approximations to the Wold representation for a broad class of stationary processes (\citet{lutkepohl2005nim}). Under Assumption~\ref{assumption:1}-[2], the innovation covariance matrix is block diagonal:
\[
 \ve{\Sigma}
 =
 \ex[\ve{\varepsilon}_t\ve{\varepsilon}_t^\prime]
 =
 \begin{bmatrix}
  \ve{\Sigma}_{xx} & \ve{0}\\
  \ve{0} & \ve{\Sigma}_{uu}
 \end{bmatrix},\quad
\ve{\Sigma}_{xx}\in\mathbb{R}^{k\times k},\quad
\ve{\Sigma}_{uu}\in\mathbb{R}^{N\times N}.
\]
The following proposition formalizes these implications and establishes the stability of the resulting VAR($p_0$) representation.

\begin{proposition}\label{proposition:1}
Suppose Assumption~\ref{assumption:1} holds. Then:
\begin{description}
 \item[(i)] Let $\ve{\varepsilon}_t:=\bar{\ve{z}}_t-\mathscr{P}_{t-1}[\bar{\ve{z}}_t]$. The centered process $\{\bar{\ve z}_t\}$ admits the Wold representation
 \begin{equation}
  \bar{\ve{z}}_t=\sum_{i=0}^{\infty}\ve{\Xi}_i\,\ve{\varepsilon}_{t-i}\quad\text{in }L^2,\qquad
  \ve{\Xi}_0=\ve{I}_m, \label{eq:VMA}
 \end{equation}
 where $\{\ve\varepsilon_t\}$ is (second-order) white noise: $\ex[\ve\varepsilon_t]=\ve0$, $\ex[\ve\varepsilon_t\ve\varepsilon_s^\prime]=\ve0$ for $s\neq t$, and $\ex[\ve\varepsilon_t\ve\varepsilon_t^\prime]=\ve{\Sigma}$ (constant in $t$). Each $\ve{\Xi}_i\in\mathbb{R}^{m\times m}$ admits the block partition
 \[
 \ve{\Xi}_i=
 \begin{bmatrix}
 \ve{\Xi}_{xx,i} & \ve{\Xi}_{xu,i}\\
 \ve{\Xi}_{ux,i} & \ve{\Xi}_{uu,i}
 \end{bmatrix},
 \quad
 \ve{\Xi}_{xx,i}\in\mathbb{R}^{k\times k},\ 
 \ve{\Xi}_{xu,i}\in\mathbb{R}^{k\times N},\ 
 \ve{\Xi}_{ux,i}\in\mathbb{R}^{N\times k},\ 
 \ve{\Xi}_{uu,i}\in\mathbb{R}^{N\times N}.
 \]

\item[(ii)] There exist matrices $\{\ve{\Psi}_j\}_{j=1}^{p_0}\subset\mathbb{R}^{m\times m}$ such that the centered process admits a VAR($p_0$) innovations representation
\begin{equation}
 \bar{\ve z}_t=\sum_{j=1}^{p_0}\ve{\Psi}_j\bar{\ve z}_{t-j} + \ve{\varepsilon}_t
 \quad\text{in }L^2. \label{eq:VARp}
\end{equation}
Each coefficient matrix $\ve{\Psi}_j$ admits the conformable block partition
\[
 \ve{\Psi}_j=
 \begin{bmatrix}
  \ve{\Psi}_{xx,j} & \ve{\Psi}_{xu,j}\\
  \ve{\Psi}_{ux,j} & \ve{\Psi}_{uu,j}
 \end{bmatrix},
 \quad
 \ve{\Psi}_{xx,j}\in\mathbb{R}^{k\times k},\
 \ve{\Psi}_{xu,j}\in\mathbb{R}^{k\times N},\ 
 \ve{\Psi}_{ux,j}\in\mathbb{R}^{N\times k},\ 
 \ve{\Psi}_{uu,j}\in\mathbb{R}^{N\times N}. 
\]
 The VAR and VMA coefficients $\{\ve{\Psi}_j\}$ and $\{\ve{\Xi}_n\}$ satisfy, for $n\ge1$,
 \[
  \ve{\Xi}_0=\ve{I}_m,\qquad 
  \ve{\Xi}_n=\sum_{j=1}^{\min\{p_0,n\}}\ve{\Psi}_j\ve{\Xi}_{n-j},\quad n\ge 1,
 \]
 equivalently $\ve{\Psi}(L)\ve{\Xi}(L)=\ve{I}_m$ with $\ve{\Psi}(L):=\ve{I}_m-\sum_{j=1}^{p_0}\ve{\Psi}_jL^j$ and $\ve{\Xi}(L):=\ve{I}_m+\sum_{n=1}^{\infty}\ve{\Xi}_nL^n$.

 \item[(iii)] The characteristic matrix polynomial associated with the VAR($p_0$) representation in part~(ii),
 \[
  \ve P(z):=\ve I_m-\sum_{j=1}^{p_0}\ve\Psi_j z^j,
 \]
 has no zeros in or on the unit disk, that is, $\det\ve P(z)\neq0$ for all $|z|\le1$. Hence the inverse filter admits the absolutely summable power series expansion
 \[
  \ve P(L)^{-1}=\ve I_m+\sum_{i=1}^{\infty}\ve\Xi_i L^i,\qquad 
  \sum_{i=0}^\infty\|\ve\Xi_i\|<\infty.
 \]

 \item[(iv)] The stacked vector process $\{(\ve y_t^\prime,\ve x_t^\prime,\ve u_t^\prime)^\prime\}$ is covariance-stationary and ergodic. In addition,
\[
\ex\|\ve x_t\|^{4+2\delta}<\infty,\qquad
\ex\|\ve u_t\|^{4+2\delta}<\infty,\qquad
\ex\|\ve y_t\|^{4+2\delta}<\infty.
\]
Since $\ve X_t$ is a deterministic reshaping of $\ve x_t$, it also follows that $\ex\|\ve X_t\|^{4+2\delta}<\infty$.
\end{description}
\end{proposition}

\begin{proof}
 See the \href{https://at-noda.com/appendix/durbin_appendix.pdf}{Online Appendix} A.4.1.
\end{proof}

Proposition~\ref{proposition:1} provides the fundamental probabilistic structure used in the subsequent analysis. Part~(i) shows that the centered joint process $\{\bar{\ve z}_t\}$ admits an orthogonal Wold representation, while part~(ii) implies that finite-predictor exactness yields a finite-order VAR innovations representation. Part~(iii) further establishes stability and absolute summability of the associated coefficients. Finally, part~(iv) delivers the covariance-stationarity, ergodicity, and moment conditions needed for the asymptotic arguments below. Accordingly, the joint system can be studied through a stable VAR representation in which contemporaneous dependence is encoded in $\ve{\Sigma}$ and dynamic dependence is encoded in the lag structure.

We adopt a reduced-form VAR($p_0$) as the benchmark because it summarizes multivariate dynamics with minimal structural assumptions and serves as a standard workhorse in empirical macroeconomics and finance (\citet{sims1980mr,stock2001var}). In asset-pricing applications, it is common for all equations to share the same regressor vector. Leading examples include the CAPM of \citet{sharpe1964cap} and \citet{lintner1965vra}, the arbitrage pricing theory (APT) of \citet{ross1976atc}, and the multi-factor models of \citet{fama1993crf,fama2015ffa,fama2016daf}. In these cases, $\ve{x}_{i,t}= \ve{x}_t$ for all $i$. This common-regressor case is nested in our general setup, and all results under Assumption~\ref{assumption:1} remain valid after a notational simplification.

\begin{remark}[Common regressors across equations]\label{remark:1}
Suppose that the regressor vector is common across equations, i.e., $\ve{x}_{i,t}:= \ve{x}_t\in\mathbb{R}^{r}$ for all $i=1,\ldots,N$. Then $k_i=r$ for each $i$, so the dimension of the stacked slope vector in the general setup is still $k=\sum_{i=1}^N k_i=Nr$. In this case, the multi-equation regression \eqref{eq_2.1}--\eqref{eq_2.2} can be written as
\[
 \ve{y}_t=\ve{\alpha} + \ve{X}_t^\prime\ve{\beta} + \ve{u}_t,
\]
where $\ve{X}_t^\prime= (\ve{I}_N\otimes \ve{x}_t^\prime) \in \mathbb{R}^{N\times k}$ and $\ve{\beta}=(\ve{\beta}_1^\prime,\ldots,\ve{\beta}_N^\prime)^\prime\in\mathbb{R}^{k}$ with $k=Nr$. Define $\ve{z}_t:=[\ve{x}_t^\prime,\ \ve{u}_t^\prime]^\prime\in\mathbb{R}^{m_c}$ with $m_c:=r+N$ and $\bar{\ve z}_t:=\ve z_t-\ve\mu_z$ as before. Under Assumption~\ref{assumption:1}, $\{\bar{\ve z}_t\}$ admits the stable VAR($p_0$) and Wold VMA($\infty$) representations
\[
 \bar{\ve{z}}_t=\sum_{j=1}^{p_0}\ve{\Psi}_j\bar{\ve{z}}_{t-j}+\ve{\varepsilon}_t,
 \qquad
 \bar{\ve{z}}_t=\sum_{i=0}^{\infty}\ve{\Xi}_i\,\ve{\varepsilon}_{t-i},
 \quad \ve{\Xi}(L)=\ve{P}(L)^{-1},
\]
where $\ve{P}(L)=\ve{I}_{m_c}-\sum_{j=1}^{p_0}\ve{\Psi}_jL^j$. The innovation covariance and VAR
coefficients admit the block partitions
\[
 \ve{\Sigma}=\ex[\ve{\varepsilon}_t\ve{\varepsilon}_t^\prime]=
 \begin{bmatrix}
  \ve{\Sigma}_{xx} & \ve{0}\\
  \ve{0} & \ve{\Sigma}_{uu}
 \end{bmatrix},\qquad
 \ve{\Psi}_j=
 \begin{bmatrix}
  \ve{\Psi}_{xx,j} & \ve{\Psi}_{xu,j}\\
  \ve{\Psi}_{ux,j} & \ve{\Psi}_{uu,j}
 \end{bmatrix},
\]
with $\ve{\Sigma}_{xx}\in\mathbb{R}^{r\times r}$, $\ve{\Sigma}_{uu}\in\mathbb{R}^{N\times N}$, $\ve{\Psi}_{xx,j}\in\mathbb{R}^{r\times r}$, $\ve{\Psi}_{xu,j}\in\mathbb{R}^{r\times N}$, $\ve{\Psi}_{ux,j}\in\mathbb{R}^{N\times r}$, and $\ve{\Psi}_{uu,j}\in\mathbb{R}^{N\times N}$.
\end{remark}

Hereafter, we establish conditions for the consistency and asymptotic normality of estimators of $(\ve{\alpha}^\prime, \ve{\beta}^\prime)^\prime$ across a variety of data-generating processes (DGPs), characterized by restrictions on $\{\ve{\Psi}_j\}_{j=1}^{p_0}$ and $\ve{\Sigma}$.

\subsection{Relationships among exogeneity conditions under dynamic regressor--disturbance dependence}

In linear regression analysis, exogeneity assumptions play a central role in determining how the stochastic relationship between regressors and error terms affects the consistency and efficiency of estimators, as well as the validity of associated test statistics. In cross-sectional settings, a contemporaneous orthogonality condition, $\ex[\ve{x}_t \ve{u}_t^\prime] = \ve{0}$, is often sufficient to characterize the consistency of the OLS estimator. In time-series environments, by contrast, contemporaneous orthogonality alone is generally insufficient to describe the forms of regressor--error dependence that are relevant for estimation and inference.

To address this issue, the literature has developed a range of exogeneity conditions. A traditional approach, following \citet{stock2019ite}, characterizes dynamic exogeneity through conditional mean restrictions linking the regressors and the error process across time. In particular, strict exogeneity requires
\begin{equation}
 \ex[\ve{u}_t \mid \sigma(\ve{x}_s : s \in \mathbb{Z})] = \ve{0},
 \label{strict_exog}
\end{equation}
so that the error term at time $t$ is mean independent of the entire past, present, and future path of the regressors. A weaker notion is present-and-past exogeneity, which requires only
\begin{equation}
 \ex[\ve{u}_t \mid \sigma(\ve{x}_t,\ve{x}_{t-1},\ve{x}_{t-2},\ldots)] = \ve{0},
 \label{present_and_past_exog}
\end{equation}
that is, the disturbance is mean independent of the contemporaneous and lagged regressors, but may still be related to future regressors. It follows immediately that strict exogeneity implies present-and-past exogeneity.

However, the error-based exogeneity conditions of \citet{stock2019ite} are not sufficiently fine-grained to distinguish orthogonality with the error process itself from orthogonality with the innovations underlying it. To make this distinction explicit, \citet{perron2026fgls} formulate exogeneity conditions in terms of the Wold innovations of the error process. In their single-equation framework, the regressors are assumed to satisfy the contemporaneous orthogonality condition $\ex[\ve{x}_t \ve{\varepsilon}_{u,t}^\prime]=\ve{0}$ as a maintained assumption, and dynamic exogeneity is then distinguished according to whether $\ve{x}_t$ is orthogonal to future or past innovations of the error process. We extend these definitions to the multiple-equation setting by using the innovation of the error process $\{\ve{\varepsilon}_{u,t}\}$ from the joint innovation process in Proposition~\ref{proposition:1}. The regressors are pre-determined if
\begin{equation}
 \ex[\ve{x}_t(\ve{\varepsilon}_{u,t+1}, \ve{\varepsilon}_{u,t+2},\ldots)^\prime]=\ve{0} \label{pre-determined}
\end{equation}
that is, $\ve{x}_t$ is uncorrelated with all future innovations of error terms. The regressors are exogenous if
\begin{equation}
 \ex[\ve{x}_t(\ve{\varepsilon}_{u,t-1},\ve{\varepsilon}_{u,t-2},\ldots)^\prime]=\ve{0} \label{exogenous}
\end{equation}
so that $\ve{x}_t$ is uncorrelated with all past innovations of error terms.

While the innovation-based conditions of \citet{perron2026fgls} are well suited to clarifying the conditions under which OLS and GLS are consistent, they do not by themselves fully capture the dynamic relationship between $\ve{x}_t$ and $\ve{u}_t$ in a joint regressor--error system. To make these forms of dynamic regressor--error dependence explicit, \citet{baillie2024rit} introduce a set of exogeneity conditions as restrictions on a joint VAR representation for $\bar{\ve{z}}_t=(\bar{\ve{x}}_t^\prime,\ve{u}_t^\prime)^\prime$.

Under the block-diagonal ($BD$) condition, the joint dynamics of regressors and error terms are block-diagonal:
\begin{equation}
 \bar{\ve{z}}_t=\sum_{j=1}^{p_0}
 \begin{bmatrix}
  \ve{\Psi}_{xx,j} & \ve{0}\\
  \ve{0} & \ve{\Psi}_{uu,j}\\
 \end{bmatrix}
 \bar{\ve{z}}_{t-j} + \ve{\varepsilon}_t,\qquad 
\ve{\Sigma}=
\begin{bmatrix}
 \ve{\Sigma}_{xx} & \ve{0}\\
 \ve{0} & \ve{\Sigma}_{uu}\\
\end{bmatrix}. 
\end{equation}
Hence, neither lagged feedback from past error terms to current regressors nor lagged feedback from past regressors to current error terms is allowed.

The GLS-exogeneity ($GEXOG$) condition relaxes the $BD$ condition by allowing feedback from past error terms to current regressors, while continuing to exclude feedback from past regressors to current error terms:
\begin{equation}
 \bar{\ve{z}}_t=\sum_{j=1}^{p_0}
 \begin{bmatrix}
  \ve{\Psi}_{xx,j} & \ve{\Psi}_{xu,j}\\
  \ve{0} & \ve{\Psi}_{uu,j}\\
 \end{bmatrix}
 \bar{\ve{z}}_{t-j} + \ve{\varepsilon}_t,\qquad 
\ve{\Sigma}=
\begin{bmatrix}
 \ve{\Sigma}_{xx} & \ve{0}\\
 \ve{0} & \ve{\Sigma}_{uu}\\
\end{bmatrix}.
\end{equation}

Finally, under the error-block-diagonal ($EBD$) condition, both directions of dynamic interaction are permitted through the VAR coefficients, while the innovation covariance matrix remains block-diagonal:
\begin{equation}
 \bar{\ve{z}}_t=\sum_{j=1}^{p_0}
 \begin{bmatrix}
  \ve{\Psi}_{xx,j} & \ve{\Psi}_{xu,j}\\
  \ve{\Psi}_{ux,j} & \ve{\Psi}_{uu,j}\\
 \end{bmatrix}
 \bar{\ve{z}}_{t-j} + \ve{\varepsilon}_t,\qquad 
\ve{\Sigma}=
\begin{bmatrix}
 \ve{\Sigma}_{xx} & \ve{0}\\
 \ve{0} & \ve{\Sigma}_{uu}\\
\end{bmatrix}. 
\end{equation}
Thus, the $BD$--$GEXOG$--$EBD$ classification makes explicit which forms of lagged regressor--error dependence are ruled out and which are allowed in the joint system, with $BD \subsetneq GEXOG \subsetneq EBD$.

Because different strands of the literature adopt different notions of exogeneity, the relationships among these conditions and their relative strength are not always transparent. Under Assumption~\ref{assumption:1}, however, the joint process $((\ve{x}_t-\ve{\mu}_x)^\prime,\ve{u}_t^\prime)^\prime$ admits a joint VAR/VMA representation, which provides a convenient framework for comparing these exogeneity conditions through the VAR coefficients $\{\ve{\Psi}_j\}$ and the innovation covariance matrix $\ve{\Sigma}$. In particular, the $BD$, $GEXOG$, and $EBD$ conditions may be viewed as structural restrictions on $(\ve{\Psi}_j,\ve{\Sigma})$ that characterize the admissible forms of dynamic feedback between regressors and error terms. The following proposition makes these relationships explicit under Assumption~\ref{assumption:1}.

\begin{proposition} \label{proposition:2}
 Under Assumption~\ref{assumption:1}, the following statements hold:
\begin{description}
 \item[(i)] Strict exogeneity and the $BD$ condition are equivalent.
 \item[(ii)] Exogenous condition is equivalent to $\ve{\Psi}_{xu,j}=\ve{0}$ for all $j=1,\ldots,p_0$.
 \item[(iii)] Pre-determined condition and the $EBD$ condition are equivalent.
 \item[(iv)] Strict exogeneity is equivalent to the conjunction of the present-and-past exogeneity condition and the exogenous condition.
\end{description}
\end{proposition}

\begin{proof}
 See the \href{https://at-noda.com/appendix/durbin_appendix.pdf}{Online Appendix} A.4.2.
\end{proof}

Proposition~\ref{proposition:2} provides a unified characterization of the exogeneity concepts developed in different strands of the literature and clarifies how these conditions are related to one another. In particular, it identifies the hierarchy among the VAR-based classes and shows how the innovation-based exogeneity conditions are positioned relative to that hierarchy.

\begin{remark}[Relationships between exogenous condition and $BD$, $GEXOG$]\label{remark:2}
Proposition~\ref{proposition:2}(ii) shows that the exogenous condition is equivalent to $\ve{\Psi}_{xu,j}=\ve{0}$ for all $j=1,\ldots,p_0$. Hence, the exogenous condition rules out lagged feedback from past disturbances to current regressors. Since the $BD$ condition imposes both $\ve{\Psi}_{xu,j}=\ve{0}$ and $\ve{\Psi}_{ux,j}=\ve{0}$ for all $j$, every DGP satisfying $BD$ also satisfies the exogenous condition. By contrast, any DGP in $GEXOG\setminus BD$ must satisfy $\ve{\Psi}_{ux,j}=\ve{0}$ for all $j$ and $\ve{\Psi}_{xu,j}\neq\ve{0}$ for at least one $j$, so it cannot satisfy the exogenous condition. Therefore, $BD$ $\subset$ exogenous condition, exogenous condition $\cap$ ($GEXOG\setminus BD$) = $\varnothing$. Equivalently, $BD = GEXOG \cap$ exogenous condition.
\end{remark}

\begin{remark}[Present-and-past exogeneity and the $BD$--$GEXOG$--$EBD$ hierarchy]\label{remark:3}
The equivalences in Proposition~\ref{proposition:2} show that the present-and-past exogeneity condition does not correspond to a single class in the VAR-based hierarchy. Rather, it overlaps with more than one class. First, present-and-past exogeneity can hold for some DGPs in $GEXOG\setminus BD$. For example, consider the scalar process
\[
 x_t=ax_{t-1}+bu_{t-1}+\varepsilon_{x,t},
 \qquad
 u_t=\varepsilon_{u,t},
\]
with suitably regular innovations. Second, present-and-past exogeneity can also hold for some DGPs in $EBD\setminus GEXOG$. For example, consider the scalar process
\[
 x_t=\frac{1}{2}x_{t-1}+\frac{3}{10}u_{t-1}+\varepsilon_{x,t},
 \qquad
 u_t=\frac{1}{5}x_{t-1}-\frac{1}{10}x_{t-2}-\frac{1}{2}u_{t-2}+\varepsilon_{u,t},
\]
where $\varepsilon_{x,t}\sim \text{i.i.d.}\ \mathcal{N}(0,\frac{11}{12})$, $\varepsilon_{u,t}\sim \text{i.i.d.}\ \mathcal{N}(0,\frac{13}{12})$. Therefore, the present-and-past exogeneity condition is neither equivalent to $BD$ nor nested within $GEXOG$. Instead, it cuts across the $BD$--$GEXOG$--$EBD$ hierarchy.
\end{remark}

Proposition~\ref{proposition:2} and Remarks~\ref{remark:2} and~\ref{remark:3} imply the relationships among the exogeneity conditions shown in the following figure.
\begin{center}
(Figure \ref{figure_exogeneity} around here)
\end{center}

While the exogenous condition and the present-and-past exogeneity condition remain useful as auxiliary concepts for relating our framework to alternative strands of the literature, they do not directly parameterize the lagged feedback structure that determines estimator consistency in the multivariate system considered below. By contrast, $BD$, $GEXOG$, and $EBD$ are defined as block restrictions on the joint VAR representation and therefore explicitly distinguish the direction and extent of dynamic regressor--disturbance dependence. For this reason, we use the $BD$--$GEXOG$--$EBD$ hierarchy as the organizing framework for assessing the validity of the multiple-equation estimators in the subsequent analysis.

\subsection{The inconsistency of the OLS estimator in multi-equation models}

In empirical applications, the ordinary least squares (OLS) estimator is routinely used to estimate the parameter vector in \eqref{eq_2.2}. For notational convenience, rewrite \eqref{eq_2.2} as
\begin{equation}
 \ve{y}_t=\ve{Z}_t^\prime\ve{\kappa}_0 + \ve{u}_t ,\quad t=1,\ldots,T, \label{eq_2.2_re}
\end{equation}
where $\ve{Z}_t^\prime=[\ve{I}_N,\ve{X}_t^\prime]\in\mathbb{R}^{N\times(N+k)}$ and $\ve{\kappa}_0=(\ve{\alpha}_0^\prime,\ve{\beta}_0^\prime)^\prime\in\mathbb{R}^{N+k}$. The OLS estimator based on \eqref{eq_2.2_re} is defined by
\begin{equation}
 \widehat{\ve{\kappa}}^{\mathrm{OLS}}
 :=
 \Big(\sum_{t=1}^T \ve{Z}_t\ve{Z}_t^\prime \Big)^{-1}
 \Big(\sum_{t=1}^T \ve{Z}_t\ve{y}_t \Big). \label{eq_ols_estimator}
\end{equation}
Although the OLS estimator in \eqref{eq_2.2_re} is widely used in time-series regressions, its consistency requires not only a full-rank condition on the regressors but also an appropriate orthogonality restriction. We begin with a standard identification condition.

\begin{assumption}\label{assumption:2}
 $\ve{Q}_Z:=\ex[\ve{Z}_t\ve{Z}_t^\prime]$ is positive definite.
\end{assumption}

\noindent
Assumption~\ref{assumption:2} is a standard full-rank condition on the regressors. It rules out exact multicollinearity and guarantees that the population normal equations admit a unique solution.

\citet{baillie2024rit} impose the $BD$ condition as a sufficient condition for OLS consistency. By contrast, \citet{perron2026fgls} use innovation-based exogeneity conditions, defined from the marginal Wold decomposition of $\{\ve{u}_t\}$, in their consistency analysis. As shown in Proposition~\ref{proposition:2}(i), under Assumption~\ref{assumption:1}, the $BD$ condition is equivalent to strict exogeneity in our framework. These conditions, however, are stronger than necessary. Relatedly, \citet{stock2019ite} emphasize present-and-past exogeneity as a key condition for OLS consistency. Covariance-based present-and-past exogeneity implies the contemporaneous orthogonality condition $\ex[\ve{X}_t\ve{u}_t]=\ve{0}$, but is not itself necessary. Thus, the exogeneity conditions used in these studies are nested within a broader class of data-generating processes under which OLS remains consistent. The next proposition characterizes the necessary and sufficient moment condition for OLS consistency in our setting and clarifies how it relates to the exogeneity classes discussed above.

\begin{proposition}\label{proposition:3}
Suppose that Assumptions~\ref{assumption:1}-\ref{assumption:2} hold. Then the OLS estimator $\widehat{\ve{\kappa}}^{\mathrm{OLS}}$ in \eqref{eq_ols_estimator} is consistent for $\ve{\kappa}_0$ if and only if $\ex[\ve{Z}_t\ve{u}_t]=\ve{0}$. Equivalently, using $\ex[\ve{u}_t]=\ve{0}$, $\widehat{\ve{\kappa}}^{\mathrm{OLS}}$ is consistent for $\ve{\kappa}_0$ if and only if $\ex[\ve{X}_t\ve{u}_t]=\ve{0}$.
\end{proposition}

\begin{proof}
 See the \href{https://at-noda.com/appendix/durbin_appendix.pdf}{Online Appendix} A.4.3.
\end{proof}

Proposition~\ref{proposition:3} shows that OLS consistency is equivalent to $\ex[\ve Z_t\ve u_t]=\ve 0$, or, since $\ex[\ve u_t]=\ve 0$, equivalently to the contemporaneous orthogonality condition $\ex[\ve X_t\ve u_t]=\ve 0$. The key issue, therefore, is how this orthogonality can arise in a dynamic environment.

Under the joint VAR/VMA representation implied by Assumption~\ref{assumption:1}, blockwise orthogonality of contemporaneous innovations does not in general guarantee the contemporaneous orthogonality condition $\ex[\ve X_t\ve u_t]=\ve 0$. Dynamic cross-dependence through the off-diagonal blocks of $\{\ve\Psi_j\}$, or equivalently through the impulse responses $\{\ve\Xi_i\}$, allows past innovations from one block to affect the current value of the other. As a result, $\ex[\bar{\ve x}_t\ve u_t^\prime]$ can be nonzero, and hence $\ex[\ve X_t\ve u_t]$ need not vanish.

The next proposition characterizes this moment condition within the joint VAR-based exogeneity classes. In particular, $BD$ is sufficient for contemporaneous orthogonality, whereas the condition need not hold under $GEXOG$. Hence, it need not hold more generally over the broader $EBD$ class either. This clarifies the precise sense in which OLS can be inconsistent in dynamic multi-equation environments.

\begin{proposition}\label{proposition:4}
Suppose Assumption~\ref{assumption:1} holds.
\begin{description}
 \item[(i)] Under $BD$, the contemporaneous orthogonality condition holds:
 $\ex[\ve X_t\ve u_t]=\ve 0$.
 \item[(ii)] The class $GEXOG$ does not in general imply the contemporaneous orthogonality condition. In particular, there exist data-generating processes in $GEXOG\setminus BD$ for which $\ex[\ve X_t\ve u_t]\neq \ve 0$.
 \item[(iii)]  Since $GEXOG\subsetneq EBD$, the class $EBD$ does not in general imply the contemporaneous orthogonality condition either. In particular, there exist data-generating processes in $EBD$ for which $\ex[\ve X_t\ve u_t]\neq \ve{0}$.
\end{description}
\end{proposition}

\begin{proof}
 See the \href{https://at-noda.com/appendix/durbin_appendix.pdf}{Online Appendix} A.4.4.
\end{proof}
Proposition~\ref{proposition:4}(ii)--(iii) do not rule out the contemporaneous orthogonality condition for particular DGPs in $GEXOG$ or $EBD$; they only show that these classes do not guarantee it uniformly.

\subsection{Two-Step Estimation and Asymptotic Variance Corrections}\label{subsec:twostep}

The previous subsection established that the exact condition for OLS consistency is the contemporaneous moment restriction $\ex[\ve X_t\ve u_t]=\ve 0$. While the $BD$ condition guarantees this restriction, the broader exogeneity classes $GEXOG$ and $EBD$ do not do so in general. Hence, in time-series environments with regressor--error dependence, OLS-based estimation and inference are not generally reliable.

This observation motivates the use of estimators that remain consistent even outside $BD$ by explicitly accounting for dynamic regressor--error dependence. However, in multiple-equation systems, parameter consistency alone is not sufficient for valid specification testing. Because disturbances may remain cross-sectionally correlated across equations, valid and efficient inference must also account for the cross-equation covariance structure.

For this reason, we consider GLS-type estimators that account for dynamic regressor--error dependence while also exploiting the cross-equation covariance structure. Because the relevant error dynamics and covariance matrices are unknown, such estimators are naturally implemented in two steps. In particular, a first step estimates the nuisance parameters governing the feasible transformation, and a second step applies feasible GLS using those estimates.

This two-step setup introduces an additional layer of uncertainty for inference. The first-step estimation error propagates into the second-step estimator, so asymptotically valid inference requires a variance formula that explicitly incorporates this additional source of uncertainty. Accordingly, this subsection develops the asymptotic distribution of such GLS-type estimators within the two-step inference framework of \citet{murphy1985eit} and \citet{pagan1986sre}.

Throughout this subsection, let $\ve\psi_{jt}(\cdot)$ ($j=1,2$) denote moment functions based on the observed data $\{(\ve y_t,\ve X_t)\}$. Let $\ve\vartheta_0$ be the nuisance parameter identified in the first step and $\ve\kappa_0$ the parameter of interest identified in the second step. The first-step estimator $\widehat{\ve\vartheta}$ is defined as a solution to
\[
  \bar{\ve\psi}_1(\ve\vartheta)
  :=
  \frac{1}{T}\sum_{t=1}^T \ve\psi_{1t}(\ve\vartheta)
  =
  \ve 0,
\]
whereas the second-step estimator $\widehat{\ve\kappa}$ is defined as a solution to
\[
  \bar{\ve\psi}_2(\ve\kappa,\widehat{\ve\vartheta})
  :=
  \frac{1}{T}\sum_{t=1}^T \ve\psi_{2t}(\ve\kappa,\widehat{\ve\vartheta})
  =
  \ve 0.
\]

Let $(\ve{\kappa}_0,\ve{\vartheta}_0)$ denote the true parameter values. Define the population Jacobian blocks, evaluated at $(\ve{\kappa}_0,\ve{\vartheta}_0)$, by
\[
  \ve{A}_{11}
  :=
  \ex\Bigg[
    \frac{\partial \ve{\psi}_{1t}(\ve{\vartheta}_0)}{\partial \ve{\vartheta}^\prime}
  \Bigg],
  \qquad
  \ve{A}_{22}
  :=
  \ex\Bigg[
    \frac{\partial \ve{\psi}_{2t}(\ve{\kappa}_0,\ve{\vartheta}_0)}{\partial \ve{\kappa}^\prime}
  \Bigg],
  \qquad
  \ve{A}_{21}
  :=
  \ex\Bigg[
    \frac{\partial \ve{\psi}_{2t}(\ve{\kappa}_0,\ve{\vartheta}_0)}{\partial \ve{\vartheta}^\prime}
  \Bigg].
\]
Since the first-step moment function $\ve{\psi}_{1t}(\ve{\vartheta})$ does not depend on $\ve{\kappa}$, there is no corresponding $\ve{A}_{12}$ block.

The two-step estimators admit the linear representations \eqref{eq:twostep_theta_expansion}--\eqref{eq:twostep_kappa_expansion}, where $\widetilde{\ve{\psi}}_{2t}$ is the adjusted second-step moment obtained by substituting the first-step linearization into the first-order expansion of the second-step moment around $(\ve\kappa_0,\ve\vartheta_0)$:
\begin{align}
  \sqrt{T}\,(\widehat{\ve{\vartheta}}-\ve{\vartheta}_0)
  &=
  -\ve{A}_{11}^{-1}\frac{1}{\sqrt{T}}\sum_{t=1}^T \ve{\psi}_{1t}(\ve{\vartheta}_0) + o_p(1),
  \label{eq:twostep_theta_expansion}\\
  \sqrt{T}\,(\widehat{\ve{\kappa}}-\ve{\kappa}_0)
  &=
  -\ve{A}_{22}^{-1}\frac{1}{\sqrt{T}}\sum_{t=1}^T \widetilde{\ve{\psi}}_{2t} + o_p(1),
  \label{eq:twostep_kappa_expansion}
\end{align}
where
\begin{equation}
  \widetilde{\ve{\psi}}_{2t}
  :=
  \ve{\psi}_{2t}(\ve{\kappa}_0,\ve{\vartheta}_0)-\ve{A}_{21}\ve{A}_{11}^{-1}\ve{\psi}_{1t}(\ve{\vartheta}_0).
  \label{eq:twostep_tildepsi2}
\end{equation}

Let $\ve\psi_{1,t}:=\ve\psi_{1t}(\ve\vartheta_0)$ and $\ve\psi_{2,t}:=\ve\psi_{2t}(\ve\kappa_0,\ve\vartheta_0)$. Define the long-run covariance (LRV) blocks of the two moment processes by
\begin{equation}\label{eq:no_hac_lrv_collapse}
  \ve{S}_{ab}:=\sum_{h=-\infty}^{\infty}\cov(\ve{\psi}_{a,t},\ve{\psi}_{b,t-h}),
  \qquad a,b\in\{1,2\}.
\end{equation}
Then
\begin{align}\label{eq:no_hac_S_tilde2tilde2}
  \ve{S}_{\widetilde{2}\widetilde{2}}
  :=&\sum_{h=-\infty}^{\infty}\cov(\widetilde{\ve{\psi}}_{2,t},\widetilde{\ve{\psi}}_{2,t-h})\nonumber\\
  =& \ve{S}_{22}
  + \ve{A}_{21}\ve{A}_{11}^{-1}\ve{S}_{11}\big(\ve{A}_{21}\ve{A}_{11}^{-1}\big)^\prime
  - \ve{A}_{21}\ve{A}_{11}^{-1}\ve{S}_{12}
  - \ve{S}_{21}\big(\ve{A}_{21}\ve{A}_{11}^{-1}\big)^\prime .
\end{align}

\begin{remark}\label{remark:4}
Equation \eqref{eq:twostep_kappa_expansion} implies that, under a central limit theorem for $T^{-1/2}\sum_{t=1}^T\widetilde{\ve\psi}_{2t}$,
\[
  \sqrt{T}\,(\widehat{\ve{\kappa}}-\ve{\kappa}_0)\xrightarrow{d}
  \mathcal{N}\big(\ve{0},\,\ve{V}\big),
  \qquad
  \ve{V}
  :=
  \ve{A}_{22}^{-1}\,\ve{S}_{\widetilde{2}\widetilde{2}}\,(\ve{A}_{22}^{-1})^\prime.
\]
\end{remark}

The naive plug-in covariance formula that treats $\widehat{\ve{\vartheta}}$ as known ignores the effect of first-step estimation error. In particular, it uses only the second-step moment $\ve{\psi}_{2t}$, together with its long-run covariance block $\ve S_{22}$, and omits the correction terms involving $\ve A_{21}\ve A_{11}^{-1}$ as well as the cross-covariance blocks $\ve S_{12}$ and $\ve S_{21}$. Remark~\ref{remark:4} shows that the second-step estimator is driven by the adjusted moment $\widetilde{\ve{\psi}}_{2t}$ rather than the original moment $\ve{\psi}_{2t}$. Accordingly, the relevant long-run covariance matrix is $\ve{S}_{\widetilde{2}\widetilde{2}}$, which generally differs from $\ve S_{22}$. The following remark is closely related to the special cases discussed in \citet{pagan1986sre}.

\begin{remark}\label{remark:5}
Under the first-order expansion \eqref{eq:twostep_kappa_expansion} and the definition \eqref{eq:twostep_tildepsi2}, the long-run covariance block $\ve{S}_{\widetilde{2}\widetilde{2}}$ admits the following simplifications under sufficient conditions.
\begin{itemize}
\item[(i)] Local insensitivity ($\ve{A}_{21}=\ve{0}$):
$\widetilde{\ve{\psi}}_{2t}=\ve{\psi}_{2t}$ and hence $\ve{S}_{\widetilde{2}\widetilde{2}}=\ve{S}_{22}$. In this case, the naive plug-in covariance that treats $\widehat{\ve\vartheta}$ as known is asymptotically valid.

\item[(ii)] No long-run cross-covariances across steps ($\ve{S}_{12}=\ve{S}_{21}=\ve{0}$):
the cross-covariance terms vanish and
\[
  \ve{S}_{\widetilde{2}\widetilde{2}}
  = \ve{S}_{22}
  + \ve{A}_{21}\ve{A}_{11}^{-1}\ve{S}_{11}\big(\ve{A}_{21}\ve{A}_{11}^{-1}\big)^\prime .
\]

\item[(iii)] Model-implied orthogonality conditions:
block-triangular (or related) moment structures, often implied by exogeneity restrictions, may yield $\ve{A}_{21}=\ve{0}$ and/or $\ve{S}_{12}=\ve{S}_{21}=\ve{0}$, leading to further simplification.
\end{itemize}
\end{remark}

\subsection{The inconsistency of OLS-based GLS estimators in multiple-equation models}

A common response to serial correlation in $\ve{u}_t$ is to adopt feasible GLS-type procedures that model and remove the error dynamics. However, although such procedures can address serial correlation under appropriate orthogonality conditions, they do not by themselves resolve the endogeneity problem highlighted above when $\ex[\ve{X}_t\ve{u}_t]\neq\ve{0}$. In particular, when the first-step OLS estimator is inconsistent, the resulting residuals do not consistently recover the true disturbance process, so subsequent GLS transformations based on those residuals need not restore consistency. Nevertheless, such procedures remain a standard workhorse in empirical time-series regressions.

In practice, these methods are typically implemented in two steps. First, one estimates \eqref{eq_2.2_re} by OLS and obtains residuals $\{\widehat{\ve{u}}_t^{OLS}\}$. Second, one fits a finite-order model for the residual dynamics (for example, equation-by-equation AR($p$) models, or a vector autoregression for the residual process), applies the \citetapos{cochrane1949als} quasi-differencing transformation based on the estimated lag polynomial, and then computes a feasible GLS estimator. We refer to such estimators as CO-type estimators. In the multiple-equation setting, \citet{nagakura2024cot} develop a multivariate CO-type estimator in which the disturbance vector is modeled as a VAR($p$) process.

Suppose first that the disturbance admits the VAR($p_0$) representation
\begin{equation}\label{eq:P5_true_VAR_u}
 \ve u_t=\sum_{j=1}^{p_0}\ve\Psi_{uu,j}\ve u_{t-j}+\ve\varepsilon_{u,t},\qquad
 \ex[\ve\varepsilon_{u,t}]=\ve 0,\quad
 \ex[\ve\varepsilon_{u,t}\ve\varepsilon_{u,t}^\prime]=\ve\Sigma_{uu},
\end{equation}
where the characteristic polynomial $\ve A_{uu}(z):=\ve I_N-\sum_{j=1}^{p_0}\ve\Psi_{uu,j}z^j$ has no zeros in or on the unit disk. Define the transformed variables $\ve{y}_{CO,t}:=\ve A_{uu}(L)\ve y_t$ and $\ve{Z}_{CO,t}^\prime:=\ve A_{uu}(L)\ve Z_t^\prime$. If $\ve A_{uu}(L)$ and $\ve\Sigma_{uu}$ were known, the corresponding infeasible GLS estimator would be
\begin{equation}\label{eq:P5_infeasible_GLS}
 \widehat{\ve\kappa}^{\tiny \mathrm{CO}}
 :=
 \Big(\sum_{t=p_0+1}^T \ve{Z}_{CO,t}\,\ve\Sigma_{uu}^{-1}\ve{Z}_{CO,t}^\prime\Big)^{-1}
 \Big(\sum_{t=p_0+1}^T \ve{Z}_{CO,t}\,\ve\Sigma_{uu}^{-1}\ve{y}_{CO,t}\Big).
\end{equation}

\begin{assumption}\label{assumption:3}
$\ve{Q}_{CO}:=\ex[\ve{Z}_{CO,t}\,\ve\Sigma_{uu}^{-1}\ve{Z}_{CO,t}^\prime]$ is positive definite.
\end{assumption}

\noindent
Assumption~\ref{assumption:3} is the GLS analogue of the full-rank condition in Assumption~\ref{assumption:2}. It rules out degeneracy of the transformed regressors and ensures that the population GLS normal equations admit a unique solution.

\begin{proposition}[Consistency region of FCO estimators]\label{proposition:5}
Suppose that Assumptions~\ref{assumption:1}--\ref{assumption:3} hold and consider the following two-step OLS-based GLS procedure (CO-type estimator):

\begin{description}
 \item[(Step 1).] Estimate \eqref{eq_2.2_re} by OLS and obtain residuals $\widehat{\ve{u}}_t^{\tiny OLS}:=\ve{y}_t-\ve{Z}_t^\prime\widehat{\ve{\kappa}}^{\tiny OLS}$.
 \item[(Step 2).] Fit a VAR($p_0$) model to $\{\widehat{\ve{u}}_t^{\tiny OLS}\}$, construct the corresponding Cochrane--Orcutt transformation, and re-estimate \eqref{eq_2.2_re} by GLS on the transformed system using $\widehat{\ve\Sigma}_{uu}^{-1}$, yielding $\widehat{\ve{\kappa}}^{\tiny \mathrm{FCO}}$.
\end{description}

Then the following statements hold.
\begin{description}
 \item[(i)] Under the $BD$ condition, $\widehat{\ve{\kappa}}^{\tiny \mathrm{FCO}}$ is consistent for $\ve{\kappa}_0$.
 \item[(ii)] Outside the $BD$ class, consistency of the FCO estimator is not guaranteed.
\end{description}
\end{proposition}

\begin{proof}
 See the \href{https://at-noda.com/appendix/durbin_appendix.pdf}{Online Appendix} A.4.5.
\end{proof}

Proposition~\ref{proposition:5} shows that the FCO estimator is guaranteed to be consistent under $BD$, where the first-step OLS estimator consistently recovers the disturbance process and hence yields a valid feasible Cochrane--Orcutt transformation. Outside the $BD$ class, however, consistency is no longer guaranteed.

Since the FCO estimator is obtained in two steps, valid inference requires an asymptotic variance formula that accounts for first-step estimation error. Because consistency of the FCO estimator is guaranteed only under $BD$, we confine attention to the asymptotic variance under $BD$ condition.

\begin{proposition}[Asymptotic irrelevance of first-step estimation error under $BD$]\label{proposition:6}
Let $T_{eff}:=T-p_0$. Suppose Assumptions~\ref{assumption:1}--\ref{assumption:3} hold, and the $BD$ condition is satisfied. Then:

\begin{description}
\item[(i)] $\sqrt{T_{eff}}\Big(\widehat{\ve\kappa}^{\mathrm{FCO}} - \widehat{\ve\kappa}^{\mathrm{CO}} \Big) =o_p(1)$. 

\item[(ii)] $\sqrt{T_{eff}} \Big(\widehat{\ve\kappa}^{\mathrm{FCO}} - \ve\kappa_0 \Big) \xrightarrow{d} \mathcal N\big(\ve 0,\ \ve{V}_{CO}\big)$, where $\ve{V}_{CO}:=\ve{Q}_{CO}^{-1}$.
\end{description}
Hence, under $BD$, $\widehat{\ve\kappa}^{\mathrm{FCO}}$ attains the GLS efficiency bound for that transformed system.
\end{proposition}

\begin{proof}
 See the \href{https://at-noda.com/appendix/durbin_appendix.pdf}{Online Appendix} A.4.6.
\end{proof}

Proposition~\ref{proposition:6} shows that, under $BD$, the FCO estimator is asymptotically equivalent to the infeasible GLS estimator for the correctly specified transformed system. Consequently, no first-step asymptotic variance correction is required. In addition, under $BD$, the FCO estimator attains the GLS efficiency bound for that transformed system. 

It follows immediately from Proposition~\ref{proposition:6} that we can conduct model specification tests for linear restrictions on $\ve{\kappa}_0$ under the $BD$ condition.

\begin{corollary}\label{corollary:1}
Let $T_{\mathrm{eff}}:=T-p_0$. Suppose Assumptions~\ref{assumption:1}--\ref{assumption:3} hold, and the $BD$ condition is satisfied. Let $\ve{R}$ be a given $q\times m$ matrix with $\mathrm{rank}(\ve{R})=q\le m$, and let $\ve{r}\in\mathbb{R}^q$. Then, under $H_0:\ve{R}\ve{\kappa}_0=\ve{r}$,
\[
 \mathcal{W}^{\mathrm{FCO}}
 :=
 T_{\mathrm{eff}}
 (\ve{R}\widehat{\ve{\kappa}}^{\mathrm{CO}}-\ve{r})^\prime
 \big[\ve{R}\widehat{\ve{V}}^{\mathrm{FCO}}\ve{R}^\prime\big]^{-1}
 (\ve{R}\widehat{\ve{\kappa}}^{\mathrm{FCO}}-\ve{r})
 \xrightarrow{d}
 \chi^2_q.
\]
\end{corollary}

\begin{proof}
 See the \href{https://at-noda.com/appendix/durbin_appendix.pdf}{Online Appendix} A.4.7.
\end{proof}

\subsection{The inconsistency of FD estimator in multiple-equation models}

The previous subsection showed that the OLS-based CO procedure is guaranteed to be consistent only under $BD$. This naturally raises the question of whether one can construct a feasible GLS estimator that remains consistent over a broader class of dynamic regressor--disturbance dependence. A leading attempt in this direction is the FD procedure proposed by \citet{perron2026fgls} for single-equation time-series regressions.

In the multiple-equation setting, the key maintained restriction underlying FD is that the disturbance depends only on its own lags. In the joint VAR for $\bar{\ve z}_t=((\ve x_t-\ve\mu_x)^\prime,\ve u_t^\prime)^\prime$, this amounts to imposing $\ve\Psi_{ux,j}=\ve 0$ for all $j$, so that lagged regressors are excluded from the $u$-equation, while lagged disturbances may still enter the regressor block through $\ve\Psi_{xu,j}\neq\ve 0$. Hence, in our notation, the maintained dynamic model of FD corresponds to the upper block-triangular region that we call $GEXOG$, together with innovation orthogonality $\ve\Sigma_{xu}=\ve 0$. Accordingly, we formulate the multivariate analogue of FD under $GEXOG$ and then characterize its consistency region.

Imposing these restrictions on $\ve{\Psi}_j$ yields
\begin{align}
  \ve{x}_t-\ve{\mu}_{x}
  =&
    \sum_{j=1}^{p_0}\ve{\Psi}_{xx,j}\big(\ve{x}_{t-j}-\ve{\mu}_{x}\big)
    + \sum_{j=1}^{p_0}\ve{\Psi}_{xu,j}\ve{u}_{t-j}
    + \ve{\varepsilon}_{x,t}, \label{eq_fglsd_x_var}\\
  \ve{u}_t
  =&
    \sum_{j=1}^{p_0}\ve{\Psi}_{uu,j}\ve{u}_{t-j}
    + \ve{\varepsilon}_{u,t}. \label{eq_fglsd_u_var}
\end{align}
where the dynamic feedback from $\ve{x}_t$ to $\ve{u}_t$ is ruled out by imposing $\ve{\Psi}_{ux,j}=\ve{0}$ for all $j$, while feedback from past disturbances to the regressors is allowed through $\ve{\Psi}_{xu,j}$.

Substituting \eqref{eq_fglsd_u_var} into \eqref{eq_2.2} and rearranging terms yields
\begin{equation}
  \ve{y}_t
  = \Bigl(\ve{I}_N-\sum_{j=1}^{p_0}\ve{\Psi}_{uu,j}\Bigr)\ve{\alpha}
     + \ve{X}_t^\prime\ve{\beta}
     - \sum_{j=1}^{p_0}\ve{\Psi}_{uu,j}\ve{X}_{t-j}^\prime\ve{\beta}
     + \sum_{j=1}^{p_0}\ve{\Psi}_{uu,j}\ve{y}_{t-j}
     + \ve{\varepsilon}_{u,t}. \label{eq_fglsd_reg}
\end{equation}
Using $\ve X_t^\prime\ve\beta=\ve B\ve x_t$ with $\ve B:=\mathrm{blkdiag}(\ve\beta_1^\prime,\ldots,\ve\beta_N^\prime)\in\mathbb R^{N\times k}$, the term $-\sum_{j=1}^{p_0}\ve\Psi_{uu,j}\ve X_{t-j}^\prime\ve\beta$ in \eqref{eq_fglsd_reg} can be rewritten as $-\sum_{j=1}^{p_0}\ve\Psi_{uu,j}\ve X_{t-j}^\prime\ve\beta =-\sum_{j=1}^{p_0}\ve\Psi_{uu,j}\ve B\,\ve x_{t-j} =\sum_{j=1}^{p_0}\ve\Delta_j\,\ve x_{t-j}$ where, at the population level, $\ve\Delta_j := -\ve\Psi_{uu,j}\ve B$, $j=1,\ldots,p_0$.

For estimation, however, we treat $\{\ve\Delta_j\}_{j=1}^{p_0}$ as unrestricted nuisance parameters, which keeps the augmented regression linear in the unknown coefficients. Under $GEXOG$, the true parameter values satisfy $\ve\Delta_j=-\ve\Psi_{uu,j}\ve B$ for each $j$, so this reparameterization does not change the population model. Rather, it embeds the $GEXOG$ restriction in a larger linear regression system, thereby avoiding the bilinear cross-parameter restriction in estimation.

With this linear reparameterization, \eqref{eq_fglsd_reg} becomes a linear regression model in which $\ve c=(\ve{I}_N-\sum_{j=1}^{p_0}\ve{\Psi}_{uu,j})\ve{\alpha}$, $\ve\beta$, $\{\ve\Psi_{uu,j}\}_{j=1}^{p_0}$, and $\{\ve\Delta_j\}_{j=1}^{p_0}$ enter linearly. Since no cross-equation parameter restrictions are imposed in this augmented representation, the first step of the FD procedure can be implemented equation by equation with OLS. Specifically, for each $i=1,\ldots,N$, we regress $y_{i,t}$ on the contemporaneous regressors in equation $i$, the lagged dependent vectors $\{\ve y_{t-j}\}_{j=1}^{p_0}$, and the corresponding lagged regressor terms. The coefficients on $\{\ve y_{t-j}\}_{j=1}^{p_0}$ across equations then deliver the row blocks of $\widehat{\ve\Psi}_{uu,1},\ldots,\widehat{\ve\Psi}_{uu,p_0}$, while those on the lagged regressor terms deliver $\widehat{\ve\Delta}_1,\ldots,\widehat{\ve\Delta}_{p_0}$. The resulting residuals, denoted by $\widehat{\ve\varepsilon}_{u,t}$, are used to estimate $\ve\Sigma_{uu}$. This leads to the following two-step FD procedure.

\begin{description}
\item[(Step 1) OLS estimation of the augmented regression.]

Estimate \eqref{eq_fglsd_reg} equation by equation by OLS. Let $\widehat{\ve{\Psi}}_{uu,1},\ldots,\widehat{\ve{\Psi}}_{uu,p_0}$ denote the matrices obtained by stacking, across equations, the estimated coefficients on $\{\ve y_{t-j}\}_{j=1}^{p_0}$. Define the first-step residuals by $\widehat{\ve{\varepsilon}}_{u,t} := \ve{y}_t - \widehat{\ve{c}} - \ve{X}_t^\prime\widehat{\ve{\beta}} - \sum_{j=1}^{p_0}\widehat{\ve{\Psi}}_{uu,j}\ve{y}_{t-j} - \sum_{j=1}^{p_0}\widehat{\ve{\Delta}}_{j}\ve{x}_{t-j}$, and estimate the innovation covariance matrix by $\widehat{\ve{\Sigma}}_{uu} := 1/(T-p_0)\sum_{t=p_0+1}^T\widehat{\ve{\varepsilon}}_{u,t}\big(\widehat{\ve{\varepsilon}}_{u,t}\big)^\prime$.

\item[(Step 2) Quasi-differencing and GLS.]

Using $\widehat{\ve{\Psi}}_{uu,1},\ldots,\widehat{\ve{\Psi}}_{uu,p_0}$ from Step 1, construct $\widehat{\ve{y}}_{\mathrm{FD},t}:=\ve{y}_t - \sum_{j=1}^{p_0}\widehat{\ve{\Psi}}_{uu,j}\ve{y}_{t-j}$, $\widehat{\ve{X}}_{\mathrm{FD},t}^\prime:=\ve{X}_t^\prime - \sum_{j=1}^{p_0}\widehat{\ve{\Psi}}_{uu,j}\ve{X}_{t-j}^\prime$, and define $\widehat{\ve{Z}}_{\mathrm{FD},t}^\prime := \bigl[\bigl(\ve{I}_N-\sum_{j=1}^{p_0}\widehat{\ve{\Psi}}_{uu,j}\bigr),\ \widehat{\ve{X}}_{\mathrm{FD},t}^\prime\bigr]$, $\ve{\kappa}:=(\ve{\alpha}^\prime,\ve{\beta}^\prime)^\prime$. The multivariate FD estimator is
\begin{equation}
  \widehat{\ve{\kappa}}^{\mathrm{FD}}
  :=
  \Biggl(\sum_{t=p_0+1}^T
           \widehat{\ve{Z}}_{\mathrm{FD},t}\,
           \widehat{\ve{\Sigma}}_{uu}^{-1}\,
           \widehat{\ve{Z}}_{\mathrm{FD},t}^\prime
  \Biggr)^{-1}
  \Biggl(\sum_{t=p_0+1}^T
           \widehat{\ve{Z}}_{\mathrm{FD},t}\,
           \widehat{\ve{\Sigma}}_{uu}^{-1}\,
           \widehat{\ve{y}}_{\mathrm{FD},t}
  \Biggr).
  \label{eq_fglsd_final}
\end{equation}
\end{description}

\begin{lemma}[Equation-by-equation properties of the augmented regression]\label{lemma:1}
Suppose Assumption~\ref{assumption:1} holds. For each $i=1,\ldots,N$, consider an augmented regression of the form $y_{i,t}=\ve w_{i,t}^\prime \ve\theta_i+\varepsilon_{u,i,t}$, where $\ve w_{i,t}$ consists of a constant, the contemporaneous regressors in equation $i$, and lagged values of $\ve y_t$ and $\ve x_t$. Then, for each $i=1,\ldots,N$, the following statements hold.

\begin{description}
\item[(i)] $\ve w_{i,t}$ is measurable with respect to $\sigma(\mathscr F_{t-1},\ve\varepsilon_{x,t})$, where $\mathscr F_{t-1}:=\sigma(\ve z_s:s\le t-1)$, and $\ex[\ve w_{i,t}\varepsilon_{u,i,t}]=\ve 0$.

\item[(ii)] $\ex\|\ve w_{i,t}\|^{2+\delta}<\infty$ and $\ex\|\ve w_{i,t}\varepsilon_{u,i,t}\|^{2+\delta}<\infty$.

\end{description}
\end{lemma}

\begin{proof}
 See the \href{https://at-noda.com/appendix/durbin_appendix.pdf}{Online Appendix} A.4.8.
\end{proof}

Lemma~\ref{lemma:1} establishes the basic stochastic properties of the augmented regression needed for equation-by-equation OLS estimation. We now impose a full-rank condition on the augmented regressors to obtain consistency of the first-step estimator.

\begin{assumption}\label{assumption:4}
For each $i=1,\ldots,N$, $\ve{Q}_{w,i}:=\ex[\ve{w}_{i,t}\ve{w}_{i,t}^\prime]$ is positive definite.
\end{assumption}

\begin{lemma}[Consistency of the equation-by-equation augmented-regression estimator]\label{lemma:2}
Let $T_{\mathrm{eff}}:=T-p_0$. Suppose Assumptions~\ref{assumption:1} and \ref{assumption:4} hold. For each $i=1,\ldots,N$, define
\[
  \widehat{\ve\theta}_i
  :=
  \Bigl(\frac{1}{T_{\mathrm{eff}}}\sum_{t=p_0+1}^T \ve w_{i,t}\ve w_{i,t}^\prime\Bigr)^{-1}
  \Bigl(\frac{1}{T_{\mathrm{eff}}}\sum_{t=p_0+1}^T \ve w_{i,t} y_{i,t}\Bigr).
\]
Then, for each $i=1,\ldots,N$, $\widehat{\ve\theta}_i \xrightarrow{p} \ve\theta_i$.
\end{lemma}

\begin{proof}
 See the \href{https://at-noda.com/appendix/durbin_appendix.pdf}{Online Appendix} A.4.9.
\end{proof}

In practice, the true lag order $p_0$ is unknown and must be estimated from the data. We can use the Bayesian information criterion (BIC) of \citet{schwarz1978edm}, which consistently selects the true lag order under standard regularity conditions.

\begin{lemma}[Consistency of BIC lag-order selection for the augmented regression]\label{lemma:3}
Suppose Assumptions~\ref{assumption:1} and \ref{assumption:4} hold, together with the regularity conditions required for consistent lag-order selection in the equation-by-equation augmented regression. Let $\mathcal P=\{1,\ldots,p_{\max}\}$ with fixed $p_{\max}\ge p_0$. For each $p\in\mathcal P$, estimate the augmented regression equation by equation using $p$ lags, and define $\widehat{\ve\Sigma}_{uu}(p)$ and $\operatorname{BIC}_T(p)$ as above. Then
\[
  \widehat p_{\mathrm{BIC}}
  :=
  \min\arg\min_{p\in\mathcal P}\operatorname{BIC}_T(p)
  \xrightarrow{p}
  p_0.
\]
\end{lemma}

\begin{proof}
 See the \href{https://at-noda.com/appendix/durbin_appendix.pdf}{Online Appendix} A.4.10.
\end{proof}

Let $\ve Z_{\mathrm{FD},t}$ denote the population filtered regressor matrix obtained by replacing $\widehat{\ve\Psi}_{uu,j}$ with $\ve\Psi_{uu,j}$ in Step 2. To ensure identification in the second-step filtered GLS regression, we impose the following full-rank condition.

\begin{assumption}\label{assumption:5}
$\ve{Q}_{\mathrm{FD}}:=\ex[\ve{Z}_{\mathrm{FD},t}\ve{\Sigma}_{uu}^{-1}\ve{Z}_{\mathrm{FD},t}^\prime]$ is positive definite.
\end{assumption}

Assumption~\ref{assumption:5} is the filtered-system analogue of the full-rank condition in Assumption~\ref{assumption:3}. It rules out degeneracy of the filtered regressors and ensures that the population GLS normal equations admit a unique solution.

\begin{proposition}[Consistency region of the FD estimator]\label{proposition:7}
Suppose that Assumption~\ref{assumption:1} and Assumptions~\ref{assumption:4}--\ref{assumption:5} hold. Then:
\begin{description}
  \item[(i)] Under the $GEXOG$ condition, $\widehat{\ve{\kappa}}^{\mathrm{FD}}$ is consistent for $\ve{\kappa}_0$. Moreover, under the $BD$ condition, $\widehat{\ve{\kappa}}^{\mathrm{FD}}$ is asymptotically equivalent to the FCO estimator $\widehat{\ve{\kappa}}^{\mathrm{FCO}}$ defined in Proposition~\ref{proposition:5}.
  
  \item[(ii)] When the $GEXOG$ condition fails, consistency of the FD estimator is not guaranteed.
\end{description}
\end{proposition}

\begin{proof}
 See the \href{https://at-noda.com/appendix/durbin_appendix.pdf}{Online Appendix} A.4.11.
\end{proof}

Proposition~\ref{proposition:7} shows that FD has a larger consistency region than FCO estimators. While CO-type procedures are guaranteed to be consistent only under $BD$, FD remains consistent under the broader $GEXOG$ condition, and under $BD$ it is asymptotically equivalent to FCO. This reflects the fact that, under $GEXOG$, the disturbance process $\{\ve u_t\}$ admits a closed VAR($p_0$) representation, so the quasi-differencing step is asymptotically correctly specified.

Although the multivariate FD estimator is consistent under $GEXOG$, it is inherently a two-step estimator. The filtered regressors used in the second-step GLS regression, as well as the feasible weighting matrix, are constructed from first-step estimates of the disturbance dynamics and innovation covariance matrix. Consequently, the second-step estimating equation inherits first-step estimation error, and the naive plug-in covariance formula that treats these generated objects as fixed is generally asymptotically invalid.

To obtain valid inference, we analyze $\widehat{\ve\kappa}^{\mathrm{FD}}$ within the general two-step framework developed in Section~\ref{subsec:twostep}. Let $\ve\vartheta^{\mathrm{FD}} := \bigl(\vecop(\ve\Psi_{uu,1})^\prime,\ldots,\vecop(\ve\Psi_{uu,p_0})^\prime\bigr)^\prime$ collect the nuisance parameters governing the disturbance dynamics. The second-step moment is given by the GLS score associated with the filtered regression generated by $\ve\vartheta^{\mathrm{FD}}$, while the first-step augmented OLS estimator admits an asymptotically linear representation. As a result, the limit distribution of $\widehat{\ve\kappa}^{\mathrm{FD}}$ is governed not by the original second-step moment alone, but by the adjusted moment
\[
  \widetilde{\ve\psi}_{2t}^{\mathrm{FD}}
  :=
  \ve\psi_{2t}^{\mathrm{FD}}
  -
  \ve A_{21}^{\mathrm{FD}}
  (\ve A_{11}^{\mathrm{FD}})^{-1}
  \ve\psi_{1t}^{\mathrm{FD}},
\]
where the Jacobian blocks are
\[
  \ve A_{11}^{\mathrm{FD}}
  :=\ex\Big[\frac{\partial \ve\psi_{1t}^{\mathrm{FD}}(\ve\vartheta_0^{\mathrm{FD}})}
  {\partial \ve\vartheta^{\mathrm{FD}\prime}}\Big],\quad
  \ve A_{22}^{\mathrm{FD}}
  :=\ex\Big[\frac{\partial \ve\psi_{2t}^{\mathrm{FD}}(\ve\kappa_0,\ve\vartheta_0^{\mathrm{FD}})}
  {\partial \ve\kappa^\prime}\Big],\quad
  \ve A_{21}^{\mathrm{FD}}
  :=\ex\Big[\frac{\partial \ve\psi_{2t}^{\mathrm{FD}}(\ve\kappa_0,\ve\vartheta_0^{\mathrm{FD}})}
  {\partial \ve\vartheta^{\mathrm{FD}\prime}}\Big].
\]
Hence the asymptotic variance takes the form $(\ve A_{22}^{\mathrm{FD}})^{-1}\ve S_{\widetilde2\widetilde2}^{\mathrm{FD}}(\ve A_{22}^{\mathrm{FD}})^{-1\prime}$, where $\ve S_{\widetilde2\widetilde2}^{\mathrm{FD}} := \sum_{h=-\infty}^{\infty}\cov\bigl(\widetilde{\ve\psi}_{2,t}^{\mathrm{FD}},\widetilde{\ve\psi}_{2,t-h}^{\mathrm{FD}}\bigr)$ is the long-run covariance matrix of the adjusted moment.

\begin{theorem}[Asymptotic distribution of the FD estimator]
\label{theorem:1}
Let $T_{\mathrm{eff}}:=T-p_0$. Suppose Assumptions~\ref{assumption:1}, \ref{assumption:4}, and \ref{assumption:5} hold, the data-generating process satisfies $GEXOG$ with correctly specified VAR order $p_0$ for $\{\ve u_t\}$, and $\ve A_{11}^{\mathrm{FD}}$ is nonsingular. Then
\[
  \sqrt{T_{\mathrm{eff}}}
  \bigl(
    \widehat{\ve\kappa}^{\mathrm{FD}}
    -
    \ve\kappa_0
  \bigr)
  \xrightarrow{d}
  \mathcal{N}\bigl(\ve 0,\ve V^{\mathrm{FD}}\bigr),\quad
\ve V^{\mathrm{FD}}
  :=
  (\ve A_{22}^{\mathrm{FD}})^{-1}
  \ve S_{\widetilde2\widetilde2}^{\mathrm{FD}}
  (\ve A_{22}^{\mathrm{FD}})^{-1\prime}.
\]
\end{theorem}

\begin{proof}
 See the \href{https://at-noda.com/appendix/durbin_appendix.pdf}{Online Appendix} A.4.12.
\end{proof}

\noindent
Explicit expressions for the Jacobian and covariance blocks entering $\ve V^{\mathrm{FD}}$ are reported in the \href{https://at-noda.com/appendix/durbin_appendix.pdf}{Online Appendix} A.2.1.

The FD estimator is a two-step estimator, since both the quasi-differenced regressors and the GLS weight are constructed from first-step estimates of the disturbance dynamics. Accordingly, whenever $\ve A_{21}^{\mathrm{FD}}\neq \ve 0$, first-step estimation error enters the first-order asymptotic variance, so a variance formula that treats these objects as known, as in \citet{perron2026fgls}, is not valid in general. In that case, valid inference requires the Murphy--Topel correction in Theorem~\ref{theorem:1}. By contrast, if $\ve A_{21}^{\mathrm{FD}}=\ve 0$ as in Remark~\ref{remark:5}(i), the first-step effect is asymptotically irrelevant at the $T_{\mathrm{eff}}^{-1/2}$ order, the correction vanishes, and $\widehat{\ve\kappa}^{\mathrm{FD}}$ attains the infeasible GLS bound.

\begin{corollary}\label{corollary:2}
Under the conditions of Theorem~\ref{theorem:1}, suppose in addition that $\ve A_{21}^{\mathrm{FD}}=\ve 0$ (equivalently, $\ve A_{21,j}^{\mathrm{FD}}=\ve 0$ for all $j=1,\ldots,p_0$). Then the first-step estimation error does not affect the first-order asymptotic distribution of $\widehat{\ve\kappa}^{\mathrm{FD}}$, so $\ve S_{\widetilde2\widetilde2}^{\mathrm{FD}}=\ve S_{22}^{\mathrm{FD}}$. With the explicit expression $\ve S_{22}^{\mathrm{FD}}=\ve Q_{\mathrm{FD}}$, we obtain 
\[
  \sqrt{T_{\mathrm{eff}}}\bigl(\widehat{\ve\kappa}^{\mathrm{FD}}-\ve\kappa_0\bigr)
  \xrightarrow{d}
  \mathcal N\bigl(\ve0,\ve Q_{\mathrm{FD}}^{-1}\bigr),
\]
so $\widehat{\ve\kappa}^{\mathrm{FD}}$ attains the Aitken (infeasible GLS) bound.
\end{corollary}

\begin{proof}
 See the \href{https://at-noda.com/appendix/durbin_appendix.pdf}{Online Appendix} A.4.13.
\end{proof}

Theorem~\ref{theorem:1} immediately yields a Wald test for linear restrictions on $\ve{\kappa}_0$ based on the FD estimator.

\begin{corollary}\label{corollary:3}
Under the conditions of Theorem~\ref{theorem:1}. Let $\ve{R}$ be a given $q\times m$ matrix with $\mathrm{rank}(\ve{R})=q\le m$, and let $\ve{r}\in\mathbb{R}^q$. Then, under $H_0:\ve{R}\ve{\kappa}_0=\ve{r}$,
\[
 \mathcal{W}^{\mathrm{FD}}
 :=
 T_{\mathrm{eff}}
 (\ve{R}\widehat{\ve{\kappa}}^{\mathrm{FD}}-\ve{r})^\prime
 \big[\ve{R}\widehat{\ve{V}}^{\mathrm{FD}}\ve{R}^\prime\big]^{-1}
 (\ve{R}\widehat{\ve{\kappa}}^{\mathrm{FD}}-\ve{r})
 \xrightarrow{d}
 \chi^2_q.
\]
\end{corollary}

\begin{proof}
 See the \href{https://at-noda.com/appendix/durbin_appendix.pdf}{Online Appendix} A.4.14.
\end{proof}

Proposition~\ref{proposition:7} clarifies the scope of the FD estimator. Although FD can attain the infeasible GLS bound when the first-step effect is asymptotically negligible, its validity still hinges on correct specification of a restricted VAR for $\{\ve u_t\}$. This requirement fails once the DGP belongs to $EBD\setminus GEXOG$: in that case, $\{\ve u_t\}$ generally does not admit a closed finite-order VAR driven solely by its own lags, so the quasi-differencing filter underlying FD becomes misspecified. Hence, like OLS and FCO, FD is not robust to bidirectional dynamic feedback between regressors and disturbances. This limitation motivates the generalized Durbin-type estimators developed in the next subsection, which remain consistent under the weaker $EBD$ condition.

\subsection{Asymptotic properties of generalized Durbin estimator in multi-equation models}

We extend Durbin regression to a multiple-equation framework and derive an estimator that remains consistent even under the $EBD$ condition. Under Assumption~\ref{assumption:1}, the centered joint process $\bar{\ve z}_t$ admits the blockwise VAR($p_0$) representation in Proposition~\ref{proposition:1}, which can be written as
\begin{align}
  \ve{x}_t-\ve{\mu}_{x} =& \sum_{j=1}^{p_0}\ve{\Psi}_{xx,j}\big(\ve{x}_{t-j}-\ve{\mu}_{x}\big) + \sum_{j=1}^{p_0}\ve{\Psi}_{xu,j}\ve{u}_{t-j} + \ve{\varepsilon}_{x,t}, \label{eq_durbin_x_var}\\
  \ve{u}_t =& \sum_{j=1}^{p_0}\ve{\Psi}_{ux,j}\big(\ve{x}_{t-j}-\ve{\mu}_{x}\big) + \sum_{j=1}^{p_0}\ve{\Psi}_{uu,j}\ve{u}_{t-j} + \ve{\varepsilon}_{u,t}. \label{eq_durbin_u_var}
\end{align}
Substituting (\ref{eq_durbin_u_var}) for $\ve{u}_t$ into (\ref{eq_2.2}) yields the following regression model:
\begin{equation}
 \ve{y}_t=\ve{\gamma}+\ve{X}_t^\prime\ve{\beta} + \sum_{j=1}^{p_0}\ve{\Psi}_{uu,j}\ve{y}_{t-j} + \sum_{j=1}^{p_0}\ve{\Lambda}_j\ve{x}_{t-j} + \ve{\varepsilon}_{u,t}, \label{eq_durbin_reg_pre}
\end{equation}
where $\ve{\gamma}=\bigl(\ve{I}_N- \sum_{j=1}^{p_0}\ve{\Psi}_{uu,j}\bigr)\ve{\alpha} - \sum_{j=1}^{p_0}\ve{\Psi}_{ux,j}\ve{\mu}_{x}$, $\ve{\Lambda}_j=\ve{\Psi}_{ux,j} -\ve{\Psi}_{uu,j}\ve{B}$ and $\ve{B}=\mathrm{blkdiag}\big(\ve{\beta}_1^\prime,\ve{\beta}_2^\prime,\ldots,\ve{\beta}_N^\prime \big)$.

For estimation, we treat $\{\ve\Lambda_j\}_{j=1}^{p_0}$ as unrestricted nuisance parameters so that the augmented regression remains linear in the unknown coefficients. This reparameterization does not alter the population regression implied by \eqref{eq_durbin_u_var}; it simply avoids imposing the bilinear cross-parameter restriction $\ve\Lambda_j=\ve\Psi_{ux,j}-\ve\Psi_{uu,j}\ve B$ during estimation.

With this linear reparameterization, \eqref{eq_durbin_reg_pre} becomes a linear regression model in which $\ve\gamma$, $\ve\beta$, $\{\ve\Psi_{uu,j}\}_{j=1}^{p_0}$, and $\{\ve\Lambda_j\}_{j=1}^{p_0}$ enter linearly. Since no cross-equation parameter restrictions are imposed in this augmented representation, the first step of the generalized Durbin procedure can be implemented equation by equation by OLS. In particular, the coefficients on $\{\ve y_{t-j}\}_{j=1}^{p_0}$ deliver the row blocks of $\widehat{\ve\Psi}_{uu,1},\ldots,\widehat{\ve\Psi}_{uu,p_0}$, while those on $\{\ve x_{t-j}\}_{j=1}^{p_0}$ deliver $\widehat{\ve\Lambda}_1,\ldots,\widehat{\ve\Lambda}_{p_0}$. These first-step estimates are then used to recover $\{\ve\Psi_{ux,j}\}_{j=1}^{p_0}$, construct the generalized Durbin transformation, and estimate the innovation covariance matrix. This leads to the following two-step generalized Durbin procedure.

\begin{description}
\item[(Step 1) OLS estimation of the augmented Durbin regression.]

Estimate \eqref{eq_durbin_reg_pre} equation by equation by OLS. Let
$\widehat{\ve\Psi}_{uu,1},\ldots,\widehat{\ve\Psi}_{uu,p_0}$ denote the matrices obtained by stacking, across equations, the estimated coefficients on $\{\ve y_{t-j}\}_{j=1}^{p_0}$, and let
$\widehat{\ve\Lambda}_1,\ldots,\widehat{\ve\Lambda}_{p_0}$ denote the corresponding coefficient matrices on $\{\ve x_{t-j}\}_{j=1}^{p_0}$. Let $\widehat{\ve\gamma}$ and $\widehat{\ve\beta}$ denote the estimated intercept and slope vectors, and define $\widehat{\ve{B}}=\mathrm{blkdiag}\big(\widehat{\ve{\beta}}_1^\prime,\widehat{\ve{\beta}}_2^\prime,\ldots,\widehat{\ve{\beta}}_N^\prime \big)$, $\widehat{\ve\Psi}_{ux,j}:= \widehat{\ve\Lambda}_j+\widehat{\ve\Psi}_{uu,j}\widehat{\ve B}$, $j=1,\ldots,p_0$. Using the first-step residuals $\widehat{\ve\varepsilon}_{u,t}:=\ve y_t-\widehat{\ve\gamma}-\ve X_t^\prime\widehat{\ve\beta} -\sum_{j=1}^{p_0}\widehat{\ve\Psi}_{uu,j}\ve y_{t-j}-\sum_{j=1}^{p_0}\widehat{\ve\Lambda}_j\ve x_{t-j}$, estimate the innovation covariance matrix by $\widehat{\ve\Sigma}_{uu}:= 1/(T-p_0) \sum_{t=p_0+1}^T \widehat{\ve\varepsilon}_{u,t}\bigl(\widehat{\ve\varepsilon}_{u,t}\bigr)^\prime$. Also define $\widehat{\ve\mu}_x:= 1/(T-p_0)\sum_{t=p_0+1}^T \ve x_t$.

\item[(Step 2) Generalized Durbin transformation and GLS.]

Using the first-step estimates, construct $\widehat{\ve{y}}_{\mathrm{GD},t}:=\ve{y}_t -\sum_{j=1}^{p_0}\widehat{\ve\Psi}_{uu,j}\ve y_{t-j} -\sum_{j=1}^{p_0}\widehat{\ve\Psi}_{ux,j}\bigl(\ve x_{t-j}-\widehat{\ve\mu}_x\bigr)$, $\widehat{\ve{X}}_{\mathrm{GD},t}^\prime:=\ve{X}_t^\prime-\sum_{j=1}^{p_0}\widehat{\ve\Psi}_{uu,j}\ve X_{t-j}^\prime$ and $\widehat{\ve{Z}}_{\mathrm{GD},t}^\prime:=\bigl[(\ve I_N-\sum_{j=1}^{p_0}\widehat{\ve\Psi}_{uu,j}),\ \widehat{\ve{X}}_{\mathrm{GD},t}^\prime\bigr]$, $\ve\kappa:=(\ve\alpha^\prime,\ve\beta^\prime)^\prime$. The multivariate generalized Durbin estimator is then defined by
\begin{equation}
  \widehat{\ve\kappa}^{\mathrm{GD}}
  :=
  \Biggl(
    \sum_{t=p_0+1}^T
    \widehat{\ve{Z}}_{\mathrm{GD},t}\,
    \widehat{\ve\Sigma}_{uu}^{-1}\,
    \widehat{\ve{Z}}_{\mathrm{GD},t}^\prime
  \Biggr)^{-1}
  \Biggl(
    \sum_{t=p_0+1}^T
    \widehat{\ve{Z}}_{\mathrm{GD},t}\,
    \widehat{\ve\Sigma}_{uu}^{-1}\,
    \widehat{\ve{y}}_{\mathrm{GD},t}
  \Biggr).
  \label{eq_generalized_durbin_estimator}
\end{equation}
\end{description}

Lemmas~\ref{lemma:1} and \ref{lemma:2} justify the first step of this procedure. In particular, for fixed lag order $p_0$, the equation-by-equation OLS estimators from the augmented regressions are consistent for the corresponding population coefficients. Hence the first-step estimates consistently recover the nuisance parameters that enter the generalized Durbin transformation. In practice, the lag order is unknown. Lemma~\ref{lemma:3} shows that the Bayesian information criterion consistently selects the true lag order under the stated regularity conditions. Therefore, the feasible procedure based on the selected lag order remains asymptotically valid.

Let $\ve{Z}_{\mathrm{GD},t}$ denote the population transformed regressor matrix obtained by replacing the first-step estimates in Step 2 with their population counterparts. To ensure identification in the second-step GLS regression, we impose the following full-rank condition.

\begin{assumption}\label{assumption:6}
$\ve{Q}_{\mathrm{GD}}:=\ex[\ve{Z}_{\mathrm{GD},t}\ve{\Sigma}_{uu}^{-1}\ve{Z}_{\mathrm{GD},t}^\prime]$ is positive definite.
\end{assumption}

\begin{theorem}[Consistency of the GD estimator]\label{theorem:2}
Suppose that Assumptions~\ref{assumption:1}, \ref{assumption:4} and \ref{assumption:6} hold. Then, $\widehat{\ve{\kappa}}^{\mathrm{GD}}$ is consistent for $\ve{\kappa}_0$.
\end{theorem}

\begin{proof}
 See the \href{https://at-noda.com/appendix/durbin_appendix.pdf}{Online Appendix} A.4.15.
\end{proof}

Theorem~\ref{theorem:2} shows that the generalized Durbin estimator remains consistent on the broader $EBD$ region. However, it is still inherently a two-step estimator: both the transformed regressors used in the second-step GLS regression and the feasible weighting matrix are constructed from first-step estimates of the disturbance dynamics and innovation covariance matrix. As a result, the second-step estimating equation inherits first-step estimation error, so a naive plug-in covariance formula that treats these generated objects as fixed is generally asymptotically invalid.

To obtain valid inference, we analyze $\widehat{\ve\kappa}^{\mathrm{GD}}$ within the general two-step framework developed in Section~\ref{subsec:twostep}. In the generalized Durbin procedure, the first step estimates the primitive nuisance parameter $\ve\vartheta^{\mathrm{GD}}:=\bigl(\ve\theta_1^\prime,\ldots,\ve\theta_N^\prime,\ve\mu_x^\prime\bigr)^\prime$, where each $\ve\theta_i$ collects the coefficients in the $i$th augmented Durbin regression. The second-step transformed regression, however, depends on a different set of nuisance objects,
\[
  \ve\eta^{\mathrm{GD}}
  :=
  \bigl(
    \vecop(\ve\Psi_{uu,1})^\prime,\ldots,\vecop(\ve\Psi_{uu,p_0})^\prime,\ 
    \vecop(\ve\Psi_{ux,1})^\prime,\ldots,\vecop(\ve\Psi_{ux,p_0})^\prime,\ 
    \ve\mu_x^\prime
  \bigr)^\prime
  =: h(\ve\vartheta^{\mathrm{GD}}),
\]
which enter the generalized Durbin transformation directly. In particular, the matrices $\{\ve\Psi_{ux,j}\}$ are recovered from the first-step estimates through the identity
\[
  \ve\Psi_{ux,j}=\ve\Lambda_j+\ve\Psi_{uu,j}\ve B,
  \qquad
  \ve B=\mathrm{blkdiag}(\ve\beta_1^\prime,\ldots,\ve\beta_N^\prime),
  \qquad j=1,\ldots,p_0.
\]

Accordingly, the second-step estimating equation inherits first-step estimation error through the mapping $h(\cdot)$. Let $\ve\psi_{2t}^{\mathrm{GD}}(\ve\kappa,\ve\eta^{\mathrm{GD}})$ denote the GLS score associated with the generalized Durbin transformed regression. Since the first-step estimator of $\ve\vartheta^{\mathrm{GD}}$ admits an asymptotically linear representation, the limit distribution of $\widehat{\ve\kappa}^{\mathrm{GD}}$ is governed not by the original second-step moment alone, but by the adjusted moment
\[
  \widetilde{\ve\psi}_{2t}^{\mathrm{GD}}
  :=
  \ve\psi_{2t}^{\mathrm{GD}}
  -
  \ve A_{21}^{\mathrm{GD}}
  (\ve A_{11}^{\mathrm{GD}})^{-1}
  \ve\psi_{1t}^{\mathrm{GD}},
\]
where the Jacobian blocks are
\[
  \ve A_{11}^{\mathrm{GD}}
  :=
  \ex\left[\frac{\partial \ve\psi_{1t}^{\mathrm{GD}}(\ve\vartheta_0^{\mathrm{GD}})}{\partial \ve\vartheta^{\mathrm{GD}\prime}}\right],
  \
  \ve A_{22}^{\mathrm{GD}}
  :=\ex\Big[\frac{\partial \ve\psi_{2t}^{\mathrm{GD}}(\ve\kappa_0,\ve\vartheta_0^{\mathrm{GD}})}
  {\partial \ve\kappa^\prime}\Big],\
  \ve A_{21}^{\mathrm{GD}}
  :=
  \ex\Big[\frac{\partial \ve\psi_{2t}^{\mathrm{GD}}(\ve\kappa_0,\ve\eta_0^{\mathrm{GD}})}
  {\partial \ve\eta^{\mathrm{GD}\prime}}
  \frac{\partial \ve\eta^{\mathrm{GD}}}{\partial \ve\vartheta^{\mathrm{GD}\prime}}\Big].
\]
Hence the asymptotic variance takes the form $(\ve A_{22}^{\mathrm{GD}})^{-1}\ve S_{\widetilde2\widetilde2}^{\mathrm{GD}}(\ve A_{22}^{\mathrm{GD}})^{-1\prime}$, where $\ve S_{\widetilde2\widetilde2}^{\mathrm{GD}} := \sum_{h=-\infty}^{\infty}\cov\bigl(\widetilde{\ve\psi}_{2,t}^{\mathrm{GD}},\widetilde{\ve\psi}_{2,t-h}^{\mathrm{GD}}\bigr)$ is the long-run covariance matrix of the adjusted moment.

\begin{theorem}[Asymptotic distribution of the GD estimator]
\label{theorem:3}
Let $T_{\mathrm{eff}}:=T-p_0$. Suppose Assumptions~\ref{assumption:1}, \ref{assumption:4}, and \ref{assumption:6} hold, and $\ve A_{11}^{\mathrm{GD}}$ is nonsingular. Then
\[
  \sqrt{T_{\mathrm{eff}}}
  \bigl(
    \widehat{\ve\kappa}^{\mathrm{GD}}
    -
    \ve\kappa_0
  \bigr)
  \xrightarrow{d}
  \mathcal{N}\bigl(\ve 0,\ve V^{\mathrm{GD}}\bigr),\quad
\ve V^{\mathrm{GD}}
  :=
  (\ve A_{22}^{\mathrm{GD}})^{-1}
  \ve S_{\widetilde2\widetilde2}^{\mathrm{GD}}
  (\ve A_{22}^{\mathrm{GD}})^{-1\prime}.
\]
\end{theorem}

\begin{proof}
 See the \href{https://at-noda.com/appendix/durbin_appendix.pdf}{Online Appendix} A.4.16.
\end{proof}
\noindent Explicit expressions for the Jacobian and covariance blocks entering $\ve V^{\mathrm{FD}}$ are reported in the \href{https://at-noda.com/appendix/durbin_appendix.pdf}{Online Appendix} A.2.2.

The generalized Durbin estimator is also a two-step estimator, since both the transformed regressors and the GLS weight used in the second step are constructed from first-step estimates of the nuisance parameters. In particular, the second-step transformation depends on the effective nuisance parameter $\ve\eta^{\mathrm{GD}}=h(\ve\vartheta^{\mathrm{GD}})$, which is itself generated from the primitive first-step estimator $\widehat{\ve\vartheta}^{\mathrm{GD}}$. Accordingly, whenever $\ve A_{21}^{\mathrm{GD}}\neq \ve 0$, first-step estimation error enters the first-order asymptotic variance, so a covariance formula that treats the transformed nuisance objects as known is not valid in general. In that case, valid inference requires the Murphy--Topel/Pagan correction in Theorem~\ref{theorem:3}. By contrast, if $\ve A_{21}^{\mathrm{GD}}=\ve 0$, the first-step effect is asymptotically irrelevant at the $T_{\mathrm{eff}}^{-1/2}$ order, the correction disappears, and $\widehat{\ve\kappa}^{\mathrm{GD}}$ attains the infeasible GLS bound.

\begin{corollary}\label{corollary:4}
Under the conditions of Theorem~\ref{theorem:3}, suppose in addition that $\ve A_{21}^{\mathrm{GD}}=\ve 0$ (equivalently, $\ve A_{21,j}^{\mathrm{GD}}=\ve 0$ for all $j=1,\ldots,p_0$). Then the first-step estimation error does not affect the first-order asymptotic distribution of $\widehat{\ve\kappa}^{\mathrm{GD}}$, so $\ve S_{\widetilde2\widetilde2}^{\mathrm{GD}}=\ve S_{22}^{\mathrm{GD}}$. With the explicit expression $\ve S_{22}^{\mathrm{GD}}=\ve Q_{\mathrm{GD}}$, we obtain 
\[
  \sqrt{T_{\mathrm{eff}}}\bigl(\widehat{\ve\kappa}^{\mathrm{GD}}-\ve\kappa_0\bigr)
  \xrightarrow{d}
  \mathcal N\bigl(\ve0,\ve Q_{\mathrm{GD}}^{-1}\bigr),
\]
so $\widehat{\ve\kappa}^{\mathrm{GD}}$ attains the Aitken (infeasible GLS) bound.
\end{corollary}

\begin{proof}
 See the \href{https://at-noda.com/appendix/durbin_appendix.pdf}{Online Appendix} A.4.17.
\end{proof}

As a direct consequence of Theorem~\ref{theorem:3}, we can conduct model specification tests for linear restrictions on $\ve{\kappa}_0$ under any of the exogeneity conditions $BD$, $GEXOG$, or $EBD$.

\begin{corollary}\label{corollary:5}
Let $T_{\mathrm{eff}}:=T-p_0$. Suppose Assumptions~\ref{assumption:1}, \ref{assumption:4}, and \ref{assumption:6} hold, and $\ve A_{11}^{\mathrm{GD}}$ is nonsingular. Let $\ve{R}$ be a given $q\times m$ matrix with $\mathrm{rank}(\ve{R})=q\le m$, and let $\ve{r}\in\mathbb{R}^q$. Then, under $H_0:\ve{R}\ve{\kappa}_0=\ve{r}$,
\[
 \mathcal{W}^{\mathrm{GD}}
 := T_{\mathrm{eff}}(\ve{R}\widehat{\ve{\kappa}}^{\mathrm{GD}}-\ve{r})^\prime
 \big[\ve{R}\widehat{\ve{V}}^{\mathrm{GD}}\ve{R}^\prime\big]^{-1}
 (\ve{R}\widehat{\ve{\kappa}}^{\mathrm{GD}}-\ve{r})
 \ \xrightarrow{d}\ \chi^2_q.
\]
\end{corollary}

\begin{proof}
 See the \href{https://at-noda.com/appendix/durbin_appendix.pdf}{Online Appendix} A.4.18.
\end{proof}

\subsection{Bootstrap-based Wald test using the generalized Durbin estimator}
\label{subsec:boot_gd}

Theorem~\ref{theorem:3} establishes the $\sqrt{T_{\mathrm{eff}}}$-asymptotic normality of the generalized Durbin estimator $\widehat{\ve\kappa}^{\mathrm{GD}}$, which in turn justifies Wald inference as shown in Corollary~\ref{corollary:5}. In finite samples, however, it is well known that Wald tests based on first-order asymptotic approximations may exhibit non-negligible size distortions; see, for example, \citet{hall1996bcv,hwang2023fsc}.

To mitigate these finite-sample distortions, we consider a bootstrap-based Wald test. Under suitable regularity conditions, bootstrap methods often improve finite-sample inference for asymptotically pivotal statistics by providing a more accurate approximation to their null distributions (\citet{hall1996bcv,davidson1999sdb,mackinnon2006bme}). In a time-series setting, however, an ordinary i.i.d.\ bootstrap is inappropriate, because resampling the observations as if they were independent fails to preserve the dynamic dependence structure in the data. A bootstrap procedure for dependent data is therefore required. In our setting, bootstrap pseudo-samples must also be generated so as to be consistent with the null-imposed bootstrap data-generating process while preserving the relevant dynamic dependence. For this purpose, a sieve bootstrap is well suited, because it generates bootstrap pseudo-samples recursively from an estimated dynamic model. Accordingly, following \citet{buhlmann1997sbt}, we approximate the relevant dependence structure by a VAR($p$), where $p$ may increase with $T$.

The bootstrap procedure is implemented as follows. We first impose the null hypothesis on the unrestricted GD estimator to obtain a restricted estimator and the associated null-imposed residuals. We then fit a VAR($\hat p$) approximation, generate bootstrap pseudo-samples by resampling the estimated innovations, and re-estimate the full feasible GD procedure in each bootstrap replication. The detailed algorithm is reported in the following table.

\begin{center}
(Table~\ref{tab:gd_var_sieve_bootstrap} around here)
\end{center}

This bootstrap procedure is designed to capture the dynamic dependence structure relevant for the generalized Durbin setting. In particular, it is implemented under the null hypothesis, preserves the dependence structure of the null-imposed joint process, and incorporates the estimation uncertainty arising from the feasible generalized Durbin procedure. The bootstrap $p$-value is then obtained from the empirical distribution of the bootstrap Wald statistics.

The formal justification of the proposed bootstrap procedure is deferred to the Appendix. In particular, the \href{https://at-noda.com/appendix/durbin_appendix.pdf}{Online Appendix} A.3 establishes the consistency of the null-imposed bootstrap construction, the validity of the fitted VAR approximation, the conditional first-order expansion of the bootstrap feasible GD estimator, and the resulting asymptotic validity of the bootstrap GD-based Wald test.

%% file: durbin_sim.tex
\section{Simulation Experiments}\label{sec:durbin_sim}
In this section, we examine the finite-sample properties of the estimators for multiple-equation systems and their associated test statistics under the $BD$, $GEXOG$, and $EBD$ conditions through Monte Carlo experiments. First, we assess the accuracy of the estimators under each exogeneity condition. Second, we investigate the size and size-adjusted power of the corresponding tests.

\subsection{Simulation Design}
In our simulation, we focus on the case in which the regressors are common across all equations, as discussed in Remark~\ref{remark:1}. This specification is particularly relevant for asset pricing models, where each portfolio equation is typically assumed to depend on the same set of risk factors. Prominent examples include the multifactor models of \citeapos{fama1993crf}{fama1993crf,fama2015ffa} and the arbitrage pricing theory of \citetapos{ross1976atc}. Motivated by this setting, we consider a multiple-equation regression model in which the regressors are common across all equations:
\begin{equation}
 \ve{y}_t=\ve{\alpha} + \big(\ve{I}_N \otimes \ve{x}_t^\prime\big) \ve{\beta} + \ve{u}_t
 =\ve{Z}_t^\prime\ve{\kappa} + \ve{u}_t,\quad t=1,\ldots,T,\label{eq_3.1}
\end{equation}
where $\ve{y}_t = (y_{1,t},\ldots,y_{N,t})^\prime \in \mathbb{R}^{N}$, $\ve{x}_t$ is a $k \times 1$ vector of common regressors ($k=k_i$ for all $i$), and $\ve{\beta}=(\ve{\beta}_1^\prime,\ldots,\ve{\beta}_N^\prime)^\prime\in\mathbb{R}^{kN}$. Let $\ve{X}_t^\prime = \ve{I}_N \otimes \ve{x}_t^\prime \in \mathbb{R}^{N \times kN}$ and $\ve{Z}_t^\prime := [\ve{I}_N,\ve{X}_t^\prime] \in \mathbb{R}^{N \times (k+1)N}$, so that $\ve{\kappa} = (\ve{\alpha}^\prime,\ve{\beta}^\prime)^\prime \in \mathbb{R}^{(k+1)N}$ collects the intercept and slope parameters. As noted in Remark~\ref{remark:1}, the specification in \eqref{eq_3.1} is a special case of \eqref{eq_2.2}.

For the centered process $\bar{\ve{z}}_t=(\ve{x}_t^\prime-\ve{\mu}_x^\prime,\ve{u}_t^\prime)^\prime$, we consider the following three VAR(1) data-generating processes under the $BD$, $GEXOG$, and $EBD$ conditions, respectively.

\begin{description}
 \item[(1)] Block-diagonal vector autoregression (VAR) ($BD$)
\begin{equation}
 \begin{split}
   \ve{y}_t&=\; \ve{\alpha} + (\ve{I}_N \otimes \ve{x}_t^\prime)\ve{\beta} + \ve{u}_t,\\
\begin{bmatrix}
 \ve{x}_t - \ve{\mu}_x\\
 \ve{u}_t
\end{bmatrix}
&=\;
\begin{bmatrix}
 \ve{\Psi}_{xx,1} & \ve{0}\\
 \ve{0} & \ve{\Psi}_{uu,1}
\end{bmatrix}
\begin{bmatrix}
 \ve{x}_{t-1} - \ve{\mu}_x\\
 \ve{u}_{t-1}
\end{bmatrix}
+
\begin{bmatrix}
 \ve{\varepsilon}_{x,t}\\
 \ve{\varepsilon}_{u,t}
\end{bmatrix},\\
\begin{bmatrix}
 \ve{\varepsilon}_{x,t}\\
 \ve{\varepsilon}_{u,t}
\end{bmatrix}
&\stackrel{i.i.d.}{\sim}\;
 \mathcal{N}
\left(
\begin{bmatrix}
 \ve{0}\\
 \ve{0}
\end{bmatrix},
\begin{bmatrix}
 \ve{\Sigma}_{xx} & \ve{0}\\
 \ve{0} & \ve{\Sigma}_{uu}
\end{bmatrix}
\right).
 \end{split}
\label{eq_sim.bd}
\end{equation}

 \item[(2)] Block-upper-triangular VAR ($GEXOG$)
\begin{equation}
 \begin{split}
   \ve{y}_t&=\; \ve{\alpha} + (\ve{I}_N \otimes \ve{x}_t^\prime)\ve{\beta} + \ve{u}_t, \\
\begin{bmatrix}
 \ve{x}_t - \ve{\mu}_x\\
 \ve{u}_t
\end{bmatrix}
&=\;
\begin{bmatrix}
 \ve{\Psi}_{xx,1} & \ve{\Psi}_{xu,1}\\
 \ve{0} & \ve{\Psi}_{uu,1}
\end{bmatrix}
\begin{bmatrix}
 \ve{x}_{t-1} - \ve{\mu}_x\\
 \ve{u}_{t-1}
\end{bmatrix}
+
\begin{bmatrix}
 \ve{\varepsilon}_{x,t}\\
 \ve{\varepsilon}_{u,t}
\end{bmatrix},\\
\begin{bmatrix}
 \ve{\varepsilon}_{x,t}\\
 \ve{\varepsilon}_{u,t}
\end{bmatrix}
&\stackrel{i.i.d.}{\sim}\;
 \mathcal{N}
\left(
\begin{bmatrix}
 \ve{0}\\
 \ve{0}
\end{bmatrix},
\begin{bmatrix}
 \ve{\Sigma}_{xx} & \ve{0}\\
 \ve{0} & \ve{\Sigma}_{uu}
\end{bmatrix}
\right).
 \end{split}
\label{eq_sim.gexog}
\end{equation}

 \item[(3)] Unrestricted VAR ($EBD$)
\begin{equation}
 \begin{split}
 \ve{y}_t&=\; \ve{\alpha} + (\ve{I}_N \otimes \ve{x}_t^\prime)\ve{\beta} + \ve{u}_t, \\
\begin{bmatrix}
 \ve{x}_t - \ve{\mu}_x\\
 \ve{u}_t
\end{bmatrix}
&=\;
\begin{bmatrix}
 \ve{\Psi}_{xx,1} & \ve{\Psi}_{xu,1}\\
 \ve{\Psi}_{ux,1} & \ve{\Psi}_{uu,1}
\end{bmatrix}
\begin{bmatrix}
 \ve{x}_{t-1} - \ve{\mu}_x\\
 \ve{u}_{t-1}
\end{bmatrix}
+
\begin{bmatrix}
 \ve{\varepsilon}_{x,t}\\
 \ve{\varepsilon}_{u,t}
\end{bmatrix},\\
\begin{bmatrix}
 \ve{\varepsilon}_{x,t}\\
 \ve{\varepsilon}_{u,t}
\end{bmatrix}
&\stackrel{i.i.d.}{\sim}\;
\mathcal{N}
\left(
\begin{bmatrix}
 \ve{0}\\
 \ve{0}
\end{bmatrix},
\begin{bmatrix}
 \ve{\Sigma}_{xx} & \ve{0}\\
 \ve{0} & \ve{\Sigma}_{uu}
\end{bmatrix}
\right).
 \end{split}
\label{eq_sim.ebd}
\end{equation}
\end{description}

In the simulation design of \citet{baillie2024rit}, lagged dependent variables are explicitly included in the regression. In contrast, our framework does not require specifying a finite lag structure for $\ve{y}_t$ \textit{a priori} in the estimating equation. Instead, under the maintained joint dynamic specification for $\ve{z}_t=(\ve{x}_t^\prime-\ve{\mu}_x^\prime,\ve{u}_t^\prime)^\prime$, together with the structural relation for $\ve{y}_t$, one obtains an equivalent Durbin-type representation. Hence, the dynamics of $\ve{y}_t$ are captured implicitly through the joint dynamics of $(\ve{x}_t,\ve{u}_t)$ rather than being specified directly in advance.

In all cases, we consider sample sizes $T \in \{100, 200, 400, 800\}$. To examine the effects of the number of regressors and the number of equations on the finite-sample accuracy of estimators and the performance of tests, we consider $k \in \{2, 4\}$ and $N \in \{5, 10\}$. In each replication, we generate $T+500$ observations and discard the first 500 observations as burn-in. For each DGP, we conduct 10,000 Monte Carlo replications.

For the VAR(1) coefficient matrix, the block restrictions depend on the DGP ($BD$/ $GEXOG$/ $EBD$) as in \eqref{eq_sim.bd}--\eqref{eq_sim.ebd}. For the diagonal blocks, we control persistence through the spectral norm. Specifically, we draw random orthogonal matrices $\ve Q_x \in \mathbb R^{k\times k}$ and $\ve Q_u \in \mathbb R^{N\times N}$ via QR decompositions of i.i.d.\ Gaussian matrices and set
\[
 \ve{\Psi}_{xx,1}=c_{xx}\,\ve Q_x,\qquad
 \ve{\Psi}_{uu,1}=c_{uu}\,\ve Q_u,
\]
so that $\|\ve{\Psi}_{xx,1}\|_2=c_{xx}$ and $\|\ve{\Psi}_{uu,1}\|_2=c_{uu}$ before any global rescaling for stationarity. For the cross blocks, we set $r=\min(k,N)$ and independently draw matrices with orthonormal columns, $\ve A_1,\ve A_2\in\mathbb R^{k\times r}$ and $\ve B_1,\ve B_2\in\mathbb R^{N\times r}$, such that
\[
 \ve A_1^\prime \ve A_1=\ve A_2^\prime \ve A_2=\ve I_r,
 \qquad
 \ve B_1^\prime \ve B_1=\ve B_2^\prime \ve B_2=\ve I_r.
\]
We then define
\[
 \ve{\Psi}_{xu,1}=\frac{c_{xu}}{\sqrt r}\,\ve A_1\ve B_1^\prime,\qquad
 \ve{\Psi}_{ux,1}=\frac{c_{ux}}{\sqrt r}\,\ve B_2\ve A_2^\prime.
\]
This scaling keeps the Frobenius norms of the cross blocks comparable across $r$. The baseline parameter values are set as $c_{xx}=0.4$, $c_{xu}=0.7$, $c_{ux}=0.3$, and $c_{uu}=0.5$. The full coefficient matrix is uniformly rescaled whenever necessary so that its spectral radius does not exceed $0.91$. Specifically, if the spectral radius of the initially generated VAR coefficient matrix, $\rho(\ve{\Psi})$, exceeds $0.91$, then all blocks of $\ve{\Psi}$ are multiplied by the common factor $0.91/\rho(\ve{\Psi})$.

For the innovation covariance matrices, we generate $\ve\Sigma_{xx}$ and $\ve\Sigma_{uu}$ separately as random positive definite matrices using a Cholesky-type construction. Specifically, for each block we draw a lower-triangular matrix $\ve L$ with unit diagonal and i.i.d.\ strictly lower-triangular entries from $\mathrm{Unif}(-\delta,\delta)$, and then set the covariance matrix equal to $\ve L\ve L^\prime$. In the implementation, we use $\delta=0.1$ for $\ve\Sigma_{xx}$ and $\delta=0.5$ for $\ve\Sigma_{uu}$. The two innovation blocks are generated independently, so that the contemporaneous innovation covariance between $\ve x_t$ and $\ve u_t$ is zero.

Given the coefficient matrix and the innovation covariance, we generate the centered vector $\ve{z}_t:=(\ve{x}_t^\prime-\ve{\mu}_x^\prime,\ve{u}_t^\prime)^\prime$ according to the VAR(1) recursion
\[
 \ve{z}_t=\ve{\Psi}\ve{z}_{t-1}+\ve{\varepsilon}_t,
 \qquad
 \ve{\varepsilon}_t\sim \mathcal N(\ve{0},\ve{\Sigma}),
 \quad
 \ve{\Sigma}=
 \begin{bmatrix}
  \ve{\Sigma}_{xx} & \ve{0}\\
  \ve{0} & \ve{\Sigma}_{uu}
 \end{bmatrix}.
\]
After generating $\ve x_t-\ve\mu_x$, we add back the mean vector $\ve\mu_x$ to obtain $\ve x_t$. Finally, the dependent variables are generated from the structural equation
\[
 \ve{y}_t=\ve{\alpha}+\big(\ve{I}_N \otimes \ve{x}_t^\prime\big)\ve{\beta}+\ve{u}_t,\quad t=1,\ldots,T.
\]
For the regressor mean, we set $\ve{\mu}_x=0.3\,\ve{1}_k$. All slope coefficients are set to one, i.e.\ $\ve{\beta}=\ve{1}_{Nk}$.

\subsection{Estimation Accuracy}\label{subsec:estimation accuracy}

Following \citet{baillie2024rit}, we examine the finite-sample accuracy of four estimators for multiple-equation systems: OLS, FCO, FD, and GD. Specifically, we compute and compare their bias and mean squared error (MSE). Because the parameters are vector-valued, we summarize bias and MSE using scalar measures based on the Euclidean norm for ease of interpretation.

Let $\ve\kappa_0=(\ve{\alpha}_0^\prime,\ve{\beta}_0^\prime)^\prime\in\mathbb R^{(k+1)N}$ denote the true parameter vector, and let $\widehat{\ve\kappa}_T$ be an estimator of $\ve\kappa_0$ based on a sample size $T$. As a scalar measure of bias, we report the Euclidean norm of the bias vector:
\[
 \mathrm{bias}(\widehat{\ve\kappa}_T) := \|\ex[\widehat{\ve\kappa}_T]-\ve\kappa_0\|_2.
\]
We further define the scalar measure of MSE by
\[
 \mathrm{MSE}(\widehat{\ve\kappa}_T)
 :=\ex\left[\|\widehat{\ve\kappa}_T-\ve\kappa_0\|_2^2\right]
 =\ex\left[(\widehat{\ve\kappa}_T-\ve\kappa_0)^\prime(\widehat{\ve\kappa}_T-\ve\kappa_0)\right].
\]
Then the bias--variance decomposition is given by
\[
 \mathrm{MSE}(\widehat{\ve\kappa}_T)
 = \mathrm{bias}(\widehat{\ve\kappa}_T)^2
 + \mathrm{tr}\left(\var(\widehat{\ve\kappa}_T)\right),
\]
where $\var(\widehat{\ve\kappa}_T)=\ex\left[(\widehat{\ve\kappa}_T-\ex[\widehat{\ve\kappa}_T])(\widehat{\ve\kappa}_T-\ex[\widehat{\ve\kappa}_T])^\prime\right]$. This decomposition allows us to assess finite-sample estimation accuracy in terms of the relative contributions of squared bias and variance.

In Monte Carlo experiments with $M$ replications, let $\widehat{\ve\kappa}_T^{(m)}$ denote the estimator obtained in replication $m$. We define the Monte Carlo average by $\bar{\ve\kappa}_T := 1/M\sum_{m=1}^M \widehat{\ve\kappa}_T^{(m)}$ and the corresponding Monte Carlo covariance matrix by $\widehat{\var}(\widehat{\ve\kappa}_T):=1/M\sum_{m=1}^M\bigl(\widehat{\ve\kappa}_T^{(m)}-\bar{\ve\kappa}_T\bigr)\bigl(\widehat{\ve\kappa}_T^{(m)}-\bar{\ve\kappa}_T\bigr)^\prime$. Then the Monte Carlo analogue of the scalar measure of bias and MSE are given by
\[
 \widehat{\mathrm{bias}}(\widehat{\ve\kappa}_T)
 :=
 \left\|\bar{\ve\kappa}_T-\ve\kappa_0\right\|_2, \quad
\widehat{\mathrm{MSE}}(\widehat{\ve\kappa}_T)
 :=
 \widehat{\mathrm{bias}}(\widehat{\ve\kappa}_T)^2
 +
 \mathrm{tr}\bigl(
 \widehat{\var}(\widehat{\ve\kappa}_T)
 \bigr).
\]

In what follows, we compare the accuracy of the estimators for multiple-equation systems across the DGPs in \eqref{eq_sim.bd}--\eqref{eq_sim.ebd}. We set the true parameters to $\ve{\alpha}_0=\ve{0}_N$ and $\ve{\beta}_0=\ve{1}_{Nk}$, so that $\ve{\kappa}_0=(\ve{0}_N^\prime,\ve{1}_{Nk}^\prime)^\prime$.

\begin{center}
(Table \ref{sim_param_accuracy_bd} around here)
\end{center}

Table \ref{sim_param_accuracy_bd} reports the bias and MSE of the GD, FD, FCO, and OLS estimators under $BD$. For all estimators, the bias is small throughout, and the MSE declines as the sample size $T$ increases, indicating improved finite-sample accuracy in larger samples. Overall, the bias also tends to become smaller as $T$ grows. These findings are in line with the asymptotic theory, under which all four estimators are consistent under $BD$. The MSEs of FD and FCO are nearly identical across all designs and are uniformly smaller than those of GD and OLS. This pattern is consistent with Proposition~\ref{proposition:7}, which establishes the asymptotic equivalence of FD and FCO under $BD$. It is also in line with Proposition~\ref{proposition:6}, which shows that FCO attains the GLS efficiency bound under $BD$.

\begin{center}
(Table \ref{sim_param_accuracy_gexog} around here)
\end{center}

Table \ref{sim_param_accuracy_gexog} reports the bias and MSE of the GD, FD, FCO, and OLS estimators under $GEXOG$. The bias of the OLS estimator remains non-negligible across all combinations of $N$ and $k$, even as the sample size $T$ increases. This pattern is in line with Proposition~\ref{proposition:4}, which establishes consistency of OLS under $BD$, while the more general $GEXOG$ condition does not in itself guarantee OLS consistency. Although Proposition~\ref{proposition:5} implies that FCO is not generally consistent under $GEXOG$, its finite-sample bias and MSE remain much smaller than those of OLS and do not deteriorate substantially in the reported designs. This finding is consistent with the broader literature showing that misspecified GLS procedures can still perform reasonably well in finite samples. For example, \citet{baillie2024rit} report under their $GEXOG$ design that traditional FGLS can substantially outperform OLS and, in some small-sample parameterizations, even yield a smaller MSE than FD. Similarly, \citet{perron2026fgls} show that even an incorrect GLS correction can achieve a lower MSE than OLS, and \citet{koreisha2001gls} provide both theoretical efficiency bounds and simulation evidence indicating that incorrect GLS procedures can be quite robust in finite samples.

\begin{center}
(Table \ref{sim_param_accuracy_ebd} around here)
\end{center}

Table \ref{sim_param_accuracy_ebd} reports the bias and MSE of the GD, FD, FCO, and OLS estimators under $EBD$. The bias of GD tends to decline as the sample size $T$ increases across all combinations of $N$ and $k$, in line with Theorem~\ref{theorem:2}, which establishes consistency of GD under $EBD$. By contrast, FD, FCO, and OLS generally exhibit larger bias than GD and do not display the same clear tendency toward zero, consistent with the fact that these estimators are not generally consistent under $EBD$. GD also attains the smallest MSE in all reported cases.

Taken together, the results in Tables \ref{sim_param_accuracy_bd}--\ref{sim_param_accuracy_ebd} show that the relative accuracy of the estimators depends crucially on the underlying exogeneity condition. Under $BD$, both FD and FCO perform well, while under $GEXOG$, FD continues to perform well and FCO can still perform reasonably well in finite samples despite the lack of general consistency. However, once the exogeneity conditions required for their consistency are violated, the finite-sample accuracy of FD and FCO deteriorates. By contrast, the GD estimator remains stable in terms of bias and MSE across all exogeneity conditions considered here. These findings are broadly consistent with the theoretical results established in the previous section.

\subsection{Size and size-adjusted power of tests}

In this subsection, we examine the size and size-adjusted power of test statistics constructed from the GD, FD, FCO, and OLS estimators. The size of a test is its probability of rejecting the null hypothesis when the null is true, and thus indicates whether the test controls the Type I error at the nominal significance level in finite samples. A test with severe size distortions may exhibit substantial over-rejection or under-rejection. By contrast, size-adjusted power measures the ability of a test to detect false null hypotheses after adjusting for differences in finite-sample size. Considering both measures allows for a fair comparison of the competing tests.

In linear asset-pricing models, the intercept vector $\ve{\alpha}$ captures pricing errors, because each $\alpha_i$ represents the component of the mean excess return on asset $i$ that is not explained by its factor exposure. Accordingly, the validity of a linear asset-pricing model is typically assessed through the joint null hypothesis $H_0:\ve{\alpha}=\ve{0}$. In empirical applications, this null hypothesis is often tested using the \citetapos{gibbons1989teg} test (the GRS test) or the HAR test of \citet{lazarus2018hir,lazarus2021spt}.

\paragraph{GRS test}
The GRS test of \citet{gibbons1989teg} is one of the most widely used test statistics for assessing the validity of asset-pricing models; see, for example, \citet{fama1993crf,fama2015ffa,fama2016daf,fama2017itf,fama2018cf,fama2020ccs} and \citet{cakici2013svm}. Here, we suppose that the regressor vector is common across equations, i.e., $\ve{x}_{i,t}:=\ve{x}_t$ for all $i=1,\ldots,N$ as in Remark~\ref{remark:1}. Define
\[
 \bar{\ve{x}}=\frac{1}{T}\sum_{t=1}^T\ve{x}_t,\quad
 \ve{S}_x=\frac{1}{T-1}\sum_{t=1}^T(\ve{x}_t-\bar{\ve{x}})(\ve{x}_t-\bar{\ve{x}})^\prime,\quad
 \widehat{\ve{\Sigma}}_{u}^{\mathrm{OLS}}=\frac{1}{T-k-1}\sum_{t=1}^T\widehat{\ve{u}}_t^{\mathrm{OLS}}\widehat{\ve{u}}_t^{\mathrm{OLS}\prime}, 
\]
where $\widehat{\ve{u}}_t^{\mathrm{OLS}}$ denotes the OLS residual vector. Then the GRS statistic is given by
\[
 \mathrm{GRS}
 =
 \frac{T(T-N-k)}{N(T-k-1)}
 \left(1+\bar{\ve{x}}^\prime\ve{S}_{x}^{-1}\bar{\ve{x}}\right)^{-1}
 \widehat{\ve{\alpha}}^{\mathrm{OLS}\prime}\widehat{\ve{\Sigma}}_{u}^{\mathrm{OLS},-1}\widehat{\ve{\alpha}}^{\mathrm{OLS}}. 
\]
Under the assumption that $\{\ve{u}_t\}_{t=1}^T$ is a sequence of independently and identically distributed multivariate normal random vectors, the GRS statistic follows an $F(N,T-N-k)$ distribution under the null hypothesis $H_0:\ve{\alpha}_0=\ve{0}$. Since these conditions exclude the serial dependence and dynamic regressor--disturbance dependence studied in this paper, the GRS test is included as a conventional benchmark rather than as a procedure justified under our general setting.\footnote{\citet{kamstra2024tra} propose a modified version of the GRS test to improve its small-sample properties. However, the performance of the original and modified GRS tests does not differ substantially once the sample size $T$ exceeds 200. For this reason, we adopt the original GRS test as the benchmark.}

\paragraph{HAR test}
When regression errors exhibit heteroskedasticity and autocorrelation, tests that do not account for these features are generally invalid. One standard approach is the heteroskedasticity and autocorrelation robust (HAR) test, which is based on a HAC estimator of the long-run covariance matrix; see \citet{newey1987sps,andrews1991hac}. Let
\[
 \ve{\Omega}
 =
 \ve{\Gamma}_0+\sum_{j=1}^{\infty}\bigl(\ve{\Gamma}_j+\ve{\Gamma}_j^\prime\bigr),
 \qquad
 \ve{\Gamma}_j=\ex[\ve{v}_t\ve{v}_{t-j}^\prime],
 \qquad
 \ve{v}_t=\ve{Z}_t\ve{u}_t,
\]
and estimate $\ve{\Omega}$ by the kernel-based HAC estimator
\[
 \widehat{\ve{\Omega}}(M)
 =
 \widehat{\ve{\Gamma}}_0
 +\sum_{j=1}^{T-1}
 k\left(\frac{j}{M}\right)
 \bigl(\widehat{\ve{\Gamma}}_j+\widehat{\ve{\Gamma}}_j^\prime\bigr),
 \qquad
 \widehat{\ve{\Gamma}}_j
 =
 \frac{1}{T}\sum_{t=j+1}^T
 \widehat{\ve{v}}_t\widehat{\ve{v}}_{t-j}^\prime,
\]
where $\widehat{\ve{v}}_t=\ve{Z}_t\widehat{\ve{u}}_t$. The corresponding HAR Wald statistic for testing $H_0:\ve{R}\ve{\kappa}=\ve{r}$ is
\[
 \mathcal{W}^{\mathrm{HAR}}
 =
 T(\ve{R}\widehat{\ve{\kappa}}^{\mathrm{OLS}}-\ve{r})^\prime
 \left[
 \ve{R}\widehat{\ve{Q}}_{Z}^{-1}
 \widehat{\ve{\Omega}}(M)
 \widehat{\ve{Q}}_{Z}^{-1}\ve{R}^\prime
 \right]^{-1}
 (\ve{R}\widehat{\ve{\kappa}}^{\mathrm{OLS}}-\ve{r}), 
\]
where $\widehat{\ve{Q}}_{Z}=T^{-1}\sum_{t=1}^T\ve{Z}_t\ve{Z}_t^\prime$. Under the conventional small-$b$ asymptotics, if $M\to\infty$ and $M/T\to0$, then under standard regularity conditions, $\mathcal{W}^{\mathrm{HAR}} \xrightarrow{d} \chi_q^2$ with $q=\rank(\ve{R})$.

A key practical issue in HAC-based inference is the choice of the bandwidth parameter $M$. In small samples, conventional HAR tests can be highly sensitive to this choice and often exhibit substantial over-rejection. To address this difficulty, \citet{kiefer2000srt} propose the fixed-$b$ approach, setting $M=bT$ with $b\in(0,1]$. In this case, the HAC estimator is no longer consistent, and the HAR statistic has a nonstandard null distribution that depends on the kernel and on $b$. This approach often yields more accurate finite-sample size control than conventional HAC-based HAR tests, i.e., \citet{kiefer2002har,jansson2004erp,kiefer2005nat,sun2014lfa}.

Building on these insights, \citet{lazarus2018hir,lazarus2021spt} provide a comprehensive comparison of fixed-$b$ methods, establish the size-power frontier, and propose loss-function-based rules for selecting $b$. They show that the frontier is achieved by the quadratic spectral (QS) kernel. Against this background, in our simulations we adopt the fixed-$b$ HAR test with the QS kernel recommended by \citet{lazarus2018hir,lazarus2021spt}, so that inference is based on the corresponding nonstandard fixed-$b$ asymptotic distribution rather than on the conventional small-$b$ HAC approximation.

\subsubsection{Rejection frequencies under the null hypothesis}
We first examine whether the empirical rejection frequency of each test under the null hypothesis is close to the nominal significance level. We then turn to size-adjusted power, which measures how well each test detects false null hypotheses after correcting for size distortions. Taken together, these results help clarify the trade-off between size control and power across the three DGPs: $BD$, $GEXOG$, and $EBD$.

Throughout this subsection, $\widetilde{\mathcal{W}}^{\mathrm{GD}}$ denotes the bootstrap Wald test based on the GD estimator, while $\mathcal{W}^{\mathrm{GD}}$, $\mathcal{W}^{\mathrm{FD}}$, and $\mathcal{W}^{\mathrm{FGCO}}$ denote the asymptotic Wald tests based on the GD, FD, and FCO estimators, respectively. In addition, $\mathcal{W}^{\mathrm{HAR}}$ and $\mathrm{GRS}$ denote the HAR and GRS tests, respectively. Under the null hypothesis $H_0:\ve{\alpha}_0=\ve{0}_N$, we report, for each test, empirical rejection frequencies at the nominal 10\%, 5\%, and 1\% significance levels.

\begin{center}
(Table \ref{sim_bd_nh} around here)
\end{center}

Table \ref{sim_bd_nh} reports empirical rejection frequencies under the null hypothesis for the $BD$ condition. Overall, the bootstrap Wald test based on the GD estimator, $\widetilde{\mathcal{W}}^{\mathrm{GD}}$, yields rejection frequencies close to the nominal levels across all reported designs, even when the sample size is as small as $T=100$, indicating satisfactory finite-sample size control. By contrast, the asymptotic Wald tests based on GD, FD, and FCO exhibit rejection frequencies well above the nominal levels in small samples, although these distortions gradually diminish as the sample size increases. The rejection frequencies of $\mathcal{W}^{\mathrm{FD}}$ and $\mathcal{W}^{\mathrm{FCO}}$ are also very similar throughout, consistent with their asymptotic equivalence under $BD$ established in Proposition~\ref{proposition:7}. The HAR test also performs reasonably well, especially when $N=5$, although for $N=10$ it tends to under-reject in small samples. By contrast, the GRS test exhibits non-negligible size distortions. In particular, in the reported designs, it tends to over-reject the null hypothesis when $N=5$ and to under-reject it when $N=10$. This pattern is consistent with the fact that the GRS test is derived under the assumption that the disturbance vector is independently and identically distributed as multivariate normal, whereas under the $BD$ condition the disturbances are serially correlated.

\begin{center}
(Table \ref{sim_gexog_nh} around here)
\end{center}

Table \ref{sim_gexog_nh} reports empirical rejection frequencies under the null hypothesis when the $GEXOG$ condition holds. As in the $BD$ case, the bootstrap Wald test based on the GD estimator, $\widetilde{\mathcal{W}}^{GD}$, continues to deliver rejection frequencies close to the nominal levels across all reported designs, regardless of the sample size $T$. In addition, the asymptotic Wald tests based on GD and FD tend to move closer to the nominal levels as the sample size increases. By contrast, the Wald test based on the FCO estimator performs less satisfactorily, especially when the number of equations is $N=10$. This is because the FCO estimator is not consistent under $GEXOG$, as shown in Proposition~\ref{proposition:5}. This tendency becomes more pronounced as the number of equations increases. The most severe size distortions arise for the HAR and GRS tests. This pattern is unsurprising, since both procedures rely on OLS-based estimation, whose consistency is not guaranteed under $GEXOG$. Taken together, the results indicate that under $GEXOG$, $\widetilde{\mathcal{W}}^{GD}$ provides the most reliable finite-sample size control among the competing procedures.

\begin{center}
(Table \ref{sim_ebd_nh} around here)
\end{center}

Table \ref{sim_ebd_nh} reports empirical rejection frequencies under the null hypothesis for the $EBD$ condition. The bootstrap Wald test based on the GD estimator, $\widetilde{\mathcal{W}}^{\mathrm{GD}}$, continues to yield rejection frequencies close to the nominal levels across all reported designs, indicating stable finite-sample size control even under the most general dependence structure considered in this paper. The asymptotic Wald test based on the GD estimator also improves as the sample size increases, although it exhibits noticeable over-rejection in smaller samples, especially when $N=10$. By contrast, the Wald tests based on the FD and FCO estimators exhibit substantial size distortions throughout, and these distortions do not vanish as the sample size increases. This pattern is consistent with the fact that these estimators are not consistent under $EBD$. The HAR and GRS tests also exhibit severe size distortions under $EBD$, with rejection frequencies far above the nominal levels in many designs. Overall, these results are consistent with the theoretical prediction that, among the competing procedures, only the GD-based procedures remain valid under $EBD$.

Taken together, the results in Tables \ref{sim_bd_nh}--\ref{sim_ebd_nh} show that the relative performance of the tests depends crucially on the underlying dependence structure. Under the more restrictive $BD$ and $GEXOG$ conditions, FGLS-based tests perform well when the assumptions required for their validity are satisfied. By contrast, across all three dependence structures, the GD-based Wald test, and especially its bootstrap version, delivers the most reliable finite-sample size control.

\subsubsection{Rejection frequencies under the alternative hypothesis}
We now examine the size-adjusted power of each test for the $BD$, $GEXOG$, and $EBD$ designs. Because finite-sample size distortions can substantially affect raw rejection frequencies, unadjusted power is not directly comparable across procedures. We therefore focus on size-adjusted power, so that differences in rejection frequencies reflect each test's ability to detect false null hypotheses rather than differences in size distortions. Specifically, we consider the alternative hypothesis $H_1:\alpha_1=0.15$ and $\alpha_j=0$ for $j\neq 1$.

\begin{center}
(Figure \ref{sim_bd_alt} around here)
\end{center}

Figure \ref{sim_bd_alt} reports the size-adjusted rejection frequencies under the alternative hypothesis for the $BD$ design. Among the competing procedures, the FGLS-based tests, especially $\mathcal{W}^{\mathrm{FD}}$ and $\mathcal{W}^{\mathrm{FCO}}$, generally attain the highest rejection frequencies across the reported designs. This pattern is consistent with the theoretical prediction that, under $BD$, FCO is asymptotically efficient (Proposition~\ref{proposition:6}) and FD is asymptotically equivalent to FCO (Proposition~\ref{proposition:7}). Accordingly, once size distortions are removed, the stronger performance of these tests can be attributed to genuine efficiency gains under correct specification rather than to spurious over-rejection under the null. In addition, $\widetilde{\mathcal{W}}^{\mathrm{GD}}$ and $\mathcal{W}^{\mathrm{GD}}$ track each other very closely in all panels, indicating that bootstrap size correction entails little, if any, loss of power under $BD$.

By contrast, $\mathcal{W}^{\mathrm{HAR}}$ is generally less powerful than the system-based procedures. This pattern is broadly consistent with the simulation evidence in \citet{baillie2024rit}, who show in a single-equation time-series regression setting that HAR-based inference can be less powerful than procedures that exploit additional dynamic structure. In addition, the GRS test exhibits the lowest power across all reported designs. This finding is also consistent with earlier studies. In particular, \citet{gungor2013tlf} show that the power of the GRS test does not necessarily increase monotonically with the number of test assets and may even decline when too many assets are included. More generally, because the exact finite-sample validity of the GRS test relies on independently and identically distributed multivariate normal disturbances, violations of these assumptions can further weaken its finite-sample performance; see, for example, \citet{affleck1989nta}.

\begin{center}
(Figure \ref{sim_gexog_alt} around here)
\end{center}

Figure~\ref{sim_gexog_alt} reports the size-adjusted power under the $GEXOG$ condition. For this design, we report size-adjusted power only for procedures that remain theoretically valid. Accordingly, we exclude $\mathcal{W}^{\mathrm{FCO}}$ and the OLS-based benchmark tests, namely $\mathcal{W}^{\mathrm{HAR}}$ and the GRS test. This exclusion reflects not only their severe size distortions under the null, but also the more fundamental point that these procedures are not valid under $GEXOG$; see Propositions~\ref{proposition:4} and \ref{proposition:5}.

Among the remaining procedures, several patterns emerge. First, the size-adjusted power of all three tests increases steadily with the sample size $T$, as expected. Second, $\mathcal{W}^{\mathrm{FD}}$ attains the highest rejection frequencies throughout the reported designs. This finding is consistent with the fact that, under $GEXOG$, the FD estimator may be asymptotically efficient; see Corollary~\ref{corollary:3}. Third, $\widetilde{\mathcal{W}}^{\mathrm{GD}}$ and $\mathcal{W}^{\mathrm{GD}}$ remain very close in all panels, indicating that bootstrap size correction entails little, if any, loss of power under $GEXOG$. As $T$ increases, the differences across the three procedures narrow, and their rejection frequencies approach one.

\begin{center}
(Figure \ref{sim_ebd_alt} around here)
\end{center}

Figure~\ref{sim_ebd_alt} reports the size-adjusted power under the $EBD$ condition. For this design, we report size-adjusted power only for the GD-based procedures, since the remaining tests are not valid under $EBD$. Two patterns stand out. First, the size-adjusted power of both $\widetilde{\mathcal{W}}^{\mathrm{GD}}$ and $\mathcal{W}^{\mathrm{GD}}$ increases steadily with the sample size $T$ in all reported designs, and approaches one as $T$ becomes large. Power is somewhat lower in the designs with $k=4$ when $T$ is small, although this difference narrows rapidly as $T$ increases. Second, the two procedures remain very close in every panel, with only negligible differences throughout. This indicates that the bootstrap implementation improves finite-sample size control under $EBD$ without any meaningful loss of power relative to the asymptotic GD test.

The simulation results show that the finite-sample performance of the competing tests depends crucially on the underlying exogeneity condition. Under the more restrictive $BD$ and $GEXOG$ conditions, procedures that exploit stronger structural restrictions, such as $\mathcal{W}^{\mathrm{FD}}$ and, when valid, $\mathcal{W}^{\mathrm{FCO}}$, can attain higher size-adjusted power. By contrast, the OLS-based tests provide meaningful inference only under $BD$; under more general dependence structures, their size distortions become severe and power comparisons based on them are no longer informative. As the dependence structure becomes more general, robustness becomes more important than efficiency gains available under more restrictive specifications. In this respect, the GD-based procedures remain stable across the $BD$, $GEXOG$, and $EBD$ designs. In particular, the bootstrap GD test consistently delivers reliable size control while preserving power close to that of the corresponding asymptotic GD test. These findings highlight the practical usefulness of the bootstrap GD test when the true exogeneity condition is unknown or may be misspecified.

%% file: durbin_emp.tex
\section{Empirical Application}\label{sec:durbin_emp}
We examine whether the Wald test based on the generalized Durbin estimator proposed in this paper can mitigate the over-rejection problem observed for Wald tests based on the FD and FCO estimators. In Section~\ref{sec:durbin_model}, we show that, under the $EBD$ condition, the generalized Durbin estimator is consistent, whereas the FD and FCO estimators are not. At the same time, our simulation results in Section~\ref{sec:durbin_sim} indicate that the Wald test based on the generalized Durbin estimator is not immune to finite-sample size distortions. We therefore employ a bootstrap bias correction to improve finite-sample size control.

To illustrate the empirical relevance of our approach, we revisit the Fama--French multifactor models studied by \citet{fama1993crf,fama2015ffa}. As noted by \citet{moriya2024tif}, there remains ongoing debate over the validity of the FF multifactor models. In empirical studies (e.g., \citet{fama1993crf,fama2015ffa,fama2016daf,fama2017itf,fama2018cf,fama2020ccs}; \citet{cakici2013svm}), it has become standard practice to apply the GRS test to assess the validity of FF multifactor models. However, our simulation evidence in Section~\ref{sec:durbin_sim} shows that both the GRS test and the HAR test can suffer severe size distortions when the $BD$ condition fails. Because $BD$ is more restrictive, and thus less empirically plausible, than $EBD$, and because the GRS and HAR tests are consistent under $BD$ but not under $EBD$, we do not employ these procedures in our empirical analysis. Instead, we compare the bootstrap bias-corrected Wald test based on the generalized Durbin estimator with Wald tests based on the FD and FCO estimators.

\subsection{Fama--French Multifactor Models}
Following \citet{fama1993crf,fama2015ffa}, we introduce the FF multifactor models. Equation (\ref{eq:FF3}) presents \citetapos{fama1993crf} three-factor (hereafter referred to as FF3) model, which is the most widely known multifactor model:
\begin{equation}
 R_{i,t}-R_{f,t}=\alpha_i+\beta^{Mkt}_{i}(R_{m,t}-R_{f,t})+\beta^{SMB}_{i}SMB_t+\beta^{HML}_{i}HML_t+\varepsilon_{i,t},\label{eq:FF3}
\end{equation}
where $R_{i,t}$ denotes the return on the portfolio $i$ at time $t$, $R_{f,t}$ is the risk-free rate at time $t$, $R_{m,t}$ is the returns on the market portfolio at time $t$, and $\varepsilon_{i,t}$ is the error term for portfolio $i$ at time $t$. The FF3 model expands the CAPM by adding size and value risk factors to capture market anomalies. The size risk factor ($SMB_t$) reflects the empirical regularity that stocks with smaller (or larger) market capitalizations tend to earn higher returns, while the value risk factor ($HML_t$) accounts for the superior performance of stocks with low (or high) price-to-book ratios. Building on the FF3 model, \citet{fama2015ffa} propose a five-factor model (hereafter referred to as FF5), as shown in Equation (\ref{eq:FF5}):
\begin{equation}
\begin{split}
 R_{i,t}-R_{f,t}=\alpha_i+\beta^{Mkt}_{i}(R_{m,t}-R_{f,t})+\beta^{SMB}_{i}SMB_t+\beta^{HML}_{i}HML_t\\
+\beta^{RMW}_{i}RMW_t+\beta^{CMA}_{i}CMA_t+\varepsilon_{i,t}.\label{eq:FF5}
\end{split}
\end{equation}
This model adds two risk factors to FF3: the profitability risk factor ($RMW_t$) and the investment risk factor ($CMA_t$). The models are said to correctly capture the behavior of stock returns if all intercept terms $\alpha_i$ for each portfolio in Equations (\ref{eq:FF3}) and (\ref{eq:FF5}) are not significantly different from zero. Therefore, to assess the validity of the FF multifactor models, we test the following null hypothesis:
\[
 H_0: \ve{\alpha}=\ve{0},\ \ H_1: {\rm{not}} \ H_0
\]
\noindent where $\ve{\alpha}=\left\{\alpha_i\right\}_{i=1}^N$ in each of Equations (\ref{eq:FF3}) and (\ref{eq:FF5}).

\subsection{Dataset}
\citet{fama1993crf,fama2015ffa} provide the returns on the benchmark portfolios and risk factors for examining the FF multifactor models. In these models, the benchmark portfolios are formed by classifying all stocks in the market into either $2\times 3=6 \ {\rm{or}} \ 5\times 5=25$ categories using two criteria (e.g., ``market capitalization (firm size) and book-to-market ratio (MB)'' as proposed by \citet{fama1993crf}). The average return of each portfolio is regarded as the representative portfolio in the market. Other sorting criteria have also been proposed, such as ``market capitalization and profitability (MO)'' and ``market capitalization and investment growth rate (MI)'' (\citet{fama2015ffa}).

\citet{lewellen2010asa} point out that many studies test the FF multifactor models using portfolios sorted by ``market capitalization (firm size)'' and ``book-to-market ratio,'' but the models may not perform as well when using portfolios sorted by other criteria. To consider the possibility that different sorting methods may affect the results, we use three types of benchmark portfolios to compare the performance of the Wald test based on the generalized Durbin estimator proposed in this paper with the GRS test.

Furthermore, the FF multifactor models are often rejected when U.S. data are used, as documented in \citet{fama2012svm,fama2017itf}. In this study, following much of the previous literature, we use U.S. monthly data to examine whether such rejections may be driven by the use of unsuitable test statistics, such as the GRS statistic. All datasets are available from \href{http://mba.tuck.dartmouth.edu/pages/faculty/ken.french/data_library.html}{Professor Kenneth French's website}. For the U.S. monthly data used in this study, the available sample spans July 1990 to December 2025. Our estimation period is restricted to the post-global financial crisis (GFC) period, from October 2008 (the month following the Lehman Brothers bankruptcy) to December 2025. We adopt this window because the estimators we employ are time-invariant and do not explicitly accommodate structural breaks.

A large literature documents that the GFC coincided with shifts in equity-market comovements and factor sensitivities. In particular, \citet{bekaert2014gce} report crisis-period changes in CAPM betas and increases in residual correlations around the Lehman episode. In addition, \citet{lehkonen2015smi} document regime-dependent changes in international stock-market integration during the GFC, suggesting that coefficients in FF multi-factor models may have shifted in that period. While the COVID-19 pandemic is widely viewed as a potential source of structural change in global equity markets, \citet{ndako2025sbg} report that, using \citeapos{bai1998etl}{bai1998etl,bai2003cam,bai2003cvm} multiple structural break tests, no additional break aligned with the onset of COVID-19 is detected across the countries and regions they analyze. Instead, their evidence points to the GFC as the dominant common event driving structural change in global stock market, and they conclude that the GFC generated the most pronounced breaks in those markets. Accordingly, our baseline sample spans October 2008 to December 2025. 

\begin{center}
(Table \ref{fgls_grs_des} around here)
\end{center}

Table \ref{fgls_grs_des} shows the descriptive statistics and the results of the unit root test for the returns on risk factors in the FF multifactor models. The descriptive statistics show that for all datasets, there are no inconsistent data in the perspective of the traditional mean-variance approach. In the estimations, each variable that appeared in the moment conditions should be stationary. We apply the augmented Dickey--Fuller (ADF) test of \citet{said1984tur} to check whether the variables satisfy the stationarity condition. The ADF test rejects the null hypothesis that each variable contains a unit root at the 5\% significance level.\footnote{The ADF test also rejects the null hypothesis that the returns on the 6 and 25 benchmark portfolios contain a unit root at the 5\% significance level, regardless of portfolio sorting method.}

\subsection{Empirical Results}
In this subsection, we evaluate the FF multifactor models using the Wald test based on the generalized Durbin estimator and compare its performance with competing Wald tests based on the FD and FCO estimators. To improve finite-sample size control, we also report a bootstrap bias-corrected Wald test based on the generalized Durbin estimator. Overall, Table \ref{fgls_grs_ff} shows that inference on the null hypothesis $H_0:\ve{\alpha}=\ve{0}$ depends materially on the estimator underlying the Wald statistic. In particular, the competing Wald tests based on the FD and FCO estimators, together with the conventional Wald test based on the generalized Durbin estimator, tend to produce smaller $p$-values and hence more frequent rejections than the bootstrap bias-corrected Wald test based on the generalized Durbin estimator, especially when the number of benchmark portfolios is large.
\begin{center}
(Table \ref{fgls_grs_ff} about here)
\end{center}

Table \ref{fgls_grs_ff} reports the test results for the FF3 and FF5 models. For the FF3 model, the evidence against the null hypothesis is noticeably stronger for the MB25 portfolios than for the MB6 portfolios. In the MB6 case, the bootstrap bias-corrected Wald test based on the generalized Durbin estimator yields $p=0.03$, whereas the other Wald statistics yield $p=0.01$. Thus, the non-bootstrap procedures imply stronger evidence against the null. For MB25, all reported statistics reject the null hypothesis at conventional significance levels; however, the bootstrap bias-corrected Wald test remains distinctly less aggressive, with $p=0.01$, whereas the other procedures yield $p$-values that are essentially zero. These results indicate that, as the cross-sectional dimension increases, the strength of evidence against the FF3 model becomes increasingly sensitive to the construction of the Wald statistic.

For the FF5 model, the results reveal a similar, though more nuanced, pattern across alternative portfolio sorts. For portfolios sorted by MB, the bootstrap bias-corrected Wald test does not reject the null at the 5\% level for MB6, although the other Wald statistics yield $p=0.02$. For MB25, all procedures reject the null, but once again the bootstrap bias-corrected Wald test is less aggressive than its competitors. For portfolios sorted by MI, none of the tests rejects the null for MI6, with all reported $p$-values lying comfortably above conventional significance thresholds. By contrast, for MI25, the bootstrap bias-corrected Wald test yields $p=0.06$ and therefore does not reject at the 5\% level, whereas the other Wald statistics yield $p$-values that are essentially zero and reject decisively. A similar pattern emerges for portfolios sorted by MO. For MO6, none of the statistics rejects the null, whereas for MO25 the bootstrap bias-corrected Wald test continues not to reject the null, with $p=0.19$, even though the competing Wald tests yield much smaller $p$-values and reject at conventional levels.

Taken together, these empirical findings suggest that the bootstrap bias-corrected Wald test based on the generalized Durbin estimator provides more conservative and comparatively stable inference across portfolio sorts and is less prone to frequent rejection in higher-dimensional systems. By contrast, the competing Wald tests based on the FD and FCO estimators, as well as the conventional Wald test based on the generalized Durbin estimator, often generate substantially more aggressive rejections in the same settings. This contrast is consistent with our simulation evidence that misspecification in the dynamic dependence between regressors and errors can induce size distortions in competing procedures, whereas the generalized Durbin approach, when combined with bootstrap bias correction, provides more reliable inference for multi-equation specification testing.

%% file: durbin_con.tex
\section{Conclusion}\label{sec:durbin_con}
This paper proposes a new model specification test for multiple-equation systems in the presence of cross-equation error dependence and dynamic regressor--error dependence. Conventional specification tests for such systems often rely on exogeneity conditions strong enough to ensure consistency of the OLS estimator. In the presence of dynamic regressor--error dependence, these exogeneity conditions may fail, in which case the OLS estimator becomes inconsistent and OLS-based model specification tests no longer deliver valid inference.

To address these limitations, we clarify the relationships among alternative exogeneity conditions and characterize how the validity of competing estimators depends on the underlying dependence structure in multiple-equation systems. Building on this analysis, we propose a generalized Durbin (GD) estimator for multiple-equation systems with an intercept and cross-equation error dependence, and show that the resulting estimator remains consistent under the weakest exogeneity condition considered in the paper. We then derive its asymptotic distribution, construct Wald tests for model specification testing, and develop a bootstrap version to improve finite-sample inference.

The simulation results show that the finite-sample performance of the competing multivariate tests depends crucially on the underlying exogeneity condition. Under the more restrictive conditions such as $BD$ and $GEXOG$, tests based on stronger structural restrictions can perform well and may achieve higher size-adjusted power when correctly specified. However, once those restrictions fail, their finite-sample performance can deteriorate substantially. By contrast, the GD-based procedures remain stable across the exogeneity conditions considered in the paper. In particular, the bootstrap GD test delivers the most reliable finite-sample size control, even when the sample size is as small as $T=100$, while preserving sufficient size-adjusted power.

Our empirical application further illustrates the practical relevance of these findings. We examine the validity of the \citeapos{fama1993crf}{fama1993crf,fama2015ffa} multifactor models and find that inference on their validity depends crucially on the choice of test statistic, especially when the number of portfolios is large. In particular, the bootstrap Wald test based on the generalized Durbin estimator often leaves the null hypothesis unrejected in cases where the competing Wald tests reject it more aggressively. These results suggest that empirical rejections based on more restrictive procedures may partly reflect finite-sample distortions or misspecification of the underlying dependence structure, rather than decisive evidence against the asset pricing model itself. This contrast is consistent with our simulation evidence that misspecification in the dynamic dependence between regressors and errors can induce size distortions in competing tests.

Overall, the proposed bootstrap GD test provides a reliable and practically useful tool for multivariate model specification testing in the presence of cross-equation dependence and dynamic regressor--error dependence. When the true exogeneity structure is unknown, it offers a robust alternative to conventional procedures whose validity depends on stronger maintained assumptions.

%% file: durbin_ack.tex
\section*{Acknowledgments}
The authors thank Tirthatanmoy Das, Katsuhito Iwai, Akitada Kasahara, Genya Kobayashi, Daisuke Nagakura, Yohei Yamamoto, Tatsuma Wada, and participants at the 100th Annual Conference of the Western Economic Association International and the Japan Society of Monetary Economics 2025 Autumn Meeting for helpful comments and suggestions. The author also gratefully acknowledges financial support from the Japan Society for the Promotion of Science, Grant-in-Aid for Scientific Research (Grant Nos. 23H00838 and 23K25535), and from the Japan Science and Technology Agency, Moonshot Research and Development Program (Grant No. JPMJMS2215). All data and programs used in this paper are available upon request.

%% file: durbin_table.tex
\setcounter{table}{0}
\renewcommand{\thetable}{\arabic{table}}

\begin{figure}[p]
 \caption{Relationships among exogeneity conditions under Assumption~\ref{assumption:1}}
 \label{figure_exogeneity}
 \centering
 \includegraphics[scale=0.8]{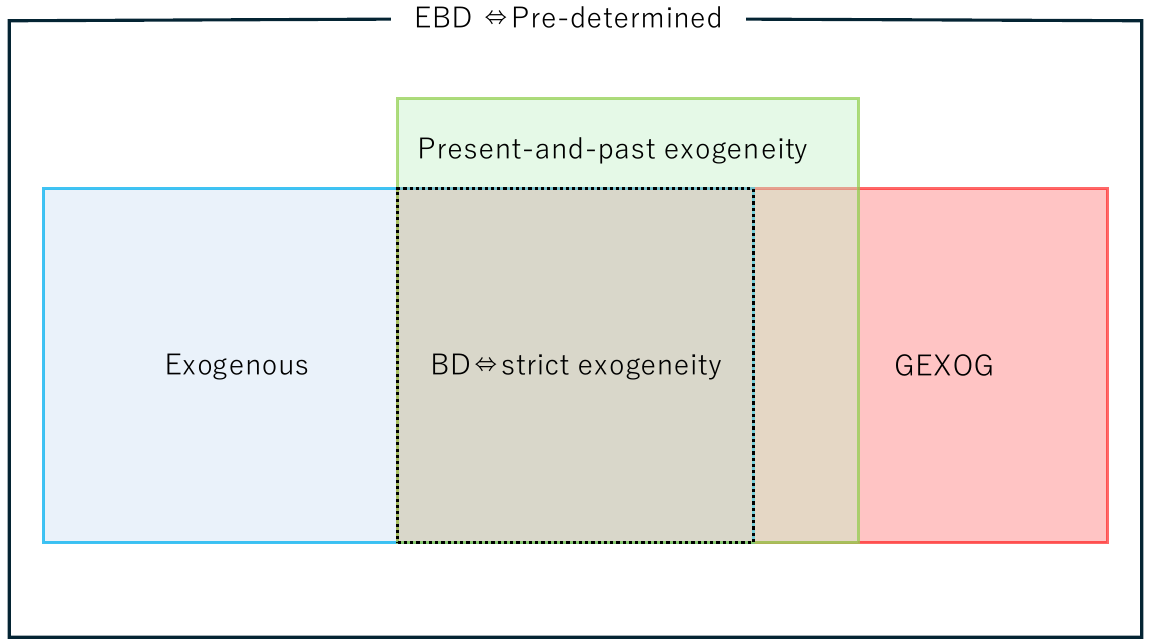}
\end{figure}

\clearpage

\begin{table}[H]
\centering
\caption{Sieve bootstrap procedure for the GD-based Wald test}
\label{tab:gd_var_sieve_bootstrap}
\hrule\vspace{1pt}\hrule
\begin{algorithmic}[1]
\State Estimate the unrestricted GD estimator $\widehat{\ve\kappa}^{\mathrm{GD}}$ and compute the Wald statistic $\mathcal W^{\mathrm{GD}}$.
\State Construct the restricted GD estimator under $H_0:\ve{R}\ve{\kappa}_0=\ve{r}$:
\[
\widetilde{\ve\kappa}^{\mathrm{GD}}
=
\widehat{\ve\kappa}^{\mathrm{GD}}
-
\widehat{\ve V}^{\mathrm{GD}}\ve R^\prime
\bigl(\ve R\widehat{\ve V}^{\mathrm{GD}}\ve R^\prime\bigr)^{-1}
\bigl(\ve R\widehat{\ve\kappa}^{\mathrm{GD}}-\ve r\bigr).
\]
\State With $\widetilde{\ve\kappa}^{\mathrm{GD}}=\bigl(\widetilde{\ve\alpha}^{\mathrm{GD}\prime},\widetilde{\ve\beta}^{\mathrm{GD}\prime}\bigr)^\prime$, compute the null-imposed error terms
\[
\ve{u}_t^0= \ve{y}_t- \widetilde{\ve\alpha}^{\mathrm{GD}}- \ve X_t^\prime \widetilde{\ve\beta}^{\mathrm{GD}},\qquad t=1,\ldots,T,
\]
and form the null-imposed centered joint process
\[
\bar{\ve z}_t^0
=
\bigl(
(\ve x_t-\widehat{\ve\mu}_x)^\prime,
\ve u_t^{0\prime}
\bigr)^\prime,
\qquad
t=1,\ldots,T.
\]
\State Let $\widehat p:=\widehat p_{\mathrm{BIC}}$ denote the lag order selected in the first step of the feasible GD procedure. Fit a VAR($\widehat p$) to $\{\bar{\ve z}_t^0\}_{t=1}^T$, yielding coefficient estimates $\{\widehat{\ve\Psi}_j^0\}_{j=1}^{\widehat p}$ and residuals
\[
\widetilde{\ve\varepsilon}_t
=
\bigl(
\widetilde{\ve\varepsilon}_{x,t}^\prime,
\widetilde{\ve\varepsilon}_{u,t}^\prime
\bigr)^\prime,
\qquad
t=\widehat p+1,\ldots,T.
\]

\For{$b=1,\ldots,B$}
    \For{$t=\widehat p+1,\ldots,T+T_{\mathrm{burn\text{-}in}}$}
        \State Draw $I_t^{(x)}$ and $I_t^{(u)}$ independently with replacement from $\{\widehat p+1,\ldots,T\}$.
        \State Set $\ve\varepsilon_{x,t}^{*(b)}=\widetilde{\ve\varepsilon}_{x,I_t^{(x)}}$, $\ve\varepsilon_{u,t}^{*(b)}=\widetilde{\ve\varepsilon}_{u,I_t^{(u)}}$, and $\ve\varepsilon_t^{*(b)}=[\ve\varepsilon_{x,t}^{*(b)\prime},\ve\varepsilon_{u,t}^{*(b)\prime}]^\prime$
    \EndFor
    
    \State Initialize $\bar{\ve z}_1^{*(b)},\ldots,\bar{\ve z}_{\widehat p}^{*(b)}$ at fixed starting values and generate
    \[
    \bar{\ve z}_t^{*(b)}
    =
    \sum_{j=1}^{\widehat p}\widehat{\ve\Psi}_j^0 \bar{\ve z}_{t-j}^{*(b)}
    +
    \ve\varepsilon_t^{*(b)},
    \qquad
    t=\widehat p+1,\ldots,T+T_{\mathrm{burn\text{-}in}}.
    \]

    \State Discard the first $T_{\mathrm{burn\text{-}in}}$ observations and retain
    \[
    \bar{\ve z}_t^{*(b)}
    =
    \bigl(
    \bar{\ve x}_t^{*(b)\prime},
    \ve u_t^{*(b)\prime}
    \bigr)^\prime,
    \qquad
    t=1,\ldots,T.
    \]

    \State Recover the bootstrap regressors as
    \[
    \ve x_t^{*(b)}
    =
    \widehat{\ve\mu}_x
    +
    \bar{\ve x}_t^{*(b)},
    \qquad
    t=1,\ldots,T,
    \]
    and generate the bootstrap sample
    \[
    \ve y_t^{*(b)}
    =
    \widetilde{\ve\alpha}^{\mathrm{GD}}
    +
    \ve X_t^{*(b)\prime}\widetilde{\ve\beta}^{\mathrm{GD}}
    +
    \ve u_t^{*(b)},
    \qquad
    t=1,\ldots,T.
    \]

    \State Re-estimate the full feasible GD procedure from $\{(\ve y_t^{*(b)},\ve X_t^{*(b)})\}_{t=1}^T$ and compute the bootstrap Wald statistic $\mathcal W^{\mathrm{GD}*(b)}$.
\EndFor
\State Compute $\widehat p_B^*=1/B\sum_{b=1}^B\mathbf 1\left\{\mathcal W^{\mathrm{GD}*(b)} \ge \mathcal W^{\mathrm{GD}}\right\}$.
\State Reject $H_0$ at level $\alpha$ if $\widehat p_B^*\le\alpha$.
\vspace{1pt}\hrule\vspace{1pt}\hrule
\end{algorithmic}
\end{table}

\clearpage
\begin{table}[p]
\caption{Bias and MSE under the $BD$ condition}\label{sim_param_accuracy_bd}
\begin{center}\resizebox{15cm}{!}{
\begin{tabular}{lllccccccccccc} \hline\hline
 &  &  &  & \multicolumn{4}{c}{$\widehat{\mathrm{bias}}(\widehat{\ve\kappa}_T)$} &  & \multicolumn{4}{c}{$\widehat{\mathrm{MSE}}(\widehat{\ve\kappa}_T)$} &  \\
 &  &  &  & GD & FD & FCO & OLS &  & GD & FD & FCO & OLS & \\\cline{5-8}\cline{10-13}
 & $N=5/K=2$ &  &  &  &  &  &  &  &  &  &  &  & \\
 &  & $T=100$ &  & $0.004$ & $0.004$ & $0.004$ & $0.004$ &  & $0.224$ & $0.175$ & $0.174$ & $0.230$ & \\
 &  & $T=200$ &  & $0.003$ & $0.002$ & $0.002$ & $0.003$ &  & $0.107$ & $0.083$ & $0.083$ & $0.112$ & \\
 &  & $T=400$ &  & $0.003$ & $0.003$ & $0.003$ & $0.003$ &  & $0.052$ & $0.041$ & $0.041$ & $0.056$ & \\
 &  & $T=800$ &  & $0.001$ & $0.001$ & $0.001$ & $0.001$ &  & $0.026$ & {\color{blue}{$0.020$}} & {\color{blue}{$0.020$}} & $0.028$ & \\\hdashline
 & $N=5/K=4$ &  &  &  &  &  &  &  &  &  &  &  & \\
 &  & $T=100$ &  & $0.006$ & $0.005$ & $0.005$ & $0.006$ &  & $0.385$ & $0.278$ & $0.276$ & $0.390$ & \\
 &  & $T=200$ &  & $0.004$ & $0.003$ & $0.003$ & $0.004$ &  & $0.179$ & $0.131$ & $0.130$ & $0.190$ & \\
 &  & $T=400$ &  & $0.002$ & $0.002$ & $0.002$ & $0.003$ &  & $0.086$ & $0.063$ & $0.063$ & $0.093$ & \\
 &  & $T=800$ &  & $0.002$ & $0.002$ & $0.002$ & $0.002$ &  & $0.042$ & {\color{blue}{$0.031$}} & {\color{blue}{$0.031$}} & $0.047$ & \\\hdashline
 & $N=10/K=2$ &  &  &  &  &  &  &  &  &  &  &  & \\
 &  & $T=100$ &  & $0.009$ & $0.007$ & $0.007$ & $0.008$ &  & $0.526$ & $0.383$ & $0.378$ & $0.515$ & \\
 &  & $T=200$ &  & $0.005$ & $0.005$ & $0.005$ & $0.005$ &  & $0.244$ & $0.176$ & $0.176$ & $0.252$ & \\
 &  & $T=400$ &  & $0.003$ & $0.003$ & $0.003$ & $0.003$ &  & $0.117$ & $0.085$ & $0.085$ & $0.124$ & \\
 &  & $T=800$ &  & $0.002$ & $0.002$ & $0.002$ & $0.002$ &  & $0.058$ & {\color{blue}{$0.042$}} & {\color{blue}{$0.042$}} & $0.062$ & \\\hdashline
 & $N=10/K=4$ &  &  &  &  &  &  &  &  &  &  &  & \\
 &  & $T=100$ &  & $0.010$ & $0.009$ & $0.009$ & $0.011$ &  & $0.926$ & $0.623$ & $0.612$ & $0.891$ & \\
 &  & $T=200$ &  & $0.006$ & $0.005$ & $0.005$ & $0.006$ &  & $0.417$ & $0.279$ & $0.278$ & $0.431$ & \\
 &  & $T=400$ &  & $0.004$ & $0.003$ & $0.003$ & $0.005$ &  & $0.198$ & $0.132$ & $0.132$ & $0.213$ & \\
 &  & $T=800$ &  & $0.003$ & $0.003$ & $0.003$ & $0.003$ &  & $0.097$ & {\color{blue}{$0.065$}} & {\color{blue}{$0.065$}} & $0.106$ & \\\hline\hline
\end{tabular}}
{\resizebox{24cm}{!}{\begin{minipage}{700pt}
\vspace{3mm}
{\underline{Note:}} R version 4.5.3 was used to compute the statistics.
\end{minipage}}}
\end{center}
\end{table}

\clearpage

\begin{table}[p]
\caption{Bias and MSE under the $GEXOG$ condition}\label{sim_param_accuracy_gexog}
\begin{center}\resizebox{15cm}{!}{
\begin{tabular}{lllccccccccccc} \hline\hline
 &  &  &  & \multicolumn{4}{c}{$\widehat{\mathrm{bias}}(\widehat{\ve\kappa}_T)$} &  & \multicolumn{4}{c}{$\widehat{\mathrm{MSE}}(\widehat{\ve\kappa}_T)$} &  \\
 &  &  &  & GD & FD & FCO & OLS &  & GD & FD & FCO & OLS & \\\cline{5-8}\cline{10-13}
 & $N=5/K=2$ &  &  &  &  &  &  &  &  &  &  &  & \\
 &  & $T=100$ &  & $0.004$ & $0.005$ & $0.010$ & $0.029$ &  & $0.225$ & $0.169$ & $0.174$ & $0.322$ & \\
 &  & $T=200$ &  & $0.003$ & $0.002$ & $0.007$ & $0.028$ &  & $0.107$ & $0.081$ & $0.088$ & $0.226$ & \\
 &  & $T=400$ &  & $0.003$ & $0.003$ & $0.007$ & $0.027$ &  & $0.052$ & $0.040$ & $0.047$ & $0.182$ & \\
 &  & $T=800$ &  & $0.001$ & $0.001$ & $0.006$ & {\color{red}{$0.028$}} &  & $0.026$ & $0.020$ & $0.027$ & {\color{red}{$0.159$}} & \\\hdashline
 & $N=5/K=4$ &  &  &  &  &  &  &  &  &  &  &  & \\
 &  & $T=100$ &  & $0.006$ & $0.005$ & $0.007$ & $0.019$ &  & $0.386$ & $0.272$ & $0.279$ & $0.515$ & \\
 &  & $T=200$ &  & $0.003$ & $0.003$ & $0.004$ & $0.018$ &  & $0.179$ & $0.128$ & $0.134$ & $0.345$ & \\
 &  & $T=400$ &  & $0.002$ & $0.002$ & $0.003$ & $0.019$ &  & $0.086$ & $0.061$ & $0.067$ & $0.265$ & \\
 &  & $T=800$ &  & $0.002$ & $0.002$ & $0.003$ & {\color{red}{$0.020$}} &  & $0.042$ & $0.030$ & $0.035$ & {\color{red}{$0.223$}} & \\\hdashline
 & $N=10/K=2$ &  &  &  &  &  &  &  &  &  &  &  & \\
 &  & $T=100$ &  & $0.009$ & $0.006$ & $0.009$ & $0.017$ &  & $0.527$ & $0.366$ & $0.366$ & $0.599$ & \\
 &  & $T=200$ &  & $0.005$ & $0.005$ & $0.006$ & $0.015$ &  & $0.245$ & $0.169$ & $0.180$ & $0.393$ & \\
 &  & $T=400$ &  & $0.003$ & $0.003$ & $0.005$ & $0.015$ &  & $0.118$ & $0.081$ & $0.094$ & $0.291$ & \\
 &  & $T=800$ &  & $0.002$ & $0.002$ & $0.004$ & {\color{red}{$0.014$}} &  & $0.058$ & $0.040$ & $0.053$ & {\color{red}{$0.243$}} & \\\hdashline
 & $N=10/K=4$ &  &  &  &  &  &  &  &  &  &  &  & \\
 &  & $T=100$ &  & $0.010$ & $0.009$ & $0.010$ & $0.016$ &  & $0.927$ & $0.603$ & $0.607$ & $1.018$ & \\
 &  & $T=200$ &  & $0.006$ & $0.005$ & $0.005$ & $0.013$ &  & $0.417$ & $0.271$ & $0.282$ & $0.636$ & \\
 &  & $T=400$ &  & $0.004$ & $0.003$ & $0.004$ & $0.012$ &  & $0.199$ & $0.128$ & $0.139$ & $0.450$ & \\
 &  & $T=800$ &  & $0.003$ & $0.003$ & $0.004$ & {\color{red}{$0.013$}} &  & $0.097$ & $0.063$ & $0.072$ & {\color{red}{$0.360$}} & \\\hline\hline
\end{tabular}}
{\resizebox{24cm}{!}{\begin{minipage}{700pt}
\vspace{3mm}
{\underline{Note:}} As for Table \ref{sim_param_accuracy_bd}.
\end{minipage}}}
\end{center}
\end{table}

\clearpage

\begin{table}[p]
\caption{Bias and MSE under the $EBD$ condition}\label{sim_param_accuracy_ebd}
\begin{center}\resizebox{15cm}{!}{
\begin{tabular}{lllccccccccccc} \hline\hline
 &  &  &  & \multicolumn{4}{c}{$\widehat{\mathrm{bias}}(\widehat{\ve\kappa}_T)$} &  & \multicolumn{4}{c}{$\widehat{\mathrm{MSE}}(\widehat{\ve\kappa}_T)$} &  \\
 &  &  &  & GD & FD & FCO & OLS &  & GD & FD & FCO & OLS & \\\cline{5-8}\cline{10-13}
 & $N=5/K=2$ &  &  &  &  &  &  &  &  &  &  &  & \\
 &  & $T=100$ &  & $0.004$ & $0.011$ & $0.022$ & $0.039$ &  & $0.228$ & $0.206$ & $0.212$ & $0.348$ & \\
 &  & $T=200$ &  & $0.003$ & $0.012$ & $0.021$ & $0.038$ &  & $0.109$ & $0.116$ & $0.125$ & $0.252$ & \\
 &  & $T=400$ &  & $0.003$ & $0.013$ & $0.021$ & $0.038$ &  & $0.053$ & $0.073$ & $0.082$ & $0.206$ & \\
 &  & $T=800$ &  & {\color{blue}{$0.001$}} & {\color{red}{$0.013$}} & {\color{red}{$0.020$}} & {\color{red}{$0.038$}} &  & {\color{blue}{$0.027$}} & {\color{red}{$0.052$}} & {\color{red}{$0.061$}} & {\color{red}{$0.183$}} & \\\hdashline
 & $N=5/K=4$ &  &  &  &  &  &  &  &  &  &  &  & \\
 &  & $T=100$ &  & $0.006$ & $0.006$ & $0.010$ & $0.025$ &  & $0.388$ & $0.310$ & $0.316$ & $0.544$ & \\
 &  & $T=200$ &  & $0.003$ & $0.004$ & $0.006$ & $0.024$ &  & $0.180$ & $0.162$ & $0.167$ & $0.371$ & \\
 &  & $T=400$ &  & $0.002$ & $0.004$ & $0.006$ & $0.025$ &  & $0.086$ & $0.094$ & $0.098$ & $0.290$ & \\
 &  & $T=800$ &  & {\color{blue}{$0.002$}} & {\color{red}{$0.003$}} & {\color{red}{$0.005$}} & {\color{red}{$0.026$}} &  & {\color{blue}{$0.043$}} & {\color{red}{$0.062$}} & {\color{red}{$0.066$}} & {\color{red}{$0.247$}} & \\\hdashline
 & $N=10/K=2$ &  &  &  &  &  &  &  &  &  &  &  & \\
 &  & $T=100$ &  & $0.008$ & $0.010$ & $0.017$ & $0.023$ &  & $0.530$ & $0.424$ & $0.414$ & $0.620$ & \\
 &  & $T=200$ &  & $0.005$ & $0.009$ & $0.015$ & $0.021$ &  & $0.246$ & $0.221$ & $0.227$ & $0.413$ & \\
 &  & $T=400$ &  & $0.003$ & $0.009$ & $0.015$ & $0.022$ &  & $0.118$ & $0.132$ & $0.140$ & $0.311$ & \\
 &  & $T=800$ &  & {\color{blue}{$0.002$}} & {\color{red}{$0.009$}} & {\color{red}{$0.014$}} & {\color{red}{$0.021$}} &  & {\color{blue}{$0.059$}} & {\color{red}{$0.089$}} & {\color{red}{$0.099$}} & {\color{red}{$0.263$}} & \\\hdashline
 & $N=10/K=4$ &  &  &  &  &  &  &  &  &  &  &  & \\
 &  & $T=100$ &  & $0.010$ & $0.009$ & $0.012$ & $0.019$ &  & $0.929$ & $0.660$ & $0.654$ & $1.046$ & \\
 &  & $T=200$ &  & $0.006$ & $0.005$ & $0.007$ & $0.016$ &  & $0.417$ & $0.323$ & $0.328$ & $0.660$ & \\
 &  & $T=400$ &  & $0.004$ & $0.004$ & $0.005$ & $0.016$ &  & $0.199$ & $0.179$ & $0.184$ & $0.473$ & \\
 &  & $T=800$ &  & {\color{blue}{$0.003$}} & {\color{red}{$0.004$}} & {\color{red}{$0.005$}} & {\color{red}{$0.017$}} &  & {\color{blue}{$0.097$}} & {\color{red}{$0.112$}} & {\color{red}{$0.117$}} & {\color{red}{$0.382$}} & \\\hline\hline
\end{tabular}}
{\resizebox{24cm}{!}{\begin{minipage}{700pt}
\vspace{3mm}
{\underline{Note:}} As for Table \ref{sim_param_accuracy_bd}.
\end{minipage}}}
\end{center}
\end{table}

\clearpage

\begin{landscape}
\begin{table}[p]
\caption{Empirical rejection rates under the null hypothesis $H_0:\bm{\alpha}_0=\mathbf{0}_{N}$ when the $BD$ holds}\label{sim_bd_nh}
\begin{center}\resizebox{25cm}{!}{
\begin{tabular}{lllccccccccccccccccccccccccc}\hline\hline
 &  &  &  & \multicolumn{3}{c}{$\widetilde{\mathcal{W}}^{GD}$}  &  & \multicolumn{3}{c}{$\mathcal{W}^{GD}$}  &  & \multicolumn{3}{c}{$\mathcal{W}^{FD}$}  &  & \multicolumn{3}{c}{$\mathcal{W}^{FCO}$}  &  & \multicolumn{3}{c}{$HAR$} &  & \multicolumn{3}{c}{$GRS$}  & \\
 &  &  &  & 10\% & 5\% & 1\% &  & 10\% & 5\% & 1\% &  & 10\% & 5\% & 1\% &  & 10\% & 5\% & 1\% &  & 10\% & 5\% & 1\% &  & 10\% & 5\% & 1\% & \\\cline{5-7}\cline{9-11}\cline{13-15}\cline{17-19}\cline{21-23}\cline{25-27}
 & $N=5/K=2$ &  &  &  &  &  &  &  &  &  &  &  &  &  &  &  &  &  &  &  &  &  &  &  &  &  & \\
 &  & $T=100$ &  & {\color{blue}{$0.095$}} & {\color{blue}{$0.048$}} & {\color{blue}{$0.010$}} &  & $0.169$ & $0.106$ & $0.034$ &  & $0.198$ & $0.128$ & $0.046$ &  & $0.185$ & $0.117$ & $0.041$ &  & {\color{blue}{$0.093$}} & {\color{blue}{$0.044$}} & {\color{blue}{$0.011$}} &  & $0.121$ & $0.076$ & $0.033$ & \\
 &  & $T=200$ &  & $0.097$ & $0.050$ & $0.010$ &  & $0.131$ & $0.072$ & $0.021$ &  & $0.145$ & $0.079$ & $0.024$ &  & $0.138$ & $0.076$ & $0.022$ &  & $0.089$ & $0.044$ & $0.009$ &  & $0.116$ & $0.078$ & $0.032$ & \\
 &  & $T=400$ &  & $0.101$ & $0.048$ & $0.009$ &  & $0.118$ & $0.060$ & $0.013$ &  & $0.123$ & $0.065$ & $0.015$ &  & $0.120$ & $0.063$ & $0.014$ &  & $0.095$ & $0.047$ & $0.010$ &  & $0.124$ & $0.081$ & $0.032$ & \\
 &  & $T=800$ &  & $0.103$ & $0.051$ & $0.009$ &  & $0.112$ & $0.056$ & $0.012$ &  & $0.114$ & $0.057$ & $0.013$ &  & $0.112$ & $0.056$ & $0.012$ &  & $0.099$ & $0.048$ & $0.010$ &  & {\color{red}{$0.122$}} & {\color{red}{$0.083$}} & {\color{red}{$0.032$}} & \\\hdashline
 & $N=5/K=4$ &  &  &  &  &  &  &  &  &  &  &  &  &  &  &  &  &  &  &  &  &  &  &  &  &  & \\
 &  & $T=100$ &  & {\color{blue}{$0.099$}} & {\color{blue}{$0.049$}} & {\color{blue}{$0.010$}} &  & $0.178$ & $0.108$ & $0.035$ &  & $0.210$ & $0.139$ & $0.053$ &  & $0.180$ & $0.114$ & $0.041$ &  & {\color{blue}{$0.098$}} & {\color{blue}{$0.048$}} & {\color{blue}{$0.012$}} &  & $0.111$ & $0.068$ & $0.024$ & \\
 &  & $T=200$ &  & $0.099$ & $0.047$ & $0.010$ &  & $0.136$ & $0.077$ & $0.022$ &  & $0.153$ & $0.088$ & $0.026$ &  & $0.140$ & $0.079$ & $0.022$ &  & $0.093$ & $0.047$ & $0.008$ &  & $0.116$ & $0.074$ & $0.029$ & \\
 &  & $T=400$ &  & $0.101$ & $0.051$ & $0.010$ &  & $0.117$ & $0.063$ & $0.015$ &  & $0.124$ & $0.066$ & $0.016$ &  & $0.117$ & $0.062$ & $0.015$ &  & $0.099$ & $0.052$ & $0.012$ &  & $0.114$ & $0.073$ & $0.026$ & \\
 &  & $T=800$ &  & $0.100$ & $0.050$ & $0.012$ &  & $0.109$ & $0.058$ & $0.014$ &  & $0.111$ & $0.061$ & $0.014$ &  & $0.108$ & $0.059$ & $0.014$ &  & $0.097$ & $0.050$ & $0.012$ &  & $0.116$ & {\color{red}{$0.071$}} & {\color{red}{$0.028$}} & \\\hdashline
 & $N=10/K=2$ &  &  &  &  &  &  &  &  &  &  &  &  &  &  &  &  &  &  &  &  &  &  &  &  &  & \\
 &  & $T=100$ &  & {\color{blue}{$0.095$}} & {\color{blue}{$0.044$}} & {\color{blue}{$0.009$}} &  & $0.291$ & $0.201$ & $0.087$ &  & $0.365$ & $0.268$ & $0.132$ &  & $0.332$ & $0.242$ & $0.114$ &  & $0.067$ & $0.029$ & $0.006$ &  & $0.085$ & $0.051$ & $0.018$ & \\
 &  & $T=200$ &  & $0.098$ & $0.047$ & $0.010$ &  & $0.183$ & $0.109$ & $0.036$ &  & $0.210$ & $0.134$ & $0.046$ &  & $0.199$ & $0.125$ & $0.042$ &  & $0.078$ & $0.036$ & $0.006$ &  & $0.087$ & $0.051$ & $0.020$ & \\
 &  & $T=400$ &  & $0.096$ & $0.051$ & $0.012$ &  & $0.135$ & $0.076$ & $0.022$ &  & $0.152$ & $0.089$ & $0.026$ &  & $0.149$ & $0.086$ & $0.024$ &  & $0.090$ & $0.045$ & $0.008$ &  & $0.083$ & $0.049$ & $0.017$ & \\
 &  & $T=800$ &  & $0.099$ & $0.050$ & $0.011$ &  & $0.118$ & $0.063$ & $0.016$ &  & $0.126$ & $0.068$ & $0.018$ &  & $0.124$ & $0.066$ & $0.017$ &  & $0.101$ & $0.052$ & $0.013$ &  & $0.083$ & $0.050$ & $0.017$ & \\\hdashline
 & $N=10/K=4$ &  &  &  &  &  &  &  &  &  &  &  &  &  &  &  &  &  &  &  &  &  &  &  &  &  & \\
 &  & $T=100$ &  & {\color{blue}{$0.095$}} & {\color{blue}{$0.046$}} & {\color{blue}{$0.009$}} &  & $0.306$ & $0.210$ & $0.087$ &  & $0.403$ & $0.299$ & $0.150$ &  & $0.336$ & $0.239$ & $0.108$ &  & $0.081$ & $0.037$ & $0.009$ &  & $0.080$ & $0.046$ & $0.014$ & \\
 &  & $T=200$ &  & $0.096$ & $0.051$ & $0.009$ &  & $0.184$ & $0.109$ & $0.036$ &  & $0.221$ & $0.140$ & $0.052$ &  & $0.196$ & $0.121$ & $0.042$ &  & $0.082$ & $0.041$ & $0.007$ &  & $0.078$ & $0.047$ & $0.016$ & \\
 &  & $T=400$ &  & $0.099$ & $0.048$ & $0.008$ &  & $0.141$ & $0.078$ & $0.018$ &  & $0.155$ & $0.088$ & $0.025$ &  & $0.146$ & $0.080$ & $0.023$ &  & $0.101$ & $0.054$ & $0.010$ &  & $0.084$ & $0.047$ & $0.016$ & \\
 &  & $T=800$ &  & $0.101$ & $0.048$ & $0.010$ &  & $0.114$ & $0.061$ & $0.012$ &  & $0.121$ & $0.063$ & $0.015$ &  & $0.117$ & $0.060$ & $0.014$ &  & $0.101$ & $0.049$ & $0.010$ &  & $0.077$ & $0.045$ & $0.014$ & \\\hline\hline
\end{tabular}}
{\resizebox{24cm}{!}{\begin{minipage}{700pt}
\vspace{3mm}
{\underline{Note:}} R version 4.5.3 was used to compute the statistics.
\end{minipage}}}
\end{center}
\end{table}
\end{landscape}

\clearpage

\begin{landscape}
\begin{table}[p]
\caption{Empirical rejection rates under the null hypothesis $H_0:\bm{\alpha}_0=\mathbf{0}_{N}$ when the $GEXOG$ holds}\label{sim_gexog_nh}
\begin{center}\resizebox{25cm}{!}{
\begin{tabular}{lllccccccccccccccccccccccccc}\hline\hline
 &  &  &  & \multicolumn{3}{c}{$\widetilde{\mathcal{W}}^{GD}$}  &  & \multicolumn{3}{c}{$\mathcal{W}^{GD}$}  &  & \multicolumn{3}{c}{$\mathcal{W}^{FD}$}  &  & \multicolumn{3}{c}{$\mathcal{W}^{FCO}$}  &  & \multicolumn{3}{c}{$HAR$} &  & \multicolumn{3}{c}{$GRS$}  & \\
 &  &  &  & 10\% & 5\% & 1\% &  & 10\% & 5\% & 1\% &  & 10\% & 5\% & 1\% &  & 10\% & 5\% & 1\% &  & 10\% & 5\% & 1\% &  & 10\% & 5\% & 1\% & \\\cline{5-7}\cline{9-11}\cline{13-15}\cline{17-19}\cline{21-23}\cline{25-27}
 & $N=5/K=2$ &  &  &  &  &  &  &  &  &  &  &  &  &  &  &  &  &  &  &  &  &  &  &  &  &  & \\
 &  & $T=100$ &  & {\color{blue}{$0.099$}} & {\color{blue}{$0.048$}} & {\color{blue}{$0.011$}} &  & $0.169$ & $0.102$ & $0.034$ &  & $0.196$ & $0.122$ & $0.046$ &  & $0.187$ & $0.118$ & $0.041$ &  & $0.150$ & $0.076$ & $0.021$ &  & $0.155$ & $0.101$ & $0.040$ & \\
 &  & $T=200$ &  & $0.096$ & $0.048$ & $0.010$ &  & $0.133$ & $0.072$ & $0.021$ &  & $0.146$ & $0.080$ & $0.023$ &  & $0.148$ & $0.084$ & $0.025$ &  & $0.219$ & $0.128$ & $0.041$ &  & $0.192$ & $0.130$ & $0.054$ & \\
 &  & $T=400$ &  & $0.101$ & $0.048$ & $0.009$ &  & $0.120$ & $0.062$ & $0.013$ &  & $0.127$ & $0.067$ & $0.015$ &  & $0.145$ & $0.080$ & $0.021$ &  & $0.425$ & $0.302$ & $0.132$ &  & $0.311$ & $0.213$ & $0.103$ & \\
 &  & $T=800$ &  & $0.103$ & $0.052$ & $0.010$ &  & $0.111$ & $0.058$ & $0.012$ &  & $0.113$ & $0.058$ & $0.013$ &  & {\color{red}{$0.156$}} & {\color{red}{$0.088$}} & {\color{red}{$0.025$}} &  & {\color{red}{$0.679$}} & {\color{red}{$0.567$}} & {\color{red}{$0.360$}} &  & {\color{red}{$0.527$}} & {\color{red}{$0.414$}} & {\color{red}{$0.225$}} & \\\hdashline
 & $N=5/K=4$ &  &  &  &  &  &  &  &  &  &  &  &  &  &  &  &  &  &  &  &  &  &  &  &  &  & \\
 &  & $T=100$ &  & {\color{blue}{$0.098$}} & {\color{blue}{$0.049$}} & {\color{blue}{$0.010$}} &  & $0.178$ & $0.108$ & $0.036$ &  & $0.216$ & $0.136$ & $0.053$ &  & $0.185$ & $0.113$ & $0.041$ &  & $0.164$ & $0.087$ & $0.026$ &  & $0.155$ & $0.098$ & $0.035$ & \\
 &  & $T=200$ &  & $0.097$ & $0.048$ & $0.010$ &  & $0.136$ & $0.077$ & $0.022$ &  & $0.152$ & $0.089$ & $0.026$ &  & $0.142$ & $0.079$ & $0.024$ &  & $0.248$ & $0.153$ & $0.047$ &  & $0.212$ & $0.137$ & $0.054$ & \\
 &  & $T=400$ &  & $0.101$ & $0.051$ & $0.011$ &  & $0.118$ & $0.065$ & $0.015$ &  & $0.121$ & $0.067$ & $0.017$ &  & $0.130$ & $0.071$ & $0.018$ &  & $0.459$ & $0.338$ & $0.158$ &  & $0.352$ & $0.250$ & $0.111$ & \\
 &  & $T=800$ &  & $0.100$ & $0.050$ & $0.012$ &  & $0.108$ & $0.058$ & $0.013$ &  & $0.110$ & $0.060$ & $0.015$ &  & $0.126$ & $0.071$ & $0.018$ &  & {\color{red}{$0.712$}} & {\color{red}{$0.613$}} & {\color{red}{$0.414$}} &  & {\color{red}{$0.592$}} & {\color{red}{$0.476$}} & {\color{red}{$0.271$}} & \\\hdashline
 & $N=10/K=2$ &  &  &  &  &  &  &  &  &  &  &  &  &  &  &  &  &  &  &  &  &  &  &  &  &  & \\
 &  & $T=100$ &  & {\color{blue}{$0.094$}} & {\color{blue}{$0.046$}} & {\color{blue}{$0.009$}} &  & $0.285$ & $0.199$ & $0.088$ &  & $0.363$ & $0.265$ & $0.134$ &  & $0.338$ & $0.250$ & $0.121$ &  & $0.099$ & $0.047$ & $0.012$ &  & $0.109$ & $0.067$ & $0.025$ & \\
 &  & $T=200$ &  & $0.098$ & $0.048$ & $0.010$ &  & $0.182$ & $0.110$ & $0.036$ &  & $0.208$ & $0.128$ & $0.048$ &  & $0.222$ & $0.140$ & $0.051$ &  & $0.196$ & $0.114$ & $0.025$ &  & $0.158$ & $0.098$ & $0.038$ & \\
 &  & $T=400$ &  & $0.098$ & $0.052$ & $0.012$ &  & $0.134$ & $0.074$ & $0.021$ &  & $0.149$ & $0.088$ & $0.026$ &  & $0.188$ & $0.113$ & $0.035$ &  & $0.417$ & $0.300$ & $0.120$ &  & $0.248$ & $0.162$ & $0.066$ & \\
 &  & $T=800$ &  & $0.100$ & $0.051$ & $0.011$ &  & $0.121$ & $0.065$ & $0.015$ &  & $0.126$ & $0.069$ & $0.018$ &  & {\color{red}{$0.194$}} & {\color{red}{$0.117$}} & {\color{red}{$0.039$}} &  & {\color{red}{$0.740$}} & {\color{red}{$0.624$}} & {\color{red}{$0.420$}} &  & {\color{red}{$0.479$}} & {\color{red}{$0.363$}} & {\color{red}{$0.188$}} & \\\hdashline
 & $N=10/K=4$ &  &  &  &  &  &  &  &  &  &  &  &  &  &  &  &  &  &  &  &  &  &  &  &  &  & \\
 &  & $T=100$ &  & {\color{blue}{$0.095$}} & {\color{blue}{$0.045$}} & {\color{blue}{$0.009$}} &  & $0.302$ & $0.206$ & $0.086$ &  & $0.401$ & $0.295$ & $0.147$ &  & $0.340$ & $0.245$ & $0.114$ &  & $0.121$ & $0.058$ & $0.013$ &  & $0.115$ & $0.071$ & $0.024$ & \\
 &  & $T=200$ &  & $0.096$ & $0.049$ & $0.009$ &  & $0.184$ & $0.111$ & $0.036$ &  & $0.219$ & $0.143$ & $0.051$ &  & $0.212$ & $0.134$ & $0.044$ &  & $0.226$ & $0.140$ & $0.033$ &  & $0.171$ & $0.105$ & $0.040$ & \\
 &  & $T=400$ &  & $0.098$ & $0.047$ & $0.008$ &  & $0.140$ & $0.078$ & $0.018$ &  & $0.154$ & $0.086$ & $0.025$ &  & $0.162$ & $0.093$ & $0.030$ &  & $0.471$ & $0.349$ & $0.156$ &  & $0.303$ & $0.206$ & $0.088$ & \\
 &  & $T=800$ &  & $0.101$ & $0.047$ & $0.010$ &  & $0.115$ & $0.061$ & $0.013$ &  & $0.118$ & $0.063$ & $0.015$ &  & {\color{red}{$0.158$}} & {\color{red}{$0.084$}} & {\color{red}{$0.023$}} &  & {\color{red}{$0.774$}} & {\color{red}{$0.674$}} & {\color{red}{$0.482$}} &  & {\color{red}{$0.560$}} & {\color{red}{$0.447$}} & {\color{red}{$0.251$}} & \\\hline\hline
\end{tabular}}
{\resizebox{24cm}{!}{\begin{minipage}{700pt}
\vspace{3mm}
{\underline{Note:}} As for Table \ref{sim_bd_nh}.
\end{minipage}}}
\end{center}
\end{table}
\end{landscape}

\clearpage

\begin{landscape}
\begin{table}[p]
\caption{Empirical rejection rates under the null hypothesis $H_0:\bm{\alpha}_0=\mathbf{0}_{N}$ when the $EBD$ holds}\label{sim_ebd_nh}
\begin{center}\resizebox{25cm}{!}{
\begin{tabular}{lllccccccccccccccccccccccccc}\hline\hline
 &  &  &  & \multicolumn{3}{c}{$\widetilde{\mathcal{W}}^{GD}$}  &  & \multicolumn{3}{c}{$\mathcal{W}^{GD}$}  &  & \multicolumn{3}{c}{$\mathcal{W}^{FD}$}  &  & \multicolumn{3}{c}{$\mathcal{W}^{FCO}$}  &  & \multicolumn{3}{c}{$HAR$} &  & \multicolumn{3}{c}{$GRS$}  & \\
 &  &  &  & 10\% & 5\% & 1\% &  & 10\% & 5\% & 1\% &  & 10\% & 5\% & 1\% &  & 10\% & 5\% & 1\% &  & 10\% & 5\% & 1\% &  & 10\% & 5\% & 1\% & \\\cline{5-7}\cline{9-11}\cline{13-15}\cline{17-19}\cline{21-23}\cline{25-27}
 & $N=5/K=2$ &  &  &  &  &  &  &  &  &  &  &  &  &  &  &  &  &  &  &  &  &  &  &  &  &  & \\
 &  & $T=100$ &  & {\color{blue}{$0.100$}} & {\color{blue}{$0.050$}} & {\color{blue}{$0.011$}} &  & $0.171$ & $0.104$ & $0.035$ &  & $0.233$ & $0.152$ & $0.059$ &  & $0.217$ & $0.135$ & $0.050$ &  & $0.159$ & $0.082$ & $0.023$ &  & $0.166$ & $0.109$ & $0.041$ & \\
 &  & $T=200$ &  & $0.096$ & $0.047$ & $0.010$ &  & $0.136$ & $0.076$ & $0.021$ &  & $0.207$ & $0.129$ & $0.043$ &  & $0.199$ & $0.122$ & $0.042$ &  & $0.241$ & $0.147$ & $0.048$ &  & $0.216$ & $0.143$ & $0.060$ & \\
 &  & $T=400$ &  & $0.100$ & $0.050$ & $0.010$ &  & $0.116$ & $0.061$ & $0.014$ &  & $0.232$ & $0.149$ & $0.048$ &  & $0.245$ & $0.157$ & $0.054$ &  & $0.469$ & $0.348$ & $0.164$ &  & $0.356$ & $0.251$ & $0.120$ & \\
 &  & $T=800$ &  & $0.103$ & $0.052$ & $0.010$ &  & $0.110$ & $0.057$ & $0.012$ &  & {\color{red}{$0.312$}} & {\color{red}{$0.212$}} & {\color{red}{$0.080$}} &  & {\color{red}{$0.345$}} & {\color{red}{$0.243$}} & {\color{red}{$0.101$}} &  & {\color{red}{$0.718$}} & {\color{red}{$0.619$}} & {\color{red}{$0.432$}} &  & {\color{red}{$0.586$}} & {\color{red}{$0.471$}} & {\color{red}{$0.281$}} & \\\hdashline
 & $N=5/K=4$ &  &  &  &  &  &  &  &  &  &  &  &  &  &  &  &  &  &  &  &  &  &  &  &  &  & \\
 &  & $T=100$ &  & {\color{blue}{$0.098$}} & {\color{blue}{$0.046$}} & {\color{blue}{$0.009$}} &  & $0.177$ & $0.107$ & $0.034$ &  & $0.242$ & $0.161$ & $0.065$ &  & $0.199$ & $0.128$ & $0.046$ &  & $0.169$ & $0.092$ & $0.030$ &  & $0.163$ & $0.103$ & $0.037$ & \\
 &  & $T=200$ &  & $0.097$ & $0.047$ & $0.011$ &  & $0.138$ & $0.077$ & $0.022$ &  & $0.198$ & $0.120$ & $0.041$ &  & $0.174$ & $0.106$ & $0.034$ &  & $0.264$ & $0.166$ & $0.054$ &  & $0.226$ & $0.148$ & $0.060$ & \\
 &  & $T=400$ &  & $0.099$ & $0.052$ & $0.010$ &  & $0.117$ & $0.065$ & $0.014$ &  & $0.204$ & $0.122$ & $0.040$ &  & $0.192$ & $0.117$ & $0.033$ &  & $0.485$ & $0.369$ & $0.182$ &  & $0.379$ & $0.276$ & $0.131$ & \\
 &  & $T=800$ &  & $0.099$ & $0.051$ & $0.011$ &  & $0.109$ & $0.058$ & $0.014$ &  & {\color{red}{$0.269$}} & {\color{red}{$0.174$}} & {\color{red}{$0.065$}} &  & {\color{red}{$0.264$}} & {\color{red}{$0.170$}} & {\color{red}{$0.063$}} &  & {\color{red}{$0.734$}} & {\color{red}{$0.645$}} & {\color{red}{$0.457$}} &  & {\color{red}{$0.626$}} & {\color{red}{$0.517$}} & {\color{red}{$0.315$}} & \\\hdashline
 & $N=10/K=2$ &  &  &  &  &  &  &  &  &  &  &  &  &  &  &  &  &  &  &  &  &  &  &  &  &  & \\
 &  & $T=100$ &  & {\color{blue}{$0.094$}} & {\color{blue}{$0.045$}} & {\color{blue}{$0.009$}} &  & $0.292$ & $0.206$ & $0.089$ &  & $0.408$ & $0.306$ & $0.159$ &  & $0.371$ & $0.271$ & $0.133$ &  & $0.104$ & $0.048$ & $0.012$ &  & $0.113$ & $0.070$ & $0.025$ & \\
 &  & $T=200$ &  & $0.098$ & $0.048$ & $0.010$ &  & $0.181$ & $0.109$ & $0.036$ &  & $0.274$ & $0.182$ & $0.073$ &  & $0.272$ & $0.179$ & $0.072$ &  & $0.210$ & $0.124$ & $0.027$ &  & $0.164$ & $0.104$ & $0.041$ & \\
 &  & $T=400$ &  & $0.098$ & $0.051$ & $0.012$ &  & $0.136$ & $0.076$ & $0.023$ &  & $0.265$ & $0.174$ & $0.064$ &  & $0.281$ & $0.191$ & $0.070$ &  & $0.454$ & $0.333$ & $0.139$ &  & $0.269$ & $0.181$ & $0.078$ & \\
 &  & $T=800$ &  & $0.102$ & $0.051$ & $0.011$ &  & $0.121$ & $0.063$ & $0.015$ &  & {\color{red}{$0.345$}} & {\color{red}{$0.241$}} & {\color{red}{$0.101$}} &  & {\color{red}{$0.397$}} & {\color{red}{$0.283$}} & {\color{red}{$0.128$}} &  & {\color{red}{$0.776$}} & {\color{red}{$0.670$}} & {\color{red}{$0.478$}} &  & {\color{red}{$0.521$}} & {\color{red}{$0.408$}} & {\color{red}{$0.220$}} & \\\hdashline
 & $N=10/K=4$ &  &  &  &  &  &  &  &  &  &  &  &  &  &  &  &  &  &  &  &  &  &  &  &  &  & \\
 &  & $T=100$ &  & {\color{blue}{$0.095$}} & {\color{blue}{$0.045$}} & {\color{blue}{$0.009$}} &  & $0.304$ & $0.207$ & $0.086$ &  & $0.426$ & $0.322$ & $0.168$ &  & $0.357$ & $0.256$ & $0.119$ &  & $0.127$ & $0.059$ & $0.016$ &  & $0.121$ & $0.074$ & $0.024$ & \\
 &  & $T=200$ &  & $0.096$ & $0.049$ & $0.009$ &  & $0.182$ & $0.109$ & $0.036$ &  & $0.275$ & $0.186$ & $0.075$ &  & $0.249$ & $0.165$ & $0.060$ &  & $0.241$ & $0.148$ & $0.037$ &  & $0.181$ & $0.111$ & $0.043$ & \\
 &  & $T=400$ &  & $0.100$ & $0.047$ & $0.008$ &  & $0.138$ & $0.078$ & $0.016$ &  & $0.240$ & $0.156$ & $0.054$ &  & $0.228$ & $0.147$ & $0.052$ &  & $0.493$ & $0.371$ & $0.170$ &  & $0.323$ & $0.226$ & $0.098$ & \\
 &  & $T=800$ &  & $0.101$ & $0.049$ & $0.009$ &  & $0.115$ & $0.060$ & $0.012$ &  & {\color{red}{$0.298$}} & {\color{red}{$0.198$}} & {\color{red}{$0.076$}} &  & {\color{red}{$0.306$}} & {\color{red}{$0.204$}} & {\color{red}{$0.080$}} &  & {\color{red}{$0.798$}} & {\color{red}{$0.700$}} & {\color{red}{$0.518$}} &  & {\color{red}{$0.598$}} & {\color{red}{$0.483$}} & {\color{red}{$0.283$}} & \\\hline\hline
\end{tabular}}
{\resizebox{24cm}{!}{\begin{minipage}{700pt}
\vspace{3mm}
{\underline{Note:}} As for Table \ref{sim_bd_nh}.
\end{minipage}}}
\end{center}
\end{table}
\end{landscape}
\clearpage

\begin{figure}[p]
 \caption{Size-adjusted power under the alternative hypothesis $H_{1}: \alpha_1=0.15$ and $\alpha_j=0$ ($j=2,\ldots,N$) when the $BD$ holds}\label{sim_bd_alt}
 \begin{center}
 \includegraphics[scale=0.5]{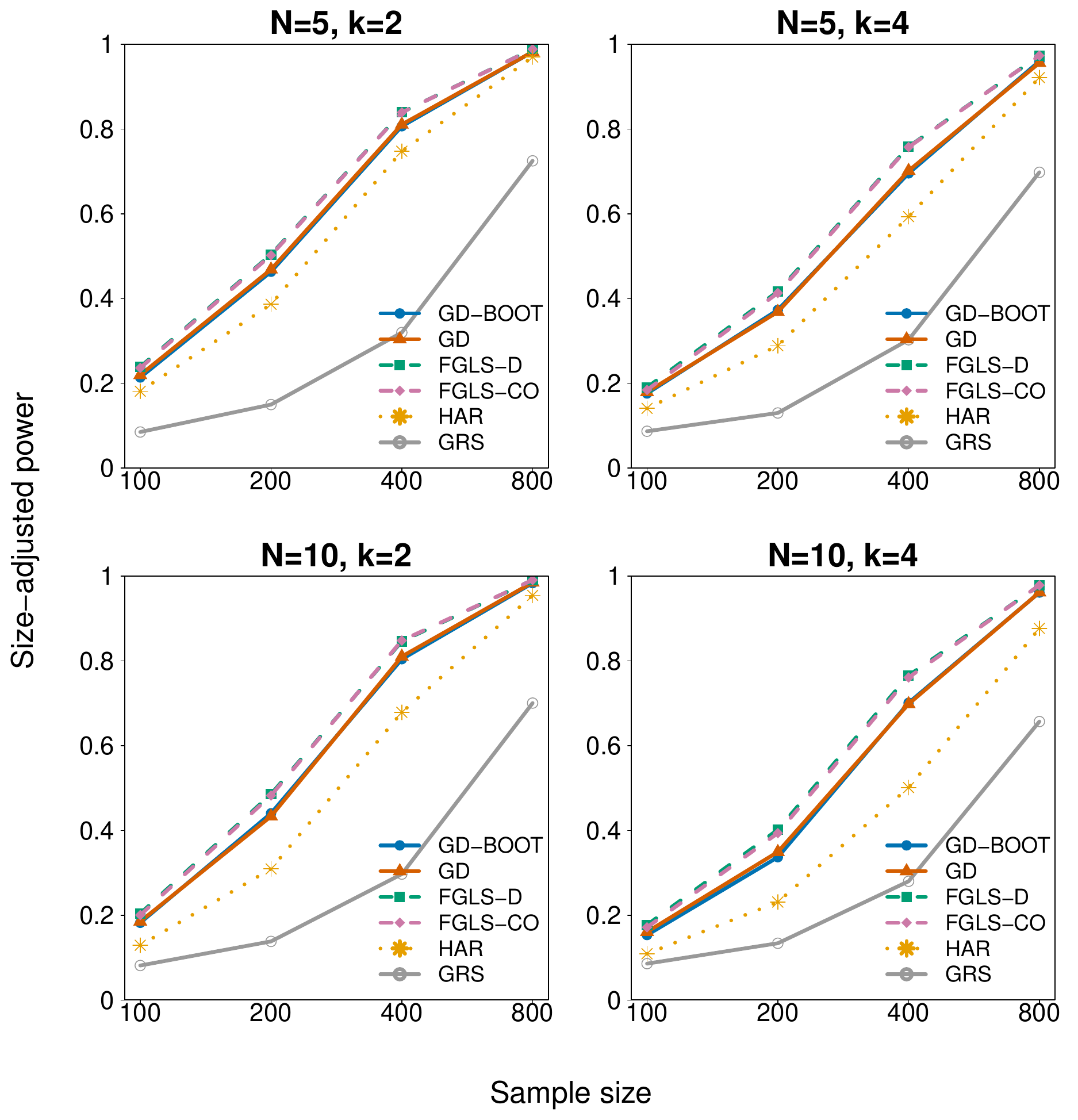}
{\resizebox{24cm}{!}{\begin{minipage}{700pt}
\vspace{3mm}
{\underline{Note:}} R version 4.5.3 was used to compute the statistics.
\end{minipage}}}
\end{center}
\end{figure}

\clearpage

\begin{figure}[p]
 \caption{Size-adjusted power under the alternative hypothesis $H_{1}: \alpha_1=0.15$ and $\alpha_j=0$ ($j=2,\ldots,N$) when the $GEXOG$ holds}\label{sim_gexog_alt}
 \begin{center}
 \includegraphics[scale=0.5]{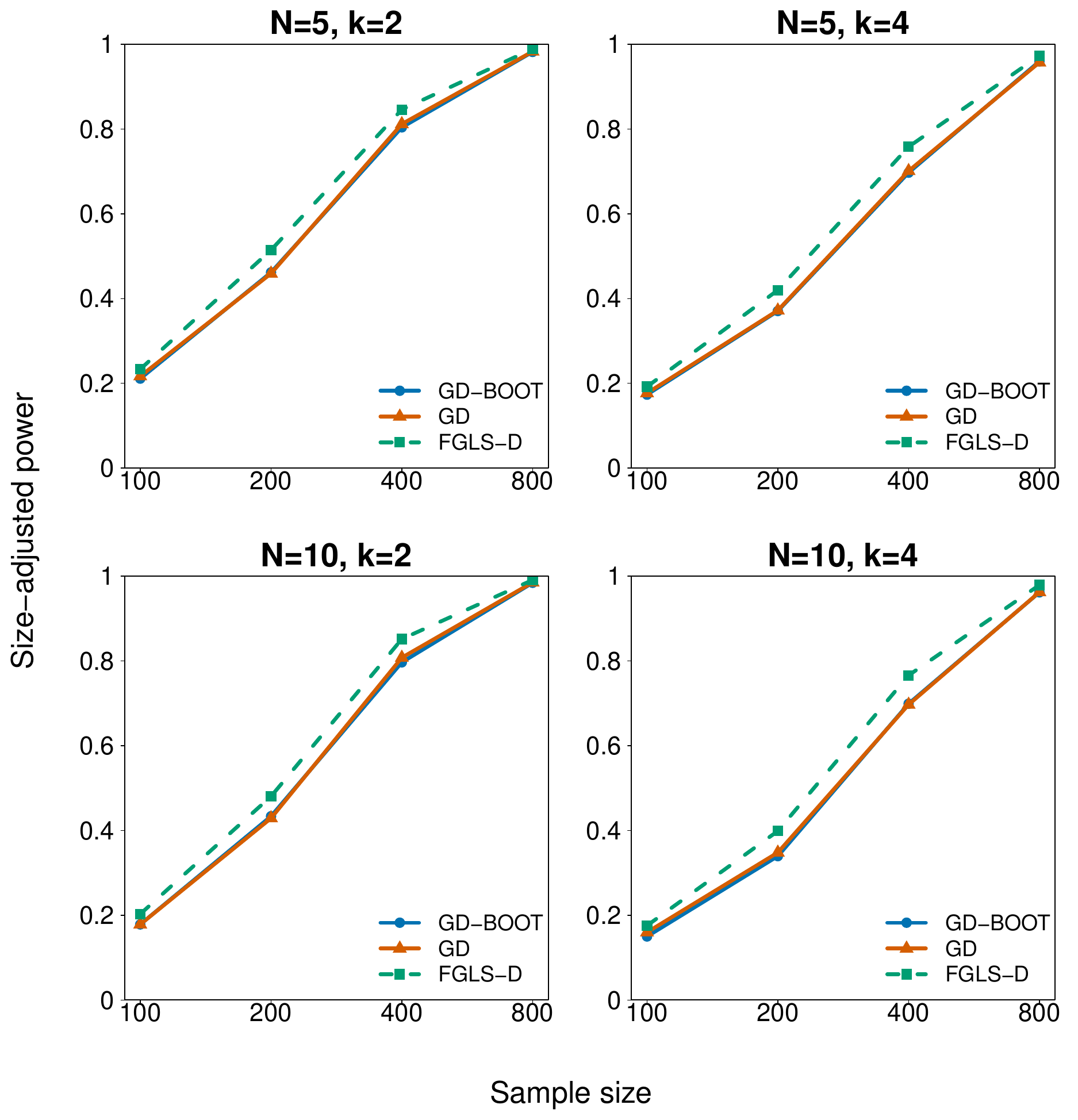}
{\resizebox{24cm}{!}{\begin{minipage}{700pt}
\vspace{3mm}
{\underline{Note:}} As for Figure \ref{sim_bd_alt}.
\end{minipage}}}
\end{center}
\end{figure}

\clearpage

\begin{figure}[p]
 \caption{Size-adjusted power under the alternative hypothesis $H_{1}: \alpha_1=0.15$ and $\alpha_j=0$ ($j=2,\ldots,N$) when the $EBD$ holds}\label{sim_ebd_alt}
 \begin{center}
 \includegraphics[scale=0.5]{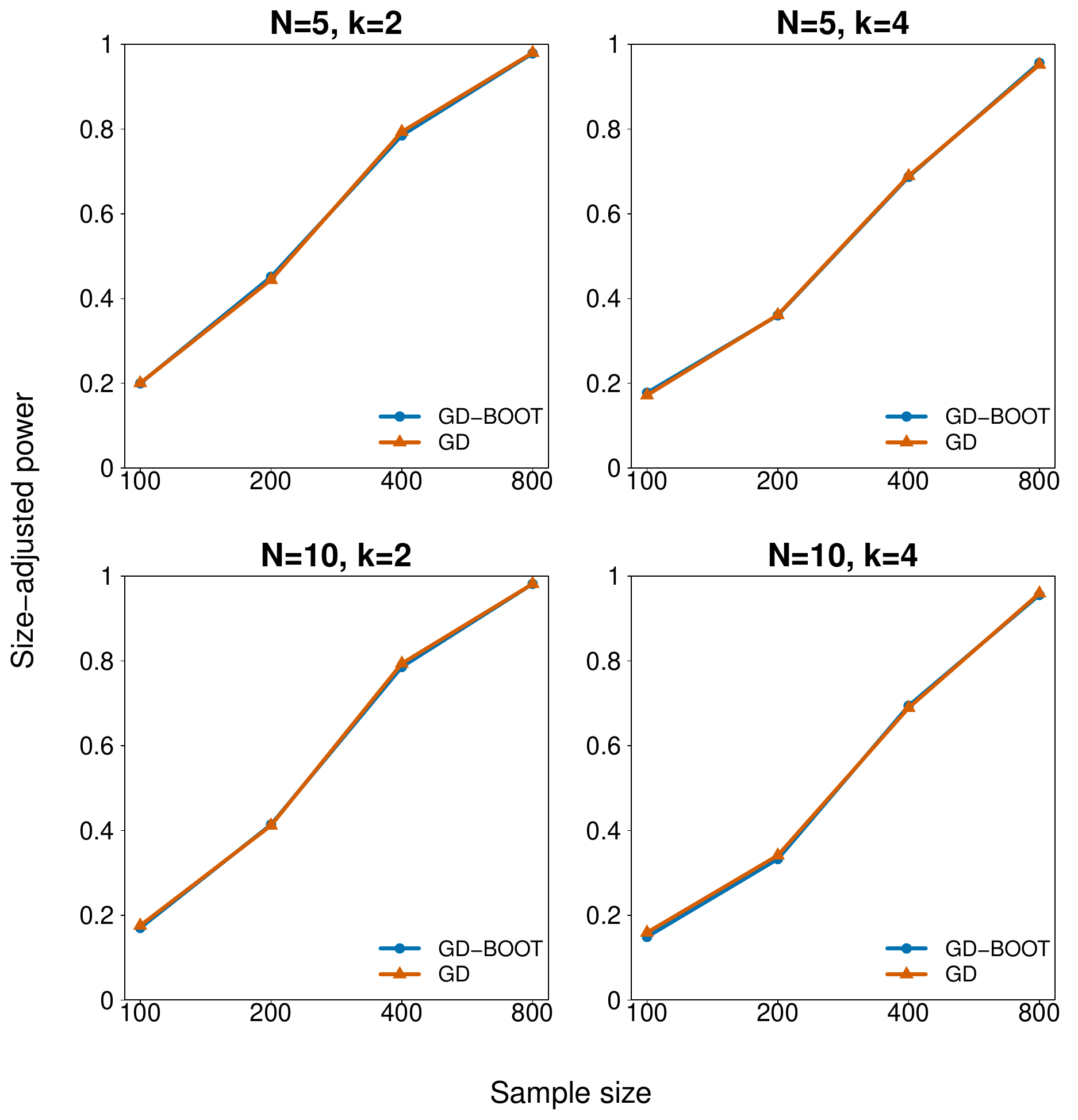}
{\resizebox{24cm}{!}{\begin{minipage}{700pt}
\vspace{3mm}
{\underline{Note:}} As for Figure \ref{sim_bd_alt}.
\end{minipage}}}
\end{center}
\end{figure}

\clearpage

\begin{table}[p]
\caption{Descriptive statistics and unit root tests}\label{fgls_grs_des}
\begin{center}\resizebox{15cm}{!}{
\begin{tabular}{lllccccccccccc}\hline\hline
 &  &  &  & Mean & SD & Min & Max &  & ADF & Lags &  & $T$ &  \\\cline{5-8}\cline{10-11}\cline{13-13}
 & FF3 &  &  &  &  &  &  &  &  &  &  &  &  \\
 &  & $R_{m}-R_{f}$ &  & $0.0103$ & $0.0463$ & $-0.1720$ & $0.1360$ &  & $-5.0487$ & $7$ &  & $207$ & \\
 &  & $SMB$ &  & $-0.0006$ & $0.0259$ & $-0.0593$ & $0.0714$ &  & $-14.9860$ & $0$ &  & $207$ & \\
 &  & $HML$ &  & $-0.0014$ & $0.0340$ & $-0.1383$ & $0.1286$ &  & $-12.1702$ & $0$ &  & $207$ & \\\hline
 & FF5 &  &  &  &  &  &  &  &  &  &  &  &  \\
 &  & $R_{m}-R_{f}$ &  & $0.0103$ & $0.0463$ & $-0.1720$ & $0.1358$ &  & $-5.0514$ & $7$ &  & $207$ & \\
 &  & $SMB$ &  & $-0.0008$ & $0.0277$ & $-0.0818$ & $0.0834$ &  & $-10.4452$ & $1$ &  & $207$ & \\
 &  & $HML$ &  & $-0.0014$ & $0.0340$ & $-0.1383$ & $0.1286$ &  & $-12.1702$ & $0$ &  & $207$ & \\
 &  & $WML$ &  & $0.0024$ & $0.0199$ & $-0.0521$ & $0.0719$ &  & $-12.1185$ & $0$ &  & $207$ & \\
 &  & $CMA$ &  & $0.0001$ & $0.0206$ & $-0.0708$ & $0.0773$ &  & $-12.7650$ & $0$ &  & $207$ & \\\hline\hline
\end{tabular}}
{\resizebox{14.5cm}{!}{\begin{minipage}{500pt}
\vspace{3mm}
{\underline{Notes:}}
 \begin{itemize}
  \item[(1)] ``$R_{m}-R_{f}$,'' ``$SMB$,'' ``$HML$,'' ``$RMW$,'' and ``$CMA$'' denote the returns on each risk factor, wich correspond to Fama-French multi-factor models.
  \item[(2)] ``ADF'' denotes the ADF test statistics and ``Lags'' denotes the lag order selected by the BIC.
  \item[(3)] In computing the ADF test, a model with a time trend and a constant is assumed. The critical value at the 5\% significance level for the ADF test is ``$-3.41$.''
  \item[(4)] R version 4.5.3 was used to compute the statistics.
 \end{itemize}
\end{minipage}}}%
\end{center}
\end{table}

\clearpage

\begin{table}[p]
\caption{$p$-values for test statistics under the null hypothesis $\bm{\alpha}_0=\mathbf{0}$ (FF3/FF5 models)}\label{fgls_grs_ff}
\begin{center}\resizebox{12cm}{!}{
\begin{tabular}{lllllcccccccc} \hline\hline
 &  &  & &  & $\widetilde{\mathcal{W}}^{GD}$ & & $\mathcal{W}^{GD}$ & & $\mathcal{W}^{FD}$ & & $\mathcal{W}^{FCO}$ &  \\\cline{6-12}
 & FF3 & &  &  &  &  &  &  &  &  &  &  \\
 &  & MB6 & &  & $0.03$ & & $0.01$ &  & $0.01$ & & $0.01$ & \\
 &  & MB25 & &  & $0.01$ & & $0.00$ &  & $0.00$ & & $0.00$ & \\\hline
 & FF5 &  & &  &  &  &  & & &  &  &  \\
 &  & MB6 & &  & $0.06$ & & $0.02$ &  & $0.02$ & & $0.02$ & \\
 &  & MB25 & &  & $0.01$ & & $0.00$ &  & $0.00$ & & $0.00$ & \\\hdashline
 &  & MI6 & &  & $0.73$ & & $0.67$ &  & $0.62$ & & $0.66$ & \\
 &  & MI25 & &  & $0.06$ & & $0.00$ &  & $0.00$ & & $0.00$ & \\\hdashline
 &  & MO6 & &  & $0.34$ & & $0.27$ &  & $0.22$ & & $0.25$ & \\
 &  & MO25 & &  & $0.19$ & & $0.00$ &  & $0.00$ & & $0.01$ & \\\hline\hline
\end{tabular}}
{\resizebox{11.5cm}{!}{\begin{minipage}{400pt}
\vspace{3mm}
{\underline{Notes:}}
 \begin{itemize}
  \item[(1)] ``$\widetilde{\mathcal{W}}^{GD}$'' denotes the bootstrap Wald test based on the GD estimator, while ``$\mathcal{W}^{GD}$,'' ``$\mathcal{W}^{FD}$,'' and ``$\mathcal{W}^{FCO}$'' denote the Wald tests based on the GD estimator, the FD estimator, and the FCO estimator, respectively.
  \item[(2)] ``MB,'' ``MO,'' and ``MI'' denotes the portfolios sorted by ``market capitalization (size of firms) and book-to-market ratio,'' ``market capitalization and profitability,'' and ``market capitalization and investment growth rate,'' respectively.
  \item[(3)] R version 4.5.3 was used to compute the statistics.
 \end{itemize}
\end{minipage}}}
\end{center}
\end{table}

\clearpage

%% file: durbin_main.bbl
\begin{thebibliography}{95}
\newcommand{\enquote}[1]{``#1''}
\expandafter\ifx\csname natexlab\endcsname\relax\def\natexlab#1{#1}\fi

\bibitem[{Affleck-Graves and McDonald(1989)}]{affleck1989nta}
Affleck-Graves, J. and McDonald, B. (1989), \enquote{Nonnormalities and Tests
  of Asset Pricing Theories,} \textit{Journal of Finance}, 44, 889--908.

\bibitem[{Andrews(1991)}]{andrews1991hac}
Andrews, D. W.~K. (1991), \enquote{Heteroskedasticity and Autocorrelation
  Consistent Covariance Matrix Estimation,} \textit{Econometrica}, 59,
  817--858.

\bibitem[{Asgharian et~al.(2023)Asgharian, Christiansen, and
  Hou}]{asgharian2023eus}
Asgharian, H., Christiansen, C., and Hou, A.~J. (2023), \enquote{The Effect of
  Uncertainty on Stock Market Volatility and Correlation,} \textit{Journal of
  Banking and Finance}, 154, 106929.

\bibitem[{Bae et~al.(2003)Bae, Karolyi, and Stulz}]{bae2003nam}
Bae, K.~H., Karolyi, G.~A., and Stulz, R.~M. (2003), \enquote{A New Approach to
  Measuring Financial Contagion,} \textit{Review of Financial Studies}, 16,
  717--763.

\bibitem[{Bai and Perron(1998)}]{bai1998etl}
Bai, J. and Perron, P. (1998), \enquote{Estimating and Testing Linear Models
  with Multiple Structural Changes,} \textit{Econometrica}, 66, 47--78.

\bibitem[{Bai and Perron(2003{\natexlab{a}})}]{bai2003cam}
--- (2003{\natexlab{a}}), \enquote{Computation and Analysis of Multiple
  Structural Change Models,} \textit{Journal of Applied Econometrics}, 18,
  1--22.

\bibitem[{Bai and Perron(2003{\natexlab{b}})}]{bai2003cvm}
--- (2003{\natexlab{b}}), \enquote{Critical Values for Multiple Structural
  Change Tests,} \textit{Econometrics Journal}, 6, 72--78.

\bibitem[{Baillie et~al.(2024)Baillie, Diebold, Kapetanios, Kim, and
  Mora}]{baillie2024rit}
Baillie, R.~T., Diebold, F.~X., Kapetanios, G., Kim, K.~H., and Mora, A.
  (2024), \enquote{On Robust Inference in Time-Series Regression,}
  \textit{Econometrics Journal}, 28, 131--173.

\bibitem[{Bali et~al.(2017)Bali, Brown, and Tang}]{bali2017ieu}
Bali, T.~G., Brown, S.~J., and Tang, Y. (2017), \enquote{Is Economic
  Uncertainty Priced in the Cross-Section of Stock Returns?} \textit{Journal of
  Financial Economics}, 126, 471--489.

\bibitem[{Bekaert et~al.(2014)Bekaert, Fratzscher, and Mehl}]{bekaert2014gce}
Bekaert, G. ad~Ehrmann, M., Fratzscher, M., and Mehl, A. (2014), \enquote{The
  Global Crisis and Equity Market Contagion,} \textit{Journal of Finance}, 69,
  2597--2649.

\bibitem[{Bewley(1983)}]{bewley1983trl}
Bewley, R.~A. (1983), \enquote{Tests of Restrictions in Large Demand Systems,}
  \textit{European Economic Review}, 20, 257--269.

\bibitem[{Boswijk and Urbain(1997)}]{boswijk1997lamt}
Boswijk, H.~P. and Urbain, J.~P. (1997), \enquote{Lagrance-Multiplier Tersts
  for Weak Exogeneity: A Synthesis,} \textit{Econometric Reviews}, 16, 21--38.

\bibitem[{Brown and Walker(1995)}]{brown1995sse}
Brown, B.~W. and Walker, M.~B. (1995), \enquote{Stochastic Specification and
  Estimation of Share Equation Systems,} \textit{Journal of Econometrics}, 66,
  175--205.

\bibitem[{B\"{u}hlmann(1997)}]{buhlmann1997sbt}
B\"{u}hlmann, P. (1997), \enquote{Sieve Bootstrap for Time Series,}
  \textit{Bernoulli}, 3, 123--148.

\bibitem[{Cakici et~al.(2013)Cakici, Fabozzi, and Tan}]{cakici2013svm}
Cakici, N., Fabozzi, F.~J., and Tan, S. (2013), \enquote{Size, Value, and
  Momentum in Emerging Market Stock Returns,} \textit{Emerging Markets Review},
  16, 46--65.

\bibitem[{Cattaneo et~al.(2019)Cattaneo, Jansson, and Ma}]{cattaneo2019tse}
Cattaneo, M.~D., Jansson, M., and Ma, X. (2019), \enquote{Two-Step Estimation
  and Inference with Possibly Many Included Covariates,} \textit{Review of
  Economic Studies}, 86, 1095--1122.

\bibitem[{Cieslak and Pang(2021)}]{cieslak2021css}
Cieslak, A. and Pang, H. (2021), \enquote{Common Shocks in Stocks and Bonds,}
  \textit{Journal of Financial Economics}, 142, 880--904.

\bibitem[{Cochrane and Orcutt(1949)}]{cochrane1949als}
Cochrane, D. and Orcutt, G. (1949), \enquote{Application of Least Squares
  Regression to Relationships Containing Auto-Correlated Error Terms,}
  \textit{Journal of the American Statistical Association}, 44, 32--61.

\bibitem[{Cont(2001)}]{cont2001epa}
Cont, R. (2001), \enquote{Empirical Properties of Asset Returns: Stylized Facts
  and Statistical Issues,} \textit{Quantitative Finance}, 1, 223--236.

\bibitem[{Cumby et~al.(1983)Cumby, Huizinga, and Obstfeld}]{cumby1983tst}
Cumby, R.~E., Huizinga, J., and Obstfeld, M. (1983), \enquote{Two-Step
  Two-Stage Least Squares Estimation in Models with Rational Expectations,}
  \textit{Journal of Econometrics}, 21, 333--355.

\bibitem[{Cushman and Zha(1997)}]{cushman1997imp}
Cushman, D.~O. and Zha, T. (1997), \enquote{Identifying Monetary Policy in a
  Small Open Economy under Flexible Exchange Rates,} \textit{Journal of
  Monetary Economics}, 39, 433--448.

\bibitem[{Davidson and MacKinnon(1999)}]{davidson1999sdb}
Davidson, R. and MacKinnon, J.~G. (1999), \enquote{The Size Distortion of
  Bootstrap Tests,} \textit{Econometric Theory}, 15, 361--376.

\bibitem[{Den~Haan and Levin(1997)}]{haan1997pgr}
Den~Haan, W.~J. and Levin, A. (1997), \enquote{A Practitioner's Guide to Robust
  Covariance Matrix Estimation,,} in \textit{Robust Inference}, Elsevier,
  vol.~15 of \textit{Handbook of Statistics}, chap.~12, pp. 299--342.

\bibitem[{Diebold and Yilmaz(2009)}]{diebold2009mfa}
Diebold, F.~X. and Yilmaz, K. (2009), \enquote{Measuring Financial Asset Return
  and Volatility Spillovers, with Application to Global Equity Markets,}
  \textit{Economic Journal}, 119, 158--171.

\bibitem[{Durbin(1970)}]{durbin1970tsc}
Durbin, J. (1970), \enquote{Testing for Serial Correlation in Least-Squares
  Regression when Some of the Regressorsare Lagged Dependent Variables,}
  \textit{Econometrica}, 38, 410--421.

\bibitem[{Engle et~al.(1983)Engle, Hendry, and Richard}]{engle1983exo}
Engle, R.~F., Hendry, D.~F., and Richard, J.~F. (1983), \enquote{Exogeneity,}
  \textit{Econometrica}, 51, 277--304.

\bibitem[{Fama and French(1993)}]{fama1993crf}
Fama, E.~F. and French, K.~R. (1993), \enquote{Common Risk Factors in the
  Returns on Stocks and Bonds,} \textit{Journal of Financial Economics}, 33,
  3--56.

\bibitem[{Fama and French(2012)}]{fama2012svm}
--- (2012), \enquote{Size, Value, and Momentum in International Stock Returns,}
  \textit{Journal of Financial Economics}, 105, 457--472.

\bibitem[{Fama and French(2015)}]{fama2015ffa}
--- (2015), \enquote{A Five-Factor Asset Pricing Model,} \textit{Journal of
  Financial Economics}, 116, 1--22.

\bibitem[{Fama and French(2016)}]{fama2016daf}
--- (2016), \enquote{Dissecting Anomalies with a Five-Factor Model,}
  \textit{Review of Financial Studies}, 29, 69--103.

\bibitem[{Fama and French(2017)}]{fama2017itf}
--- (2017), \enquote{International Tests of a Five-Factor Asset Pricing Model,}
  \textit{Journal of Financial Economics}, 123, 441--463.

\bibitem[{Fama and French(2018)}]{fama2018cf}
--- (2018), \enquote{Choosing Factors,} \textit{Journal of Financial
  Economics}, 128, 234--252.

\bibitem[{Fama and French(2020)}]{fama2020ccs}
--- (2020), \enquote{Comparing Cross-Section and Time-SeriesFactor Models,}
  \textit{Review of Financial Studies}, 33, 1891--1926.

\bibitem[{Fiebig(2001)}]{fiebig2001cte}
Fiebig, D.~G. (2001), \enquote{Seemingly Unrelated Regression,} in \textit{A
  Conpanion to Theoretical Econometrics}, ed. Baltagi, B.~H., Blackwell,
  Massachusetts, chap.~5.

\bibitem[{Gibbons et~al.(1989)Gibbons, Ross, and Shanken}]{gibbons1989teg}
Gibbons, M.~R., Ross, S.~A., and Shanken, J. (1989), \enquote{A Test of the
  Efficiency of a Given Portfolio,} \textit{Econometrica}, 57, 1121--1152.

\bibitem[{Gungor and Luger(2013)}]{gungor2013tlf}
Gungor, S. and Luger, R. (2013), \enquote{Testing Linear Factor Pricing Models
  with Large Cross Sections: A Distribution-Free Approach,} \textit{Journal of
  Business \& Economic Statistics}, 31, 66--77.

\bibitem[{Haavelmo(1943)}]{haavelmo1943sis}
Haavelmo, T. (1943), \enquote{The Statistical Implications of a System of
  Simultaneous Equations,} \textit{Econometrica}, 11, 1--12.

\bibitem[{Hall and Horowitz(1996)}]{hall1996bcv}
Hall, P. and Horowitz, J.~L. (1996), \enquote{Bootstrap Critical Values for
  Tests Based on Generalized-Method-of-Moments Estimators,}
  \textit{Econometrica}, 64, 891--916.

\bibitem[{Hansen(1982)}]{hansen1982lsp}
Hansen, L.~P. (1982), \enquote{{Large Sample Properties of Generalized Method
  of Moments Estimators},} \textit{Econometrica}, 50, 1029--1054.

\bibitem[{Hansen and Sargent(1982)}]{hansen1982ivp}
Hansen, L.~P. and Sargent, T.~J. (1982), \enquote{Instrumental Variables
  Procedures for Estimating Linear Rational Expectations Models,}
  \textit{Journal of Monetary Economics}, 9, 263--296.

\bibitem[{Harvey and Siddique(2002)}]{harvey2002csa}
Harvey, C.~R. and Siddique, A. (2002), \enquote{Conditional Skewness in Asset
  Pricing Tests,} \textit{Journal of Finance}, 55, 1263--1295.

\bibitem[{Hayashi and Sims(1983)}]{hayashi1983nee}
Hayashi, F. and Sims, C. (1983), \enquote{Nearly Efficient Estimation of Time
  Series Models with Predetermined, but notExogenous, Instruments,}
  \textit{Econometrica}, 51, 783--798.

\bibitem[{Henley and Peirson(1994)}]{henley1994tue}
Henley, A. and Peirson, J. (1994), \enquote{Time-of-Use Electricity Pricing,}
  \textit{Economics Letters}, 45, 421--426.

\bibitem[{Huang et~al.(2012)Huang, Liu, Rhee, and Wu}]{huang2012edr}
Huang, W., Liu, Q., Rhee, G., and Wu, F. (2012), \enquote{Extreme Downside Risk
  and Expected Stock Returns,} \textit{Journal of Banking \& Finance}, 36,
  1492--1502.

\bibitem[{Hwang and Vald\'{e}s(2023)}]{hwang2023fsc}
Hwang, J. and Vald\'{e}s, G. (2023), \enquote{Finite-Sample Corrected Inference
  for Two-Step GMM in Time Series,} \textit{JournalofEconometrics}, 234,
  327--352.

\bibitem[{Jalloul and Miescu(2023)}]{jalloul2023emc}
Jalloul, M. and Miescu, M. (2023), \enquote{Equity Market Connectedness across
  Regimes of Geopolitical Risks: Historical Evidence and Theory,}
  \textit{Journal of International Money and Finance}, 137, 102910.

\bibitem[{Jansson(2004)}]{jansson2004erp}
Jansson, M. (2004), \enquote{The Error in Rejection Probability of Simple
  Autocorrelation Robust Tests,} \textit{Econometrica}, 72, 937--946.

\bibitem[{Kamstra and Shi(2024)}]{kamstra2024tra}
Kamstra, M.~J. and Shi, R. (2024), \enquote{Testing and Ranking of Asset
  Pricing Models Using the GRS Statistic,} \textit{Journal of Risk and
  Financial Management}, 17, 168.

\bibitem[{Kiefer and Vogelsang(2002)}]{kiefer2002har}
Kiefer, N.~M. and Vogelsang, T.~J. (2002),
  \enquote{Heteroskedasticity-Autocorrelation Robust Standard Errors Using the
  Bartlett Kernel without Truncation,} \textit{Econometrica}, 70, 2093--2095.

\bibitem[{Kiefer and Vogelsang(2005)}]{kiefer2005nat}
--- (2005), \enquote{A New Asymptotic Theory for
  Heteroskedasticity-Autocorrelation Robust Tests,} \textit{Econometric
  Theory}, 21, 1130--1164.

\bibitem[{Kiefer et~al.(2000)Kiefer, Vogelsang, and Bunzel}]{kiefer2000srt}
Kiefer, N.~M., Vogelsang, T.~J., and Bunzel, H. (2000), \enquote{Simple Robust
  Testing of Regression Hypotheses,} \textit{Econometrica}, 68, 695--714.

\bibitem[{Koopmans(1945)}]{koopmans1945ses}
Koopmans, T. (1945), \enquote{Statistical Estimation of Simultaneous Economic
  Relations,} \textit{Journal of the American Statistical Association}, 40,
  448--466.

\bibitem[{Koreisha and Fang(2001)}]{koreisha2001gls}
Koreisha, S. and Fang, Y. (2001), \enquote{Generalized Least Squares with
  Misspecified Serial Correlation Structures,} \textit{Journal of the Royal
  Statistical Society}, 63, 515--531.

\bibitem[{Kumbhakar and Heshmati(1996)}]{kumbhakar1996tct}
Kumbhakar, S.~C. and Heshmati, A. (1996), \enquote{Technical Change and Total
  Factor Productivity Growth in Swedish Manufacturing Industries,}
  \textit{Econometric Reviews}, 15, 275--298.

\bibitem[{Laitinen(1978)}]{laitinen1978wdh}
Laitinen, K. (1978), \enquote{Why is Demand Homogeneity So often Rejected?}
  \textit{Economics Letters}, 1, 187--191.

\bibitem[{Lazarus et~al.(2021)Lazarus, Lewis, and Stock}]{lazarus2021spt}
Lazarus, E., Lewis, D.~J., and Stock, J.~H. (2021), \enquote{The Size-Power
  Tradeoff in HAR Inference,} \textit{Econometrica}, 89, 2497--2516.

\bibitem[{Lazarus et~al.(2018)Lazarus, Lewis, Stock, and
  Watson}]{lazarus2018hir}
Lazarus, E., Lewis, D.~J., Stock, J.~H., and Watson, M.~W. (2018), \enquote{HAR
  Inference: Recommendations for Practice,} \textit{Journal of Business \&
  Economic Statistics}, 36, 541--559.

\bibitem[{Lehkonen(2015)}]{lehkonen2015smi}
Lehkonen, H. (2015), \enquote{Stock Market Integration and the GlobalFinancial
  Crisis,} \textit{Review of Finance}, 19, 2039--2094.

\bibitem[{Lewellen et~al.(2010)Lewellen, Nagel, and Shanken}]{lewellen2010asa}
Lewellen, J., Nagel, S., and Shanken, J. (2010), \enquote{A Skeptical Appraisal
  of Asset Pricing Tests,} \textit{Journal of Financial Economics}, 96,
  175--194.

\bibitem[{Lintner(1965)}]{lintner1965vra}
Lintner, J. (1965), \enquote{The Valuation of Risky Assets and the Selection of
  Risky Investments in Stock Portfolios and Budget Constraints,} \textit{Review
  of Economics and Statistics}, 47, 13--37.

\bibitem[{Longin and Solnik(2001)}]{longin2001eci}
Longin, F. and Solnik, B. (2001), \enquote{Extreme Correlation of International
  Equity Markets,} \textit{Journal of Finance}, 56, 649--676.

\bibitem[{L{\"u}tkepohl(2005)}]{lutkepohl2005nim}
L{\"u}tkepohl, H. (2005), \textit{New Introduction to Multiple Time Series
  Analysis}, Springer, Berlin, Germany.

\bibitem[{MacKinnon(2006)}]{mackinnon2006bme}
MacKinnon, J.~D. (2006), \enquote{Bootstrap Methods in Econometrics,}
  \textit{The Economic Record}, 82, S2--S18.

\bibitem[{Mayer(2020)}]{mayer2020cts}
Mayer, A. (2020), \enquote{(Consistently) Testing Strict Exogeneity against the
  Alternative of Predeterminedness in Linear Time-Series Models,}
  \textit{Economics Letters}, 193, 109335.

\bibitem[{Mikusheva and Solvsten(2025)}]{mikusheva2025lrw}
Mikusheva, A. and Solvsten, M. (2025), \enquote{Linear Regression with Weak
  Exogeneity,} \textit{Quantitative Economics}, 16, 367--403.

\bibitem[{Moriya and Noda(2024)}]{moriya2024tif}
Moriya, K. and Noda, A. (2024), \enquote{Time Instability of the Fama-French
  Multifactor Models: An International Evidence,} [arXiv:2208.01270], Available
  at https://arxiv.org/abs/2208.01270.

\bibitem[{M\"{u}ller(2014)}]{muller2014hcs}
M\"{u}ller, U. (2014), \enquote{HAC Corrections for Strongly AutocorrelatedTime
  Series,} \textit{Journal of Business \& Economic Statistics}, 32, 311--322.

\bibitem[{Murphy and Topel(1985)}]{murphy1985eit}
Murphy, K.~M. and Topel, R.~H. (1985), \enquote{Estimation and Inference in
  Two-Step Econometric Models,} \textit{Journal of Business \& Economic
  Statistics}, 3, 370--379.

\bibitem[{Nagakura(2024)}]{nagakura2024cot}
Nagakura, D. (2024), \enquote{Cochrane--Orcutt Type Estimator for Multivariate
  Linear Regression Model with Serially Correlated Errors,} Available at SSRN:
  https://ssrn.com/abstract=4951695.

\bibitem[{Ndako et~al.(2025)Ndako, Kumeka, Adedoyin, and Asongu}]{ndako2025sbg}
Ndako, J.~A., Kumeka, T.~T., Adedoyin, F.~F., and Asongu, S, A. (2025),
  \enquote{Structural Breaks in Global Stock Markets: Are They Caused by
  Pandemics, Protests or Other Factors?} \textit{Transnational Corporations
  Review}, 17, 200147.

\bibitem[{Newey(1984)}]{newey1984mmi}
Newey, W.~K. (1984), \enquote{A Method of Moments Interpretation of Sequential
  Estimators,} \textit{Economics Letters}, 14, 201--206.

\bibitem[{Newey and West(1987)}]{newey1987sps}
Newey, W.~K. and West, K.~D. (1987), \enquote{A Simple, Positive Semi-Definite,
  Heteroskedasticity and Autocorrelation Consistent Covariance Matrix,}
  \textit{Econometrica}, 55, 703--708.

\bibitem[{Newey and West(1994)}]{newey1994als}
--- (1994), \enquote{Automatic Lag Selection in Covariance Matrix Estimation,}
  \textit{Review of Economic Studies}, 61, 631--653.

\bibitem[{Pagan(1986)}]{pagan1986sre}
Pagan, A. (1986), \enquote{Two Stage and Related Estimators and Their
  Applications,} \textit{Review of Economic Studies}, 53, 517--538.

\bibitem[{Perron and Gonz\'{a}lez-Coya(2026)}]{perron2026fgls}
Perron, P. and Gonz\'{a}lez-Coya, E. (2026), \enquote{Feasible GLS for Time
  Series Regression,} Working Paper. Department of Economics, Boston
  University.

\bibitem[{Phillips et~al.(2006)Phillips, Sun, and Jin}]{phillips2006sde}
Phillips, P. C.~B., Sun, Y., and Jin, S. (2006), \enquote{Spectral Density
  Estimation and Robust Hypothesis Testing Using Steep Origin Kernels without
  Truncation,} \textit{International Economic Review}, 47, 837--894.

\bibitem[{Phillips et~al.(2007)Phillips, Sun, and Jin}]{phillips2007lrv}
--- (2007), \enquote{Long Run Variance Estimation and Robust Regression Testing
  Using Sharp Origin Kernels with No Truncation,} \textit{Journal of
  Statistical Planning and Inference}, 137, 985--1023.

\bibitem[{Ray and Savin(2008)}]{ray2008pha}
Ray, S. and Savin, N.~E. (2008), \enquote{The Performance of Heteroskedasticity
  and Autocorrelation Robust Tests: A Monte Carlo Study with an Application to
  the Three-Factor Fama: French Asset-Pricing Model,} \textit{Journal of
  Applied Econometrics}, 23, 91--109.

\bibitem[{Ray et~al.(2009)Ray, Savin, and Tiwari}]{ray2009tcr}
Ray, S., Savin, N.~E., and Tiwari, A. (2009), \enquote{Testing the CAPM
  Revisited,} \textit{Journal of Empirical Finance}, 16, 721--733.

\bibitem[{Ross(1976)}]{ross1976atc}
Ross, S. (1976), \enquote{The Arbitrage Theory of Capital Asset Pricing,}
  \textit{Journal of Economic Theory}, 13, 341--360.

\bibitem[{Said and Dickey(1984)}]{said1984tur}
Said, S. and Dickey, D. (1984), \enquote{Testing for Unit Roots in
  Autoregressive-Moving Average Models of Unknown Order,} \textit{Biometrika},
  71, 599--607.

\bibitem[{Sarafidis and Wansbeek(2012)}]{sarafidis2012csd}
Sarafidis, V. and Wansbeek, T. (2012), \enquote{Cross-Sectional Dependence in
  Panel Data Analysis,} \textit{Econometric Reviews}, 31, 483--531.

\bibitem[{Schwarz(1978)}]{schwarz1978edm}
Schwarz, G. (1978), \enquote{Estimating the Dimension of a Model,}
  \textit{Annals of Statistics}, 6, 461--464.

\bibitem[{Sharpe(1964)}]{sharpe1964cap}
Sharpe, W.~F. (1964), \enquote{Capital Asset Prices: A Theory of Market
  Equilibrium under Conditions of Risk,} \textit{Journal of Finance}, 19,
  425--442.

\bibitem[{Sims(1980)}]{sims1980mr}
Sims, C.~A. (1980), \enquote{Macroeconomics and Reality,}
  \textit{Econometrica}, 48, 1--48.

\bibitem[{Stock and Watson(2001)}]{stock2001var}
Stock, J.~H. and Watson, M.~W. (2001), \enquote{Vector Autoregressions,}
  \textit{Journal of Economic Perspectives}, 15, 101--115.

\bibitem[{Stock and Watson(2019)}]{stock2019ite}
--- (2019), \textit{Introduction to Econometrics}, Pearson, fourth edition ed.

\bibitem[{Sun(2014)}]{sun2014lfa}
Sun, Y. (2014), \enquote{Let's Fix It: Fixed-b Aymptotics versus Small-b
  Asymptotics in Heteroskedasticity and Autocorrelation Robust Inference,}
  \textit{Journal of Econometrics}, 178, 659--677.

\bibitem[{Sun et~al.(2008)Sun, Phillips, and Jin}]{sun2008obs}
Sun, Y., Phillips, P. C.~B., and Jin, S. (2008), \enquote{Optimal Bandwidth
  Selection in Heteroskedasticity-Autocorrelation Robust Testing,}
  \textit{Econometrica}, 76, 175--194.

\bibitem[{Umar et~al.(2022)Umar, Polat, Choi, and Teplova}]{umar2022iru}
Umar, Z., Polat, O., Choi, S.~Y., and Teplova, T. (2022), \enquote{The Impact
  of the Russia-Ukraine Conflict on the Connectedness Offinancial Markets,}
  \textit{Finance Research Letters}, 48, 102976.

\bibitem[{Wagner and Hong(2020)}]{wagner2020fmo}
Wagner, M .and~Grabarczyk, P. and Hong, S.~H. (2020), \enquote{Fully Modified
  OLS Estimation and Inference for Seemingly Unrelated Cointegrating Polynomial
  Regressions and the Environmental Kuznets Curve for Carbondioxide Emissions,}
  \textit{Journal of Econometrics}, 214, 216--255.

\bibitem[{Zellner(1962)}]{zellner1962eme}
Zellner, A. (1962), \enquote{An Efficient Method of Estimating Seemingly
  Unrelated Regressions and Tests forAggregation Bias,} \textit{Journal of the
  American Statistical Association}, 57, 348--368.

\bibitem[{Zha(1999)}]{zha1999brs}
Zha, T. (1999), \enquote{Block Recursion and Structural Vector
  Autoregressions,} \textit{Journal of Econometrics}, 90, 291--316.

\bibitem[{Zhang et~al.(2020)Zhang, Hu, and Ji}]{zhang2020fmg}
Zhang, D., Hu, M., and Ji, Q. (2020), \enquote{Financial Markets under the
  Global Pandemic of COVID-19,} \textit{Finance Research Letters}, 36, 101528.

\bibitem[{Zhou(1993)}]{zhou1993apt}
Zhou, G. (1993), \enquote{Asset-Pricing Tests under Alternative Distributions,}
  \textit{Journal of Finance}, 48, 1927--1942.

\end{thebibliography}
